\documentclass[a4paper,11pt]{article}
\pdfoutput=1
\usepackage{jcappub}
\usepackage[T1]{fontenc}
\usepackage{physics}
\usepackage{graphicx}
\usepackage{subcaption}
\usepackage{color}

\newcommand{\comment}[1]{}

\title{Exploring  circular polarization in the CMB due to conventional sources of cosmic birefringence}

\author{Paulo Montero-Camacho}
\author{and Christopher M. Hirata}
\affiliation{Center for Cosmology and AstroParticle Physics, The Ohio State University, 191 West Woodruff Lane, Columbus, Ohio 43210, USA}
\emailAdd{monterocamacho.1@osu.edu}
\emailAdd{hirata.10@osu.edu}

\abstract{The circular polarization of the cosmic microwave background (CMB) is usually taken to be zero since it is not generated by Thomson scattering. Here we explore the actual level of circular polarization in the CMB generated by conventional cosmological sources of birefringence. We consider two classes of mechanisms for birefringence. One is alignment of the matter to produce an anisotropic susceptibility tensor: the hydrogen spins can be aligned either by density perturbations or CMB anisotropies themselves. The other is anisotropy of the radiation field coupled to the non-linear response of the medium to electromagnetic fields: this can occur either via photon-photon scattering (non-linear response of the vacuum); atomic hyperpolarizability (non-linear response of neutral atoms); or plasma delay (non-linear response of free electrons). The strongest effect comes from photon-photon scattering from recombination at a level of $ \sim 10^{-14} \ \textup{K}$. Our results are consistent with a negligible circular polarization of the CMB in comparison with the linear polarization or the sensitivity of current and near-term experiments.}


\newcommand{\lambdabar}{{\mkern0.75mu\mathchar '26\mkern -9.75mu\lambda}}

\begin{document}
\maketitle
\flushbottom

\section{Introduction}
\label{sec:intro}

The past two decades have seen enormous progress in cosmology thanks to rapid advances in the observational data. Most of our knowledge of the early Universe comes from the cosmic microwave background (CMB), including in particular the temperature and linear polarization anisotropies. Using these observations cosmologists have been able to constrain inflationary models, tightly constrain the composition of the Universe (e.g.\ the ratio of photons, baryons, dark matter, and neutrinos), pin down the epoch of reionization, realize that our Universe is very close to spatially flat, and even explore the late Universe via secondary anisotropies. However, the CMB anisotropies, while enormously useful, suffer from the limitation of cosmic variance: with only one sky, we can access a finite number of independent modes in the CMB, and thus there is a fundamental minimum statistical error \cite{Durrer}. The {\em Planck} mission has already reached the cosmic variance limit in temperature over a wide range of angular scales \cite{ade2016planck}.

The limitation posed by cosmic variance might be reduced by exploring other aspects of the CMB; these also often carry not only {\em more} information, but information on otherwise inaccessible physics. For instance, one could examine the spectral distortions of the CMB, i.e.\ the deviations from a perfect blackbody spectrum. These encode information on the thermal history of the Universe -- e.g.\ energy injection from epochs well before last scattering, which would be otherwise invisible -- as well as very small-scale perturbations that are Silk-damped in the CMB anisotropies and destroyed by non-linear evolution in galaxy surveys \cite{Chluba}. Today, the ``definitive'' constraint on spectral distortions is from the COBE/FIRAS experiment, however proposed experiments using modern technology (such as PIXIE \cite{kogut2011primordial} and PRISM \cite{andre2014prism}) could make dramatic improvements. One could also consider sources of frequency-dependent anisotropy. For example, Rayleigh scattering from neutral atoms leads to small frequency dependent distortions to the CMB power spectrum \cite{yu2001rayleigh,lewis2013rayleigh}. In terms of observations,  future experiments like PRISM and PIXIE may be able to detect this signal \cite{alipour2015effects}. Another idea is to use the time evolution of the CMB anisotropies \cite{LangeandPage}, which probes the radial direction at the surface of last scattering instead of giving simply a 2D slice.

This paper considers the circular polarization of the CMB. In radiative transfer problems, circular polarization is often the result of a two-step process: first, linear polarization is generated (e.g.\ by selective absorption or emission, or scattering), and then a phase delay is induced between the $x$ and $y$ axes (e.g.\ passage through a birefringent medium; reflection off a tilted surface). In this case, circular polarization traces the overall geometry of the setup as well as properties of the medium. Examples include circular polarization of starlight \cite{Martin} or of diffuse radiation in star-forming regions \cite{Kwon}, and the circular polarization observed from solar system planets \cite{swedlund1972circular, kemp1971circular}. An alternative source of circular polarization (relevant for both foregrounds and new physics) involves magnetized media, where the intrinsic emission can be circularly polarized due to the preferred handedness of particle trajectories, as occurs in masers \cite{Deguchi, watson2001relationship}, GRB afterglows \cite{Wiersema, matsumiya2003circular}, and (we expect) in the diffuse synchrotron emission from our Galaxy. In AGN jets, {\em both} mechanisms are considered as candidates as a wide range of physical conditions and observed polarization properties can occur \cite{Brunthaler, Homan, homan2001parsec, beckert2002circular}. Finally, there is ongoing discussion on the origin of circularly polarized radio emission from pulsars \cite{melrose2003causes, karastergiou2003v,mitra2009unraveling}. The circular polarization of the CMB is usually assumed to be zero in the context of standard cosmology, though it could be a channel for new physics. 

The interest in the circular polarization of the CMB has been steadily growing over the recent past. This rising enthusiasm in circular polarization has also motivated the search and study of new possible sources of circular polarization in the CMB, e.g.\ Zarei et~al.\ \cite{Zarei} where the authors show that the CMB picks up a small circular polarization using a background magnetic field and by considering physics outside of the standard model. Mohammadi \cite{Mohammadi} argues that CMB photons can acquire circular polarization because the anisotropies of the cosmic neutrino background acts as a birefringent medium, although we do not expect such a process from conventional physics.\footnote{
The problem of a photon beam ($A$) experiencing birefringence by interacting with a neutrino beam ($B$) is similar in concept to the photon-photon scattering problem treated in \S\ref{sec:photon}, except that a $W$ and a charged lepton appear in the loop \cite{Mohammadi}. Without loss of generality, the interaction can be treated in the center-of-mass frame, where beams $A$ and $B$ are collinear. Under the little group of rotations that fix the momentum vectors, the birefringence terms that convert linear to circular polarization in beam $A$ ($n_Q$ and $n_U$ in the notation of our \S\ref{sec:circular}) are spin 2. If beam $B$ consists of spin-$s$ particles, its density matrix contains components of spin up to $2s$; to have the needed spin 2 component, the spin of beam $B$ must be at least 1. Thus consideration of the symmetry group allows circular polarization to be sourced when beam $B$ is a photon beam (spin 1), but not when $B$ is a neutrino beam (spin $\frac12$). This argument remains true regardless of the neutrino masses and PMNS matrix, and whether the neutrino has a Majorana or Dirac mass term. The linear-to-circular conversion from an anisotropic neutrino background is thus not expected in conventional physics.
} De \& Tashiro \cite{De} suggest that the CMB photons could become circularly polarized from propagating through a magnetized plasma by means of Faraday conversion. King \& Lubin\ \cite{King} explore additional sources of circular polarization in the CMB and conclude that Population III stars are the strongest source. They also discuss the detectability of this signal. In addition, there has been considerable recent effort directed towards observation of circular polarization; see Refs.~\cite{Mainini, nagy2017new} for recent upper limits on the CMB circular polarization. 

Here we explore the possibility of circular polarization in the CMB due to cosmic birefringence at high redshift. (Astrophysical mechanisms and foregrounds at low redshift are outside the main scope of this paper, though we do comment on them briefly.) We restrict ourselves to conventional physics in the standard cosmological scenario. We focus on order-of-magnitude estimates because our main goal is to identify if an effect is strong enough to require further examination; nevertheless, we keep factors of $2$, $\pi$, etc.\ in the derivation of the indices of refraction because such effects can ``multiply out'' to give large factors even if these are in principle order unity. We study different possible sources for cosmic birefringence and estimate their respective levels of circular polarization near recombination.

Conversion of linear to circular polarization can occur in a medium where the two axes ($x$ and $y$) have different indices of refraction. There are two classes of ways this can occur in cosmology. One is a medium where the structure of the matter has a preferred axis; in the gas phase, this would come from alignment of the atomic spins by an external radiation field. We will consider this mechanism both during the recombination epoch (where the dominant alignment comes from the CMB anisotropies acting on excited hydrogen atoms through the H$\alpha$ line) and in the epoch of Cosmic Dawn (where the alignment comes from scattering of 21 cm radiation). The other involves the fact that the response of a medium to electromagnetic fields is not linear. In non-linear electrodynamics, the presence of an anisotropic radiation background\footnote{An external magnetic field with net alignment would also do -- it involves the same couplings but is a DC rather than AC field. This is not present in the standard picture of the early Universe, but is relevant to secondary sources of circular polarization and foregrounds.} makes the medium birefringent. We consider three sources of non-linearity: the non-linearity induced by the ionized plasma component (where the non-linearity arises from the finite displacements of the electrons); the non-linearity induced by the atoms (where the non-linearity arises from the fact that the hydrogen atom potential is not a harmonic oscillator); and the non-linearity of the vacuum (photon-photon scattering, where the non-linearity arises from virtual electron-positron pairs).

Of these candidate primordial mechanisms, we find that photon-photon scattering at recombination produces the strongest circular polarization. This mechanism has received the greatest attention in the recent past \cite{motie2012euler, sawyer2015photon, ejlli2016magneto,sadegh}, although we have had to correct some of the calculations in the literature.

This paper is organized as follows. We start by studying some general facts of circular polarization in the CMB in \S\ref{sec:circular}. We then proceed to consider each effect in turn; in every case, we make an order of magnitude estimate and then proceed to a more detailed calculation. We examine the effect of birefringence due to spin-polarization of the hydrogen atoms in \S\ref{sec:21z20} (for the Cosmic Dawn epoch) and \S\ref{sec:21z1100} (for the recombination epoch). We consider photon-photon scattering at recombination in \S\ref{sec:photon}. In \S\ref{sec:nonlinear}, we estimate the cosmic birefringence due to the static non-linear polarizability of hydrogen.  Then we proceed to explore the birefringence produced by plasma delay in \S\ref{sec:plasma}. We conclude in \S\ref{sec:conclusion}. We use SI units throughout.

Unless stated otherwise\footnote{The line of sight computation for photon-photon scattering in \S\ref{sec:photon} relies on an older version of the \textit{Planck} cosmology.} throughout this paper  we used the (Plick) \textit{Planck} cosmology \cite{ade2016planck}. Specifically, $H_0 = 67.26$, $\Omega_{\rm b} h^2 = 0.02222$, $\Omega_{\rm cdm} h^2 = 0.1199 $, $\Omega_{\rm m} = 0.316 $ and $z_{\rm re} = 8.8$.

\section{General aspects of circular polarization}
\label{sec:circular}

In this section, we explore some aspects of circular polarization in the CMB that are common to all of the mechanisms in this paper.

Most CMB polarization is produced by Thomson scattering, which interconverts quadrupolar anisotropy and linear polarization. However, symmetry considerations prevent Thomson scattering from converting either temperature or linear polarization perturbations into circular polarization; this would have to come from propagation effects. Here the effect of interest is birefringence. The general description of birefringence, for light propagating on the $z$-axis in a medium of low density, is to introduce an index of refraction tensor, 
\begin{equation}
n_{ij} = \delta_{ij} + \frac12(\chi_{e,ij} + \chi_{m,ij}),
\end{equation}
where $\chi_e$ and $\chi_m$ are the electric and magnetic susceptibilities respectively and only the $x$ and $y$ components of the tensor are considered (there is no longitudinal polarization). This can be decomposed as
\begin{equation}
n = \left( \begin{array}{cc} n_I+n_Q & n_U+in_V \\ n_U-in_V & n_I-n_Q \end{array} \right).
\end{equation}
Here $n_Q$ is (half of) the difference of indices of refraction on the $x$ and $y$-axes, and $n_U$ represents the half-difference of indices of refraction on the two diagonal axes. The component $n_I$ is the polarization-averaged index of refraction, which is not of interest as it induces no phase shift. Finally, $n_V$ is the difference in indices of refraction between the two circular polarizations; it is non-zero only for media that are not time-reversible (e.g.\ magnetized), and is responsible for Faraday rotation; it does {\em not} convert linear to circular polarization. Since the background cosmology is homogeneous and isotropic, $n_Q$ and $n_U$ must originate from perturbations in either the matter or the radiation.

If we write $\Delta n$ as the difference of the two eigenvalues of $n$, we have a phase shift between the two eigenvectors given by
\begin{equation}
\label{eq:phase}
\Delta \phi = \phi_x - \phi_y = \int \frac{\omega}{c} \Delta n\, dr_{\rm proper} \approx \frac{\omega}{c} \Delta n \frac{\Delta\chi}{1+z} .
\end{equation}
Here $\omega$ is the proper frequency of the CMB photons, $z$ is the redshift and $\chi$ is the comoving distance (see Table~\ref{tab:table1} for a glossary of physical quantities used throughout this work). Conversion from pure linear to pure circular polarization occurs in the idealized circumstance that $n_V$ is negligible, the phase shift is $\pm\pi/2$ and the incident linearly polarized wave makes an angle of $\pi/4$ to the principal directions of the birefringent material. In the cosmological context, the phase shifts are $\ll 1$ and instead we are concerned with conversion from pure linear polarization to mostly linear polarization with a small admixture of circular polarization. The circular polarization so induced is $V = 2 U \Delta \phi$ \cite{Cooray}, where $U$ is the input diagonal linear polarization in the frame chosen to align with the principal axes of the medium ($n_Q>0$, $n_U=0$).
In the more general case, where we allow anisotropy on an arbitrary axis, the equation of radiative transfer \cite{beckert2002circular}, becomes
\begin{equation}
\frac{dV}{dr_{\rm proper}} = \frac{2\omega}c(n_QU - n_UQ)
~~~{\rm or}~~~
V = \int \frac{2\omega}c(n_QU - n_UQ)\,\frac{d\chi}{1+z}.
\label{eq:Vr}
\end{equation}

\begin{figure}[t]
\begin{subfigure}{.5\textwidth}
  \centering
  \includegraphics[width=1\linewidth]{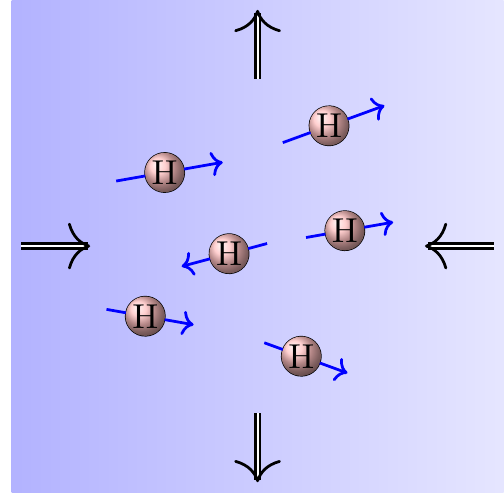}
  \caption{Matter-related mechanism: \S\S\ref{sec:21z20} and \ref{sec:21z1100}.}
  \label{fig:mabi}
\end{subfigure}
\begin{subfigure}{.5\textwidth}
  \centering
  \includegraphics[width=1\linewidth]{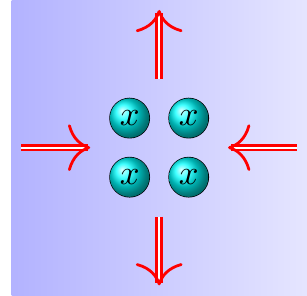}
  \caption{Radiation-related mechanism: \S\S\ref{sec:photon}, \ref{sec:nonlinear}, and \ref{sec:plasma}.}
  \label{fig:radbi}
\end{subfigure}
\caption{Two classes of cosmic  birefringence sources. In Fig.~\ref{fig:mabi} the inhomogeneous matter, i.e. spin polarized hydrogen atoms, is responsible for the difference in indices of refraction. The blue arrows represent the spins of the hydrogen atoms. Here the black arrows stand for the radiation field, which can be either the CMB or the 21 cm radiation from other hydrogen atoms. In contrast, inhomogeneous radiation produces birefringence through nonlinear response in Fig.~\ref{fig:radbi}, here the red arrows stand for the CMB quadrupole and the $x$ in the circles corresponds to the source of the nonlinearity, e.g. vacuum, bounded electrons or free electrons.}
\label{fig:bifig}
\end{figure}

We proceed to classify the two kind of cosmic birefringence that will be treated in this work in terms of the type of mechanism that generates the anisotropy $n_Q,n_U\neq 0$. In Fig.~\ref{fig:bifig} we illustrate the two classes of sources: matter-related and radiation-related. Our first two mechanisms are matter-related and both involve the spin-polarization of the hydrogen atoms; however, the mechanism for the alignment of the spins is not the same for both cases. In the Cosmic Dawn era (\S\ref{sec:21z20}) the atoms are aligned by 21 cm radiation, whereas in the recombination era (\S\ref{sec:21z1100}) the spins are aligned by Balmer line (mostly H$\alpha$) radiation. For the radiation-related cases, the anisotropic radiation field is always the perturbed CMB, but the nonlinear behavior in the radiation can be generated by the vacuum by electron positron pairs as in the case of photon-photon scattering (\S\ref{sec:photon}), by bounded electrons in hydrogen atoms like in the static nonlinear polarizability of hydrogen (\S\ref{sec:nonlinear}) or by free electrons as in the plasma delay (\S\ref{sec:plasma}).

Having discussed the origin of the circular polarization, briefly highlighted its importance in astrophysics, and simplified our problem to identifying possible sources of birefringence, we proceed to study spin polarized hydrogen atoms as a source of birefringence in \S\ref{sec:21z20}.

\begin{table}[!t]
\centering
\footnotesize
\caption{\label{tab:table1}Glossary of physical quantities used in this paper.}
\begin{tabular}{ccl}
\hline \hline
Symbol & SI unit & Physical quantity \\
\hline
$z$ & 1 & Redshift \\
$\omega$ & s$^{-1}$ & Angular frequency of the CMB photons \\
$\chi$ & m & Comoving distance \\
$\mu_e$ & J T$^{-1}$ & Magnetic moment of the electron \\
$n_{\rm H}$ & m$^{-3}$ & Number density of hydrogen atoms \\
$n_{nl}$ & m$^{-3}$ & Number density of hydrogen atoms for the state $nl$\\
$\omega_{\rm hf}$ & s$^{-1}$ & Angular frequency of the hyperfine transition \\
$\omega_{me}$ & s$^{-1}$ & Electron mass in angular frequency units, $m_ec^2/\hbar$ \\
$\omega_{Ly\alpha}$ & s$^{-1}$ & Angular frequency of the Lyman$-\alpha$ photons \\
$\rho_{nm}$ & 1 & Components of the density matrix \\
${\cal P}_{jm}$ & 1 & Irreducible components of the density matrix \\
$\bar{E}^{(0)}_n$ & J & Energy eigenvalues of the unperturbed hamiltonian \\
$\mu_i$ & J T$^{-1}$ & Magnetic dipole moment in the $i$-direction \\
$F$ & 1 & Total angular momentum quantum number (nuclear + orbital + spin)\\
$M$ & 1 & Total magnetic quantum number \\
$\alpha_{ij}$ & C m$^2$ V$^{-1}$ & Components of the atomic electric polarizability tensor \\
$\alpha^{\rm m}_{ij}$ & J T$^{-2}$ & Components of the atomic magnetic polarizability tensor \\
$Q$, $U$,  $V$ & K & Stokes parameters (in temperature units) \\
$T = T_{\rm CMB} = T_{\gamma} $ & K & Temperature of the CMB \\
$T_s$ & K & Spin temperature \\
$\tau$ & 1 & Optical depth \\
$C_l$ & 1 & Angular power spectrum of $\Delta T/T$ \\
$P_\delta$ & m$^3$ & Matter power spectrum \\
$P_\zeta$ & m$^3$ & Primordial curvature power spectrum \\
$H$ & s$^{-1}$ & Hubble expansion rate \\
$J_\alpha$ & m$^{-2}$ s$^{-1}$ Hz$^{-1}$ sr$^{-1}$ & Flux of Lyman$-\alpha$ photons on the blue side of the line \\
$\tilde{S}_{\alpha,(2)}$ & 1 & Correction factor for the Lyman-$\alpha$ line shape \\ 
$\Gamma_{2p} $ & s$^{-1}$ & Einstein A-coefficient for the Lyman$-\alpha$ \\
$\gamma_{2p}$ & Hz & HWHM of the Lyman$-\alpha$ line \\
$A_{x \rightarrow y}$ & s$^{-1}$ & Einstein A-coefficient for the transition from state x to state y \\
$x_e$ & 1 & Ionization fraction \\
$X_i$ & 1 & Fraction of hydrogen atoms in level $i$ \\
$\Lambda$ & s$^{-1}$ & Decay rate for the two-photon decay \\ 
$\alpha_{\rm B}$ & m$^3$ s$^{-1}$ & Case B recombination coefficient (used in the Peebles model) \\
${\cal U}^{B}_{\rm ani}$ & J m$^{-3}$ & Anisotropic magnetic energy density of the CMB \\
${\cal U}^{E}_{\rm ani}$ & J m$^{-3}$ & Anisotropic electric energy density of the CMB \\
$A_e$ & T$^{-2}$ & Euler-Heisenberg interaction constant \\
$\chi^{\rm iso}_e$ & 1 & Isotropic contribution to the electric susceptibility \\
$\chi^{\rm iso}_m$ & 1 & Isotropic contribution to the magnetic susceptibility \\
$\chi^{\rm ani,e}_{ij}$ & 1 & Anisotropic contribution to the electric susceptibility \\
$\chi^{\rm ani,m}_{ij}$ & 1 & Anisotropic contribution to the magnetic susceptibility \\
$ _{s}Y_{lm}$ & 1 & Spin-weighted spherical harmonic \\
$a_{\rm rad}$ & J m$^{-3}$ K$^{-4}$ & Radiation energy density constant \\
$\alpha^A_{\textup{NL},ij}$ & C m$^2$ V$^{-1}$ & Anisotropic non-linear components of the polarizability tensor \\
$d{\cal U}/d\vec\Omega$ & J m$^{-3}$ sr$^{-1}$ & Energy density in ambient electromagnetic waves per unit solid angle \\
$\gamma$ & J m$^4$ V$^{-4}$ & Second-order hyperpolarizability of the hydrogen atom \\
\hline \hline
\end{tabular}
\end{table}

\section{Birefringence from spin polarized hydrogen atoms from the Cosmic Dawn epoch}
\label{sec:21z20}
In this section we will explore spin polarized hydrogen atoms as a possible source of cosmic birefringence. Hydrogen atoms in the Cosmic Dawn epoch can become spin-polarized in the presence of an anisotropic radiation background, which here is primarily the 21 cm radiation field generated by other nearby atoms \cite{Venumadhav}. Since the line is narrow and is formed in an expanding medium, the line profile-weighted intensity and the probability of re-absorption depend on the Einstein coefficients, the local density, and the velocity gradient \cite{1960mes..book.....S}.\footnote{The line width itself does not matter as long as it is narrow. This is because as the line profile gets narrower,  there are two competing effects: (i) the path length over which the photon can be reabsorbed gets smaller; but (ii) the cross section increases over a small range of frequencies centered on $\nu_{21}$ as appropriate for a $\delta$-function. Thus there is still a finite probability to reabsorb the redshifted 21 cm radiation by the surrounding neutral hydrogen atoms.} In an unperturbed universe, the resulting radiation in the 21 cm line is isotropic, but in the presence of velocity shear (direction-dependent velocity gradient) it has a quadrupole anisotropy. For pedagogical purposes we elaborate on the mechanism that leads to this anisotropic radiation in appendix \ref{ap:vgrad}. This anisotropic radiation unevenly populates the triplet state of the hyperfine $F=1$ level as illustrated in Figures 1 and 2 in Ref.~\cite{Venumadhav}.

Thus the 21 cm anisotropic radiation is continuously sourcing the anisotropy of the spins of the hydrogen atoms. The alignment of the hydrogen spins in steady state is the result of the balance between the aligning effect (anisotropic component of the 21 cm radiation) and the randomizing effect from collisions, Lyman-$\alpha$ radiation and the isotropic component of 21 cm radiation.
Spin-polarized hydrogen is then a source of birefringence because the magnetic polarizability of a hydrogen atom in the $F=1$ level is different on the axes parallel to and perpendicular to the total angular momentum.

As a starting point we will compute an order of magnitude estimate for both the birefringence and the phase shift generated because of this effect. Note that this mechanism only applies to hydrogen -- the other abundant element in the early Universe, helium, is a spin singlet and cannot be polarized.

\subsection{Order of magnitude}

We start with a semiclassical order of magnitude estimate of this effect at $1+ z=20$. The difference in indices of refraction should be proportional to the number density of hydrogen atoms $n_{\rm H}$, the magnetic permeability and the magnetic polarizability, $\Delta n \sim n_{\rm H} \mu_0 \Delta \alpha_{\rm m}$. This polarizability should depend on the polarization of the hydrogen atoms, which is described by irreducible spherical tensor components of the density matrix: using the conventions of Ref.~\cite{Venumadhav}, there is a spin-0 part (net population of the $F=1$ level), spin-1 part (net vectorial polarization of the atomic spins), and a spin-2 part (``headless vector'' alignment of the atomic spins). The relevant part is only the spin-2 part of the density matrix ${\cal P}_{2m}$, which has the correct symmetry properties to generate $n_Q$ and $n_U$ since $Q$ and $U$ are spin-2 quantities.

The anisotropy of the magnetic polarizability should scale linearly with the atomic polarization: $\Delta n \sim n_{\rm H} \mu_0 {\cal P}_{2m} \Delta \alpha_{\rm m} |_{\rm pp}$, where $\Delta \alpha_{\rm m} |_{\rm pp}$ is the anisotropic polarizability for perfectly polarized atoms. This perfectly polarized polarizability can be approximated by the change in magnetic moment from the magnetic field in the time $\sim1/\omega$, where $\omega$ is the CMB photon frequency. The change in angular momentum is $\sim \mu_e B/\omega$, where $\mu_e$ is the magnetic moment of the electron, and the change in magnetic moment is different by a factor of $\sim \mu_e/\hbar$. Hence $\Delta \alpha_{\rm m} |_{\rm  pp} \sim \mu_e^2/\hbar\omega =  \alpha c^5 \epsilon_0/(\omega^2_{\rm me} \omega)$, where $\alpha$ is the fine structure constant, $c$ is the speed  of light, $\epsilon_0$ is the vacuum permittivity and $\omega_{\rm me} = m_ec^2/\hbar$ is the mass of the electron in frequency units. Then the classical birefringence is given by $\Delta n \sim \alpha n_{\rm H} c^3/(\omega^2_{\rm me} \omega)\times {\mathcal P}_{2m}$. 

\begin{figure}[t]
\begin{subfigure}{.6\textwidth}
  \centering
  \includegraphics[width=1\linewidth]{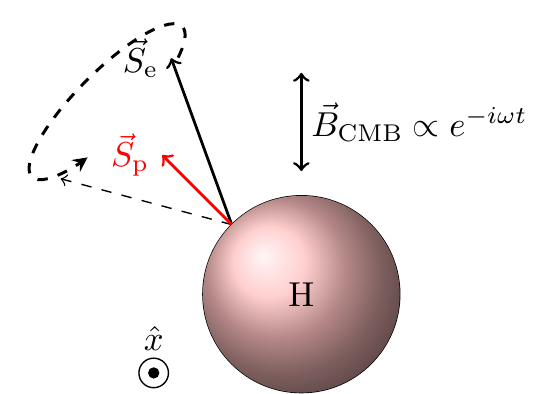}
\caption{Spin polarized hydrogen atom.}
\label{fig:fig1a}
\end{subfigure}%
\begin{subfigure}{.4\textwidth}
\centering
\includegraphics[width=.8\linewidth]{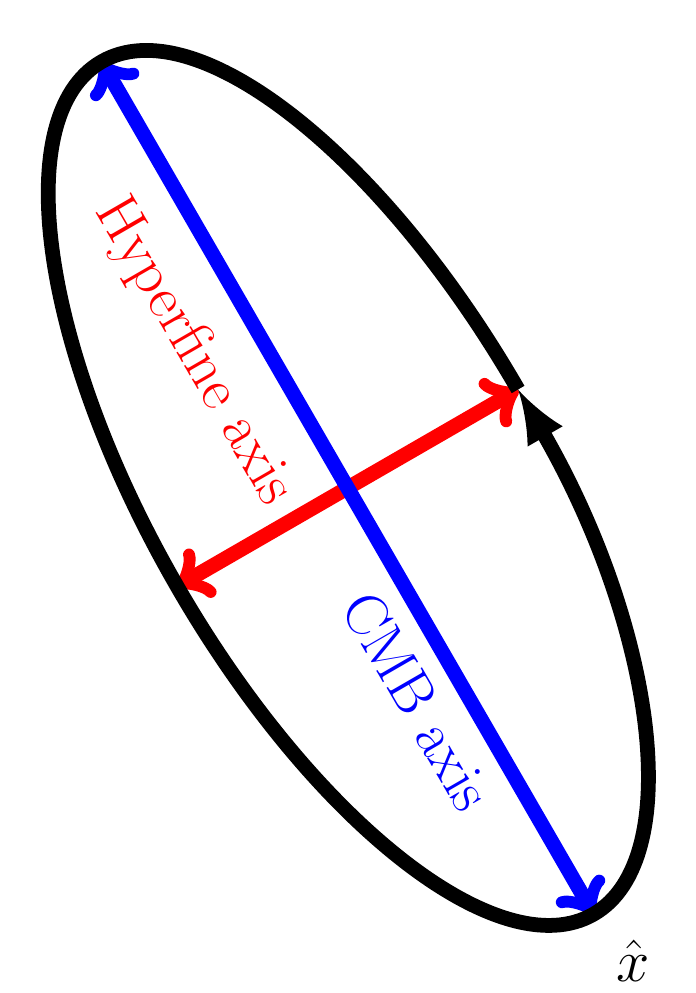}
\caption{Simplified trajectory for the spin of the electron.}
\label{fig:fig1b}
\end{subfigure}

  \caption{Spin polarized hydrogen atom with CMB magnetic field. Note that the CMB field is oscillating with frequency $\omega$, which is considerably larger than the hyperfine frequency. We have used red to illustrate the hyperfine contribution, i.e. the spin of the proton. In the absence of hyperfine structure the change in the magnetic moment is perpendicular to the CMB magnetic field, see the blue line in figure \ref{fig:fig1b}. Note that we show the precession of the electron spin with respect to the proton spin ignoring the CMB magnetic field in figure \ref{fig:fig1a}, however, it should be pointed out that the proton spin also precesses around the electron spin. We illustrated the hyperfine contribution with the red line in figure \ref{fig:fig1b}.}
  \label{fig:fig1}
\end{figure}

Nevertheless, this simple estimate is incorrect, since the change in angular momentum (and hence magnetic moment) is perpendicular to the magnetic field and hence the diagonal (or, more generally, symmetric) contribution to $\alpha^{\rm m}_{ij}$ vanishes (see Fig.~\ref{fig:fig1}). Hence one gets no contribution to $n_Q$ or $n_U$. However, if we consider the coupling to the spin of the proton i.e.\ the hyperfine structure, then the electron spin would now try to precess around the proton spin, and vice versa. This interaction results in a small component of the magnetic moment of the electron aligned with the magnetic field (see \S IVB of Ref.~\cite{HMV} for a detailed description of how this works). Hence, the electron magnetic moment follows an ellipse with major axis sourced by the CMB anisotropies and minor axis sourced by the hyperfine splitting. Therefore, one should multiply the previous estimate for the birefringence by the ratio of the axes $\sim \omega_{\rm hf}/\omega$ to include the correct geometrical factor. Evaluating the different parameters at redshift $1+z = 20$,
\begin{equation}
\label{eq:n}
\Delta n \sim \frac{\alpha n_{\rm H} c^3}{(\omega_{\rm me} \omega)^2} \omega_{\rm hf} {\cal P}_{2m} \sim  10^{-36} \textup{,}
\end{equation}
where we have used the results of Ref.~\cite{Venumadhav} to estimate the alignment tensor at the desired redshift of $1+z=20$  (${\cal  P}_{2m} \sim 10^{-5} $) and $\omega \approx 6\times 10^{12} \ \textup{rad s}^{-1}$.

Applying Eq.~(\ref{eq:phase}) for a photon with $\lambda \sim 1 \ \textup{mm}$ today ($\lambda \sim 5 \times 10^{-5} \ \textup{m}$ at $1+z = 20$), and with a proper distance of approximately 100 Mpc (the Hubble length at $1+z=20$), we find a phase shift of 
\begin{equation}
\label{eq:phi}
\Delta \phi \sim 10^{-7} \textup{.}
\end{equation}
Note that the spin alignment direction will vary along the line of sight, so this should be thought of as an upper limit to the integrated phase shift.

\subsection{Detailed calculation}
\label{ssec:21z20}

We now turn to the detailed calculation of the CMB circular polarization induced by propagation through spin-polarized neutral gas. There are three steps to this calculation. First, we apply quantum mechanics to determine the anisotropic magnetic polarizability of the atoms and hence index of refraction of the gas. Then we invoke previous computations of the spin-polarization in the Cosmic Dawn epoch to relate this to the cosmological density perturbations. Finally we use the statistics of large-scale structure (specifically the Limber equation, which deals with projected quantities) to compute the RMS circular polarization.

In the Schr\"odinger picture, the Hamiltonian for a hydrogen atom in an oscillating magnetic field is given by
\begin{equation}
\label{eq:hamil}
{\cal H} = {\cal H}^{(0)} + {\cal H}^{(1)}={\cal H}^{(0)} - B_{0j} \mu_j e^{-i \omega t} - B^*_{0j} \mu_j e^{i \omega t} \textup{,}
\end{equation}
where $\omega$ is the frequency from the CMB photons and $\mu_j$ is the magnetic dipole moment.

We suppose that the unperturbed state of the atom is $|\Psi^{(0)}\rangle$, and that this is an eigenstate of ${\cal H}^{(0)}$ with energy $E_0$. Using first-order perturbation theory, we find that the perturbation to the state is
\begin{equation}
|\Psi^{(1)}\rangle
= \sum_n \left[ \frac{B_{0j} e^{-i\omega t}}{E_n^{(0)} - E_0 - \hbar\omega}
 + \frac{B_{0j}^\ast e^{i\omega t}}{E_n^{(0)} - E_0 + \hbar\omega} \right]\langle n | \mu_j | \Psi^{(0)} \rangle
 ,
\end{equation}
where $E_n^{(0)}$ is the unperturbed energy of state $|n\rangle$. The perturbed dipole moment is
\begin{equation}
\langle \mu_i \rangle =
\langle \Psi^{(0)} | \mu_i | \Psi^{(0)} \rangle
+ \sum_n \langle \Psi^{(0)} | \mu_i | n \rangle \langle n | \mu_j | \Psi^{(0)} \rangle \left[ \frac{B_{0j} e^{-i\omega t}}{E_n^{(0)} - E_0 - \hbar\omega}
 + \frac{B_{0j}^\ast e^{i\omega t}}{E_n^{(0)} - E_0 + \hbar\omega} \right]
 + {\rm c.c.},
\label{eq:exp}
\end{equation}
where ``c.c.'' indicates a complex conjugate.
We can then identify the magnetic polarizability tensor at positive frequency, $\alpha^{\rm m}_{ij}$, as the coefficient of $B_{0j}e^{-i\omega t}$ in $\mu_i$:
\begin{equation}
\alpha^{\rm m}_{ij} = \sum_n \left[
\frac{\langle \Psi^{(0)} | \mu_i | n \rangle \langle n | \mu_j | \Psi^{(0)} \rangle }{E_n^{(0)} - E_0 - \hbar\omega}
+\frac{ \langle \Psi^{(0)} | \mu_i | n \rangle^\ast \langle n | \mu_j | \Psi^{(0)} \rangle^\ast}{E_n^{(0)} - E_0 + \hbar\omega} \right].
\end{equation}
Finally, we are interested in the average response over an ensemble of atoms in the $F=1$ level, described by a density matrix $\rho_{MM'}$. To carry out this average, we make the replacement $| \Psi^{(0)} \rangle\langle \Psi^{(0)}| \rightarrow \rho$:
\begin{eqnarray}
\alpha^{\rm m}_{ij} & = &  \sum_{n,M,M'} \left[ \frac{\rho_{MM'}
\langle 1,M'|\mu_j|n\rangle \langle n|\mu_i|1,M\rangle}{{E}^{(0)}_n-E^{(0)}_{F=1} - \hbar\omega}
 +  \frac{\rho_{M'M} \langle 1,M'|\mu_j|n\rangle^\ast \langle n|\mu_i |1,M\rangle^\ast}{{E}^{(0)}_n-E^{(0)}_{F=1} + \hbar\omega} \right].
\label{eq:aa+}
\end{eqnarray}

The magnetic dipole moment has three main ingredients -- contributions from orbital motion, electron spin and nuclear spin:
\begin{equation}
\label{eq:mu}
{\boldsymbol\mu} = -\frac{e}{2m_e}{\boldsymbol L} -\frac{e}{m_e}{\boldsymbol S} + \frac{eg_p}{2m_p} {\boldsymbol I} 
\approx -\frac{e}{m_e}{\boldsymbol S},
\end{equation}
where we have dropped the first term since we are working with electrons in $s$-orbitals and the third term since $m_p\gg m_e$. Using Eq.~(\ref{eq:mu}) and the Clebsch-Gordan coefficients, we can construct any of the matrix elements needed for the polarizability tensor.

We furthermore use that $E^{(0)}_{F=0}-E^{(0)}_{F=1} = -\hbar\omega_{\rm hf}$. Then
\begin{equation}
\label{eq:a+x}
\alpha^{\rm m}_{xx} \approx \frac{e^2\hbar}{8m^2_e}
\left( \frac1{-\omega_{\rm hf} - \omega} + \frac1{-\omega_{\rm hf} + \omega} \right)
 (\rho_{11} - \rho_{-1,1} + \rho_{-1,-1} - \rho_{1,-1}) \textup{.}
\end{equation}
Taking the limit of $\omega\gg\omega_{\rm hf}$, the $xx$ component of the susceptibility is given by
\begin{equation}
\label{eq:chixx}
 \chi_{xx}  \approx  n_{\rm H} \frac{e^2\mu_0}{4m^2_e} \frac{\omega_{\rm hf}}{\omega^2}(\rho_{11} - \rho_{-1,1} + \rho_{-1,-1} - \rho_{1,-1}) .
\end{equation}
Similarly,
\begin{equation}
\label{eq:chiyy}
 \chi_{yy} \approx n_{\rm H} \frac{e^2\mu_0}{4m^2_e} \frac{\omega_{\rm hf}}{\omega^2}(\rho_{11} + \rho_{-1,1} + \rho_{-1,-1} + \rho_{1,-1}) .
\end{equation}
At last, we can compute the birefringence using Eq.~(\ref{eq:chixx}) and Eq.~(\ref{eq:chiyy}):
\begin{equation}
\label{eq:21z20n}
2n_Q \equiv n_{xx} - n_{yy} \approx  -2\pi \frac{\alpha n_{\rm H} c^3 \omega_{\rm hf}}{\omega^2_{\rm me} \omega^2} (\rho_{-1,1} + \rho_{1,-1}) =  -\frac{2\pi}{\sqrt{3}} \frac{\alpha n_{\rm H} c^3 \omega_{\rm hf}}{\omega^2_{\rm me} \omega^2} ({\cal P}_{22} + {\cal P}_{2,-2}) \textup{,}
\end{equation}
where $\omega_{\rm me}$ is the mass of the electron in frequency units, and in the last step we changed from the density matrix to the irreducible components of the density matrix following Ref.~\cite{Venumadhav}.

The alignment tensor for hydrogen atoms in the range of redshifts we are currently interested in  is given in the Fourier domain \cite{Venumadhav} as
\begin{equation}
\label{eq:21zali}
{\mathcal P}_{2m}(\hat{\boldsymbol k}) = \frac{1}{20\sqrt{2}} \frac{T_*}{T}\left(1 -\frac{T}{T_s}\right) \frac{\tau}{1 + x_{\alpha,(2)}+x_{c,(2)}} \delta_{\rm m}(\hat{\boldsymbol k}) \sqrt{\frac{4 \pi}{5}} Y_{2m}(\hat{\boldsymbol k}),
\end{equation}
where $\hat{\boldsymbol k}$ is the direction of the wavevector of the radiation, $\tau$ is the optical depth of the neutral hydrogen gas, $T_s$ is the spin temperature\footnote{The spin temperature parametrizes the difference of population of hydrogen atoms in the hyperfine $F=1$ and $F=0$ levels.}, $T_* = 68\,$mK is the hyperfine splitting in temperature units, $\delta_{\rm m}$ is the density contrast and the $x_x$ parametrize the rates of depolarization by optical pumping and collisions. We neglect the primordial magnetic field term since we are focusing specifically on conventional physics in this study. The derivation of Eq.~(\ref{eq:21zali}) is presented in Ref.~\cite{Venumadhav}, but the reader who wishes to follow the basic ingredients without as much mathematical formalism may consult the abbreviated treatment in Appendix~\ref{ap:vgrad}.

We may now compute the phase shift using Eq.~(\ref{eq:phase}):
\begin{eqnarray}
\frac{d\phi}{d\ln a}
&=& \frac{c}{aH} \frac{d\phi}{d\chi}
\nonumber \\
&=& \frac{\omega}{H}(n_{xx} - n_{yy})
\nonumber \\
&=& -\frac{2\pi}{\sqrt{3}} \frac{\alpha  c^3 \omega_{\rm hf}}{\omega^2_{\rm me}} \frac{n_{\rm H}}{\omega H} ({\cal P}_{2,2} + {\cal P}_{2,-2}) 
\nonumber \\
\label{eq:phasez20}
&=& -1.4 \times 10^{-7} \left( \frac{1+z}{20} \right)^{1/2} \left( \frac{{\cal P}_{2,2} + {\cal P}_{2,-2}}{10^{-5}} \right) \left( \frac{100 \ \textup{GHz}}{\nu_{\rm today}} \right) \textup{,}
\end{eqnarray} 
where we obtain our estimate for the alignment tensor from the expression computed by Venumadhav et al. in \cite{Venumadhav}. Note that the pre-factor in Eq.~(\ref{eq:phasez20}) is an over-estimate of the net phase shift since the sign of ${\cal P}_{2,m}$ will change along the line of sight.

To obtain the net circular polarization due to the passage through the neutral medium, we need to write out the full line-of-sight integral and then perform a statistical study. Therefore in what follows we are no longer constrained to a single redshift. The conversion integral, Eq.~(\ref{eq:Vr}), can be written as
\begin{equation}
\label{eq:m2}
V= \frac{2 \omega_0}{c} \int U n_Q dr -  \frac{2 \omega_0}{c} \int Q n_U dr .
\end{equation}
Substituting in Eqs.~(\ref{eq:21z20n}) and (\ref{eq:21zali}), we find that
\begin{equation}
V =  U\phi_Q - Q\phi_U,
~~~{\rm where}~~~
\phi_Q = p\int W(r) [Y_{22} (\hat{\boldsymbol k}) + Y_{2,-2}(\hat{\boldsymbol k})] \delta_{\rm m} (r \hat{\boldsymbol n}) dr
\label{eq:m3}
\end{equation}
(and similarly for $\phi_U$).
Here $\hat{\boldsymbol k}$ is the direction of the wave-vector of the radiation and
\begin{equation}
p = \frac{- 2\pi}{\sqrt{3}} \frac{\alpha \omega_0 c^2 \omega_{\rm hf}}{\omega^2_{\rm me}} 
~~~{\rm and}~~~
W(r) = \frac{1}{ 20 \sqrt{2}} \sqrt{\frac{4 \pi}{5}} \frac{n_{\rm H}}{\omega^2} \frac{T_{*}}{T_{\rm CMB}} \left(1 - \frac{T_{\rm CMB}}{T_s} \right) \frac{\tau}{1 + x_{\alpha,(2)} +x_{c,(2)}}\, \textup{.}
\end{equation}
[Note that Eq.~(\ref{eq:m3}) contains both a real-space line of sight integral and density field, and a Fourier-space operator, $Y_{2m}(\hat{\boldsymbol k})$, that depends on the direction of the Fourier modes. The latter is intended to operate on the density field in the sense of taking the inverse Fourier transform, multiplying by the stated function of ${\boldsymbol k}$, and Fourier-transforming back.]

The optical depth for resonant 21 cm absorption is given by Eq.~(87) of \cite{Venumadhav}
\begin{equation}
\label{eq:opti}
\tau(z) = \frac{\pi^2 c^3 n_{\rm H}(z) A_{\rm hf} (3 - 4 {\mathcal P}_{00}(z))}{H(z) \omega_{\rm hf}^3[1 + (1/H(z))(dv_{\parallel}/dr_{\parallel})]} \approx \frac{3 c^3 n_{H}(z) A_{\rm hf}}{4\omega^3_{\rm hf} H(z)} \frac{T_*}{T_s} \, \textup{,}
\end{equation}
where the $A_{\rm hf}$ is the Einstein coefficient and ${\mathcal P}_{00} \approx 3/4 - 3T_{*}/(16T_{s})$ is the occupancy of the excited state. Note that the Hubble flow in the denominator indicates the path length available for a 21 cm photon to resonate with  nearby neutral hydrogen atoms before its frequency is shifted out of resonance by the Hubble flow. Finally, $\tau$ is used here multiplying a perturbation, so to linear order we may use its value in the unperturbed Universe.      

The total variance of $V$ is related to that of the phase shifts via
\begin{equation}
V_{\rm rms}^2 =  \langle (U\phi_Q - Q\phi_U)^2\rangle
= 2 \langle Q^2\rangle \langle \phi_Q^2\rangle, 
\label{eq:vartot}
\end{equation}
where we have used the symmetry relations that $\langle Q^2\rangle = \langle U^2\rangle$, $\langle QU\rangle=0$, and similarly for the phase shifts.

Here $\langle Q^2\rangle$ is simply the linear polarization of the primary CMB; it is given by
\begin{equation}
\langle Q^2\rangle = \frac1{4\pi} \int \ell^2 (C_\ell^{EE} + C_\ell^{BB} ) \,\frac{d\ell}{\ell}.
\label{eq:Q2}
\end{equation}
The variance of the conversion angle $\langle \phi_Q^2\rangle$ can be obtained from its power spectrum:
\begin{equation}
C_\ell^{\phi_Q} = p^2 \int W^2(r) P_{\delta} \left(k = \frac{\ell}{r}\right) \frac{dr}{r^2} |G(\hat{\boldsymbol \ell})|^2,
\label{eq:m4}
\end{equation}
where $P_\delta$ is the matter power spectrum, $\hat{\boldsymbol\ell}$ is the direction in the sky, and $G(\hat{\boldsymbol\ell}) = Y_{22}(\hat{\boldsymbol\ell}) + Y_{2,-2}(\hat{\boldsymbol\ell})$. Normally we average $C_\ell^{\phi_Q}$ over azimuthal directions, so we make the replacement
\begin{equation}
|G(\hat{\boldsymbol \ell})|^2 \rightarrow \frac1{2\pi} \int_0^{2\pi} \left|G\left( \frac\pi2,\phi\right) \right|^2\,d\phi
 = \frac{15}{16\pi}.
\end{equation}
\begin{figure}[t]
\begin{subfigure}{.5\textwidth}
  \centering
  \includegraphics[width=\linewidth]{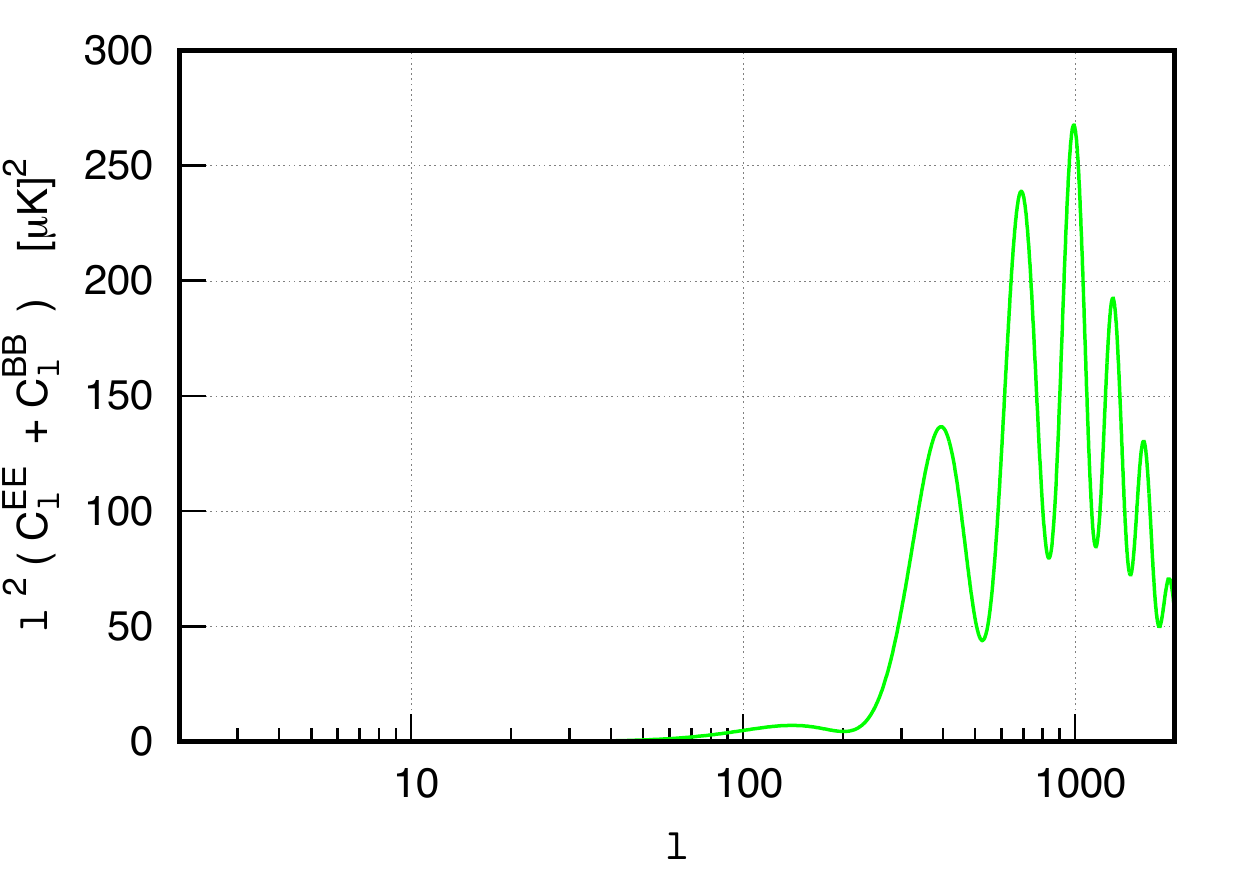}
\caption{$\ell$-integrand.}
\label{fig:plotl}
\end{subfigure}%
\begin{subfigure}{.5\textwidth}
\centering
\includegraphics[width=\linewidth]{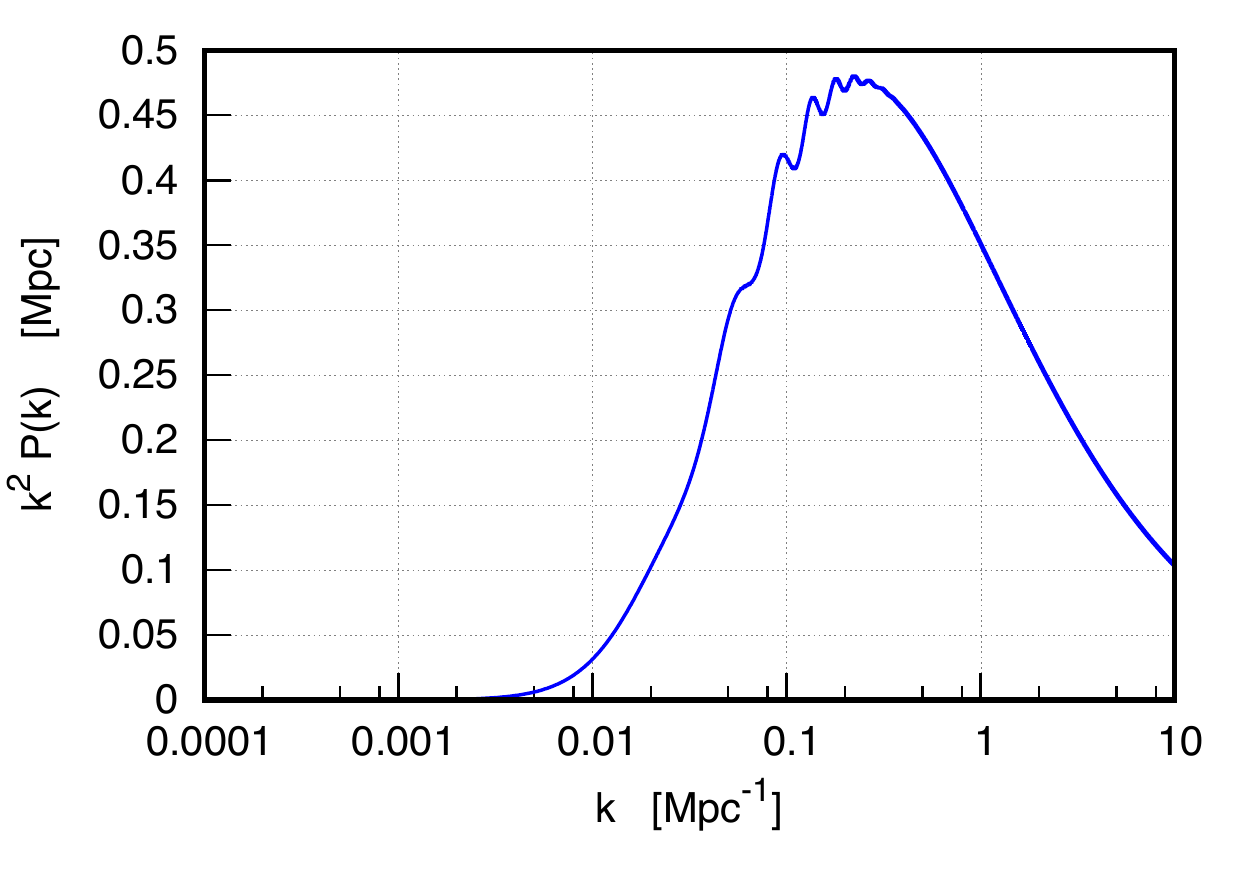}
\caption{$k$-integrand.}
\label{fig:plotk}
\end{subfigure}
  \caption{Integrands of Eq.~(\ref{eq:VRMS2}). Essentially, figure \ref{fig:plotl} represents the contribution of the linear polarization of the CMB to the rms of the circular polarization. Note that the peak of the distribution is approximately at $\ell= 1000$. In figure \ref{fig:plotk} we have evaluated the power spectrum at $z=20$. This is the contribution due to the spin polarized hydrogen atoms. Note that the peak of the distribution is near $k \sim 0.1 ~ \textup{Mpc}^{-1}$.}
  \label{fig:figplot}
\end{figure}
\begin{figure}[t]
\centering
\includegraphics[width=0.5\linewidth]{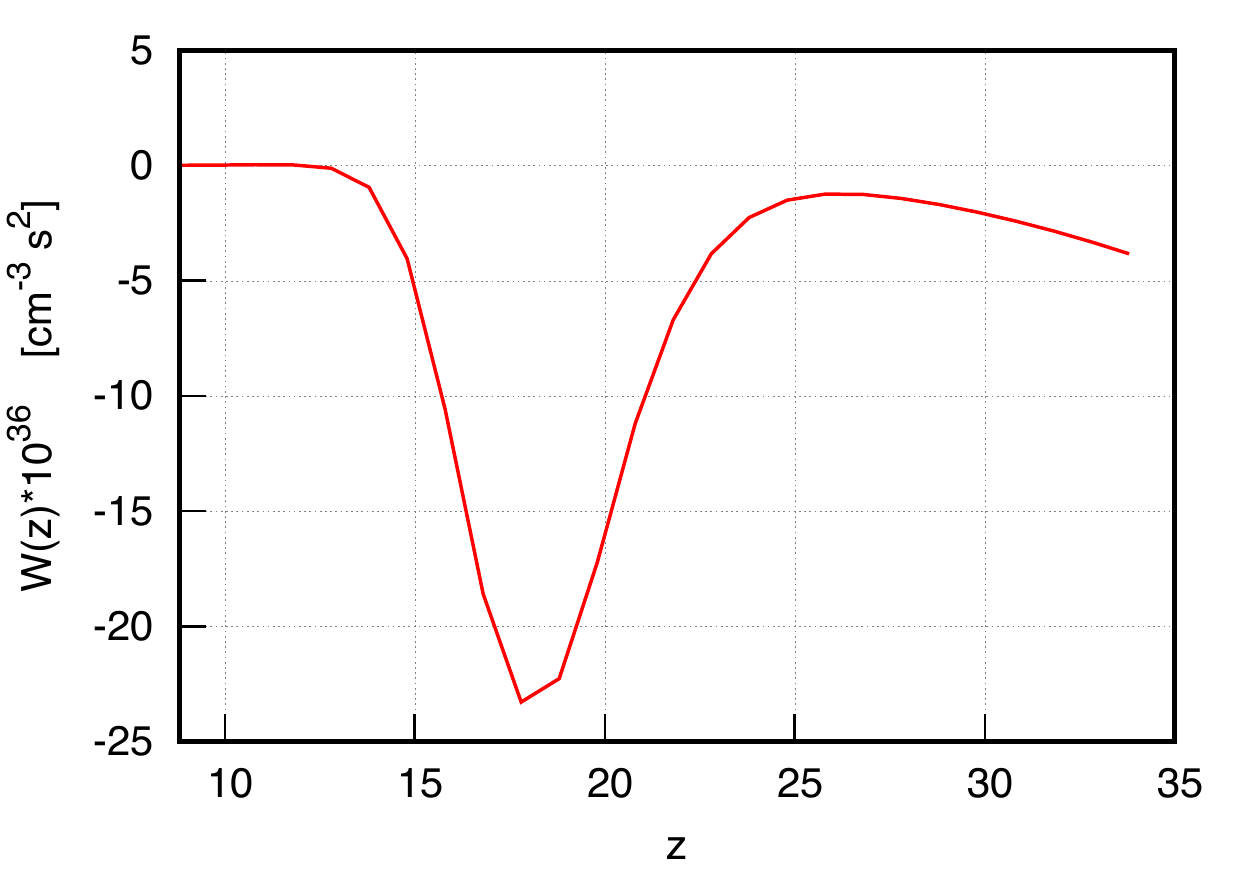}
\caption{Window function, note that we multiply the y axis by a factor of $10^{36}$ for convenience.}
\label{fig:window}
\end{figure}
Changing variables from proper distance to redshift using $dr = c\,dz/H(z)$ gives
\begin{equation}
C_\ell^{\phi_Q} = \frac{15}{16\pi} p^2 \int W^2(r) P_{\delta} \left(k = \frac{\ell}{r}, z\right) \frac{c\,dz}{r^2\,H(z)},
\end{equation}
and hence
\begin{equation}
\langle \phi_Q^2\rangle = \frac1{2\pi} \int \ell^2 C_\ell^{\phi_Q}\,\frac{d\ell}{\ell}
= \frac{15}{32\pi^2} p^2 \int W^2(r) \left[ \int k^2 P_{\delta} \left(k = \frac{\ell}{r}, z\right) \frac{dk}{k}\right] \frac{c\,dz}{H(z)}.
\end{equation}
Combining this with Eq.~(\ref{eq:Q2}), we conclude that the RMS circular polarization is
\begin{equation}
V_{\rm rms}^2 = \frac{15p^2 }{64\pi^3} \left[ \int \ell^2 (C_\ell^{EE} + C_\ell^{BB} ) \,\frac{d\ell}{\ell} \right]
\int W^2(r) \left[ \int k^2 P_{\delta} \left(k = \frac{\ell}{r}, z\right) \frac{dk}{k}\right] \frac{c\,dz}{H(z)}.
\label{eq:VRMS2}
\end{equation}
Here in principle the integrals over $\ell$ and $k$ range over all scales, although they are dominated by the peak of the CMB polarization power spectrum ($\ell\sim 10^3$) and the $k^2$-weighted matter power spectrum ($k\sim 0.5\,$Mpc$^{-1}$), respectively. This is shown in Figures \ref{fig:plotl} and \ref{fig:plotk}. The redshift integral extends over the Cosmic Dawn epoch -- from the beginning of the Lyman-$\alpha$ coupling (when $T_s$ drops below $T_\gamma$) through reionization (when there are no more neutral hydrogen atoms); here we take the range $8.8<z<34$.

We first focus in the $\ell$-integral. We obtain the angular power spectrum for the linear polarization from CLASS \cite{blas2011cosmic},
\begin{equation}
\label{eq:vrmsl}
V^2_{\rm rms}=\frac{15p^2}{64 \pi^3} \left[ 252 ~ (\mu \textup{K})^2 \right] \int W^2(r) \left[ \int k^2 P_{\delta} \left(k = \frac{\ell}{r}, z\right) \frac{dk}{k}\right] \frac{c\,dz}{H(z)}.
\end{equation}

In order to compute the RMS of the circular polarization from Eq.~(\ref{eq:vrmsl}) we employ 21cmFAST \cite{mesinger2007efficient, mesinger201121cmfast}, with the Plick cosmology from \textit{Planck} and mostly all the default parameters from the code\footnote{We use Population III stars as the sources for early heating.}, to directly obtain the spin temperature and the Lyman-$\alpha$ flux (which is needed for the rate of depolarization by optical pumping) as functions of redshift. Then one can compute the optical depth of the neutral hydrogen gas with Eq.~(87) of \cite{Venumadhav} (i.e. Eq.~(\ref{eq:opti})). Moreover, using Eqs.~(98-99) of \cite{Venumadhav} and the Lyman-$\alpha$ flux, we compute the rate of randomizing spins by collisions $x_{{\rm c}, (2)}$ and Lyman-$\alpha$ photons $x_{\alpha, (2)}$. In addition, the matter power spectrum was extracted from CLASS with the {\slshape {Planck}} cosmological parameters (see Figure \ref{fig:plotk}) also we plotted the window function in terms of redshift in Figure \ref{fig:window}. We get
\begin{equation}
\label{eq:vrms}
V_{\rm rms} = 3.1 \times 10^{-16} \ \textup{K}.
\end{equation}

Our resulting circular polarization amplitude may be sensitive to the reheating and reionization history chosen in the 21cmFAST realization, since these control the window function $W$. The peak of $|W|$ occurs at the era when the Lyman-$\alpha$ coupling turns on ($x_\alpha$ of order $\sim 1$) since the window function is suppressed both in the limit of $x_\alpha\ll 1$ (where $T_s\approx T_{\rm CMB}$) {\em and} the limit where $x_\alpha\gg 1$ (where $T_s$ and $\tau$ are constant but $x_{\alpha,(2)}$ is large). In our fiducial model this happens at $z=z_{\rm trans}\sim 17$, and this is before the redshift $z_{\rm heat}$ when X-ray heating is significant. If we vary the Lyman-$\alpha$ coupling transition redshift $z_{\rm trans}$, but retain the assumption that $z_{\rm trans}>z_{\rm heat}$ -- so that the gas kinetic temperature $T_{\rm k}\propto (1+z)^2$ in accordance with adiabatic expansion -- then we have $W(z_{\rm trans})
\propto (1+z_{\rm trans})^{1/2}$, so over reasonable (factor of $\sim 2$) variations in the Lyman-$\alpha$ coupling redshift there are only minor changes in the implied window function and circular polarization properties. If X-ray heating took place before Lyman-$\alpha$ coupling, then $1-T_{\rm CMB}/T_s$ is suppressed (it may flip sign, but since $1/T_s$ is bounded, $|1-T_{\rm CMB}/T_s|$ can never be as strong in emission as it is for unheated gas in absorption), and also $\tau$ is suppressed -- thus the window function $W$ can be suppressed. We thus conclude that it would be difficult to increase the circular polarization signal by more than a factor of a few with standard physics\footnote{The EDGES experiment \cite{2018Natur.555...67B} has recently reported a stronger absorption dip in 21 cm radiation than would be expected even for no X-ray heating and $T_s\approx T_{\rm k}$ (but see, e.g., Ref.~\cite{2018arXiv180501421H}). If confirmed, this would imply a larger optical depth $\tau$ and greater circular polarization than conventional scenarios. Since this paper is focused only on mechanisms that operate in the conventional cosmological model without adding new physics, we do not consider this further here.}, but with X-ray heating it could be significantly suppressed. In any case, the Cosmic Dawn circular polarization signal is small both compared to potentially observable signals and to other sources of circular polarization.

Having done the analysis of the circular polarization produced by the spin polarized hydrogen atoms at the Cosmic Dawn epoch, $8.8<z<34$; we proceed to explore the same effect near recombination, $1+z_{rec} \sim 1000$. 

\section{Birefringence from spin polarized hydrogen atoms at recombination}
\label{sec:21z1100}

Birefringence from spin-polarized atoms should exist not just during reionization but also during the epoch of recombination. The main difference is that the spin alignment is not coming from the 21 cm radiation, but rather from the much more intense Balmer radiation (primarily H$\alpha$) present during recombination.

\subsection{Order of magnitude}

Aside from the actual mechanism behind the alignment of the spins, which is mainly present in the expression for ${\mathcal P}_{2m}$, the dependence of the cosmic birefringence on the physical parameters from the last section should remain intact. Thus we can use Eq.~(\ref{eq:n}) here with only two caveats. First, redshift factors must be evaluated at recombination, and we must construct an expression for the alignment tensor
\begin{equation}
n_x - n_y \approx \frac{\alpha n_{\rm H} c^3 \omega_{\rm hf}}{(\omega_{\rm me} \omega)^2} {\cal P}_{2m} \textup{.}
\label{eq:nsec2}
\end{equation}
The main issue is how to estimate ${\cal P}_{2m}$. This is determined by a balance between the isotropic component of the radiation bath, which acts to randomize the orientations of hydrogen spins (with a rate $R$) and the anisotropic components, which act to align the spins (with a rate $S_{2m}$):
\begin{equation}
\dot{\cal P}_{2m} = -R{\cal P}_{2m} + S_{2m}.
\label{eq:RS}
\end{equation}
The major randomizing processes that we would consider are Lyman-$\alpha$ scattering and 21 cm absorption/emission.\footnote{Rayleigh scattering does not flip the hydrogen spins, since unlike Lyman-$\alpha$ scattering it is non-resonant. Thus the energy denominators $1/(E_{\rm init} + h\nu - E_{\rm exc})$ in the scattering amplitude are approximately the same for all spin states, and the scattering amplitude is a Kronecker delta in the spin states. A similar argument applies to the $1s-2s$ two photon transition.} The rate of Lyman-$\alpha$ scattering per hydrogen atom is similar to the 2p$\rightarrow$1s decay rate, since 2p is mostly populated and de-populated via Lyman-$\alpha$: $R({\rm Ly}\alpha) \sim A_{{\rm Ly}\alpha}x_{2p} \sim (10^9\,{\rm s}^{-1})(10^{-14}) \sim 10^{-5}\,$s$^{-1}$. The rate of 21 cm absorption/emission is the natural rate $3\times 10^{-15}\,$s$^{-1}$ enhanced by the stimulated emission factor for the Rayleigh-Jeans tail of the CMB, $T_\gamma/T_\star \sim (3000\,{\rm K})/(68\,{\rm mK}) = 4\times 10^4$: thus $R(21\,{\rm cm})\sim 10^{-10}\,$s$^{-1}$. Thus, in terms of spin-randomizing processes, we can neglect the 21 cm interaction in comparison with Lyman-$\alpha$.

\begin{figure}[t]
\centering
\includegraphics[width=0.5\linewidth]{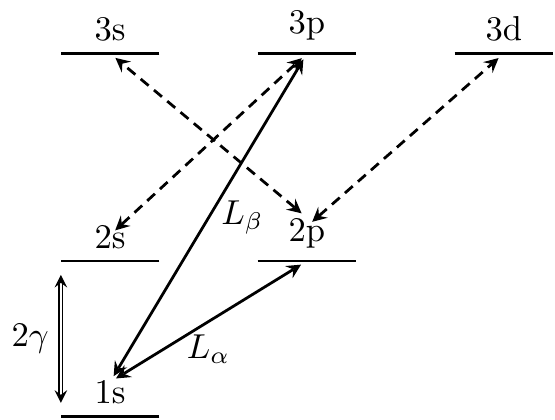}
\caption{Relevant processes for estimating the alignment tensor. The contribution from Lyman-$\beta$ and Lyman-$\alpha$ are represented by solid lines. The dashed lines represents the H$\alpha$ transitions. The $2\gamma$ transition will play a minor role in the Peebles model calculation.}
\label{fig:processes}
\end{figure}

Let us now turn to the contributions to the alignment, $S_{2m}$: one can expect the atoms to be polarized if the radiation field in any of the major atomic lines is anisotropic. We would first consider the Lyman series, since this interacts directly with the ground state, however the optical depth is so large (of order $10^9$ for Lyman-$\alpha$) that the radiation is extremely isotropic. Thus we instead consider the Balmer series, which is optically thin and has anisotropies equal to the continuum background radiation, i.e.\ of order $\Theta_{2m}\sim 10^{-5}$. At $z\sim 1100$, the CMB blackbody peak is at $\lambda\sim 1$ $\mu$m, and hence of the Balmer lines, H$\alpha$ will dominate the atom-CMB interactions. The alignment mechanism would be 2s$\rightarrow$3p$\rightarrow$1s (see Fig.~\ref{fig:processes}). The alignment rate would be $S_{2m}\sim A_{{\rm H}\alpha}x_{2s}e^{-h\nu_{{\rm H}\alpha}/k_{\rm B}T}\Theta_{2m}\sim (10^7\,{\rm s}^{-1})(10^{-14})(10^{-3})(10^{-5})\sim 10^{-15}\,$s$^{-1}$. We should also consider the anisotropy in the 21 cm line, coming from the Rayleigh-Jeans tail of the CMB anisotropies. If the CMB has an anisotropy $\Theta_{2m}$, then we would expect that in steady state (i.e.\ setting $\dot{\cal P}_{2m}=0$ in Eq.~\ref{eq:RS}) and considering only the 21 cm line, that the solution would be a spin temperature that differs by $\Delta T_s = \Theta_{2m}T_s$ depending on which excited state is used to define the spin temperature. This leads to ${\cal P}_{2m}({\rm steady\,state})\sim \Theta_{2m}T_\star/T_s\sim 2\times 10^{-5}\Theta_{2m}$ (since $T_s\approx T_\gamma$ in the recombination epoch), and hence $S_{2m}(21\,{\rm cm})\sim 2\times 10^{-5}R(21\,{\rm cm})\Theta_{2m}\sim 2\times 10^{-20}\,$s$^{-1}$. Thus we are justified in ignoring alignment by 21 cm in comparison to H$\alpha$.


Figure~\ref{fig:processes} shows the key radiative processes involving hydrogen atoms in energy levels up through the third shell. We have seen that the dominant randomizing term ($R$ in Eq.~\ref{eq:RS}) is through Lyman-$\alpha$ scattering with $R\sim 10^{-5}$ s$^{-1}$, while the dominant aligning term ($S_{2m}$) involves H$\alpha$ absorption by the $n=2$ level, with the $n=3$ intermediate level, followed by Ly$\beta$ emission to the $n=1$ level, here $S_{2m}\sim 10^{-15}$ s$^{-1}$.
This leads to ${\mathcal P}_{2m} \sim 10^{-10}$.
We would then obtain a birefringence of
\begin{equation}
\label{eq:21zrn}
 n_x - n_y \approx 10^{-40} \textup{,} 
\end{equation}
where we have evaluated the redshift dependent parameters at $1+z =1000$, namely $n_H \approx 2 \times 10^8 \ \textup{m}^{-3}$ and $\omega \sim 10^{15} \ \textup{rad s}^{-1}$ (approximate angular frequency of a CMB photon).
Using Eq.~(\ref{eq:phase}) for a photon of $ 1 \ \mu \textup{m}$ at the desired redshift, and a proper path length of the horizon size at recombination $\sim 100$ kpc (physical), we would have a phase shift of 
\begin{equation}
\label{eq:21zrp}
\Delta \phi \approx 10^{-12} \textup{.} 
\end{equation}

\subsection{Detailed calculation}
\label{ssec:21z1100}

We now perform a more detailed calculation of the circular polarization induced to the CMB by spin-polarized neutral hydrogen atoms at recombination, with an emphasis on estimating the alignment tensor. The key radiative processes are shown in Fig.~\ref{fig:processes}; we note that ${\cal P}_{2m}$ is both sourced by the CMB (hence H$\alpha$) anisotropy, but that there is also the randomizing effect of Lyman-$\alpha$ scattering.

First, let us deal with the randomization by Lyman-$\alpha$ photons. From \cite{Venumadhav}, we find that
\begin{equation}
\label{eq:y:1}
\dot{\mathcal{P}}_{2m} |_{{\rm Ly}\alpha} =-0.601 \times 6\pi \lambda^2_{{\rm Ly}\alpha}\gamma_{2p}\tilde{S}_{\alpha,(2)}J_\alpha \mathcal{P}_{2m} \textup{,}
\end{equation}
where $\lambda_{Ly\alpha} = 1215 \ \textup{\r{A}}$ is the wavelength of a Lyman-$\alpha$ photon, $J_\alpha$ is the flux of Lyman alpha photons on the blue side of the line, $\tilde{S}_{\alpha,(2)}$ corresponds to correction factors for the detailed frequency dependence of Lyman-$\alpha$ flux (we approximate this as 1 since the corrections are small at high temperature), and  $\gamma_{2p}$ is the half-width at half maximum (HWHM) of the 2p states, given by
\begin{equation}
\gamma_{2p} = \frac{ \Gamma_{2p}}{4 \pi} 
\approx 5.0\times 10^7 \ \textup{s}^{-1} \textup{,}
\end{equation}
where $\Gamma_{2p}$ is taken to be the natural width (i.e.\ the Einstein coefficient for 2p$\rightarrow$1s decay).
%
We use the cosmological recombination code {\sc HyRec} \cite{ali2011hyrec} to compute the Lyman-$\alpha$ flux, $J_\alpha = 0.353 \ {\rm m}^{-2}~{\rm s}^{-1}~{\rm Hz}^{-1}~{\rm sr}^{-1}$ at $1+z = 1000$. Therefore, the randomizing effect is given by
\begin{equation}
\label{eq:ly}
\dot{\mathcal{P}}_{2m} |_{{\rm Ly}\alpha} \approx -(2.97 \times 10^{-6} ~{\rm s}^{-1}) \  \mathcal{P}_{2m} \textup{.}
\end{equation}

The most direct aligning process is a combination of H$\alpha$ excitation and Lyman-$\beta$ decay: 2s$\rightarrow$3p$\rightarrow$1s. As previously noted, the ``source'' $S_{2m}$ for this process involves the abundance of atoms in 2s; the excitation rate to 3p (the Einstein coefficient $A_{3p\rightarrow 2s}$ times the photon phase space density $e^{-h\nu_{{\rm H}\alpha}/k_{\rm B}T}$); the branching fraction for 3p decays to 1s, $A_{3p\rightarrow 1s}/(A_{3p\rightarrow 1s}+A_{3p\rightarrow2s})$; and the CMB anisotropy $\Theta_{2m}$. We also expect a factor of $h\nu_{{\rm H}\alpha}/k_{\rm B}T$ in the Wien tail for conversion of the fractional temperature perturbation into a fractional anisotropy perturbation. The hardest part is the numerical pre-factor, which contains a long chain of Clebsch-Gordan coefficients, and is computed in Appendix~\ref{app:CG}.
Incorporating this into Eq.~(\ref{eq:ly}) gives:
\begin{equation}
\label{eq:rely}
\frac{d}{dt} (x_{1s} {\cal P}_{2m}) = -(2.97 \times 10^{-6} ~ {\rm s}^{-1})  {\cal P}_{2m} + \frac{1}{16\sqrt{10\pi}} x_{2s}  \frac{A_{3p \rightarrow 2s} A_{3p \rightarrow 1s}}{A_{3p \rightarrow 1s} + A_{3p \rightarrow 2s}} \Theta_{2m} \frac{ h \nu_{{\rm H}\alpha}}{k_{\rm B} T} e^{- {h \nu_{{\rm H}\alpha}}/{k_{\rm B} T}} \textup{.}
\end{equation}
Note the small pre-factor $\tfrac{1}{16\sqrt{10\pi}} \approx 0.011$ in the source term for polarized atoms, which arises from ``order unity'' effects that turn out to be small. We note that there are other excitation chains that start with anisotropic H$\alpha$ radiation and end with polarized atoms in 1s (e.g.\ 2p$\rightarrow$3d$\rightarrow$2p$\rightarrow$1s; 2s$\rightarrow$3p$\rightarrow$2s$\rightarrow$1s$+2\gamma$) but we have considered here only the most direct chain since there is some de-polarization at each step and we expect even smaller contributions from longer chains that end in 1s. We also ignore H$\beta$ and higher lines as they are farther into the Wien tail (hence less abundant) than H$\alpha$. We will thus proceed using Eq.~(\ref{eq:rely}).

Now we only need to obtain $x_{2s}$ to determine the strength of the alignment. The fraction of hydrogen atoms in the 2s level can be computed using the Peebles model of recombination \cite{peebles1968recombination}: 
\begin{equation}
\label{eq:x2}
x_{2} = 4\frac{\alpha_{\rm B} n_{\rm H} x^2_{e} +(\Lambda + \Lambda_\alpha) x_1 e^{-\hbar \omega_{Ly\alpha}/T k_{\rm B}} }{\Lambda + \Lambda_\alpha + 4 \beta} \textup{,}
\end{equation}
where $\omega_{Ly\alpha}$ is the frequency of the  Lyman-$\alpha$ photons, $\Lambda =8.2 \ \textup{s}^{-1}$ is the decay rate for the two-photon decay, $x_e \simeq 0.049423$ is the ionized fraction at $z = 1000$ obtained using {\sc HyRec}, $\Lambda_\alpha \approx 20.2 ~ \textup{s}^{-1}$ is the Lyman-$\alpha$ decay rate with optical depth suppression, $\alpha_{\rm B}$ is the case B recombination coefficient, and $\beta$ is the thermal photoionization rate from excited states. Then
\begin{equation}
\label{eq:x21}
x_2 \approx 5.209 \times 10^{-15} \textup{.}
\end{equation}
Hence the fraction of hydrogen atoms in the $2s$ level is  $x_{2s}  \approx \frac14 x_2 = 1.302 \times 10^{-15}$.
We thus rewrite Eq.~(\ref{eq:rely}) as
\begin{equation}
\frac{d}{dt} (x_{1s} {\cal P}_{2m}) = - (2.97 \times 10^{-6} \ {\rm s}^{-1}) \  {\cal P}_{2m} + (7.29 \times 10^{-13} \ {\rm s}^{-1} )\ \Theta_{2m} \textup{.}
\end{equation}
We furthermore ignore the left hand side since the change with time is small (probably of order $H{\cal P}_{2m}$, where the Hubble rate $H\sim 10^{-13}\,$s$^{-1}$), so the alignment tensor is given by ${\mathcal P}_{2m} \approx 2.45 \times 10^{-12}$.

Finally, we follow the steps used in Eq.~(\ref{eq:phasez20}) to compute the relative phase shift per Hubble time generated by the spin-polarized hydrogen atoms at recombination:
\begin{eqnarray}
\frac{d\phi}{d\ln a}
&=& \frac{\omega}{H}(n_{xx} - n_{yy})
\nonumber \\
&=& \frac{2\pi}{\sqrt{3}} \frac{\alpha  c^3 \omega_{\rm hf}}{\omega^2_{\rm me}} \frac{n_{\rm H}}{\omega H} ({\cal P}_{2,2} + {\cal P}_{2,-2}) 
\nonumber \\
\label{eq:phasez1000}
&=& 1.9 \times 10^{-13} \left( \frac{1+z}{1000} \right)^{1/2} \left( \frac{{\cal P}_{2,2} + {\cal P}_{2,-2}}{2.45 \ \times 10^{-12}} \right) \left( \frac{100 \ \textup{GHz}}{\rm \nu_{today}} \right) \textup{.} 
\end{eqnarray} 

Note that Eq.~(\ref{eq:phasez1000}) is not in agreement with the initial order-of-magnitude estimate, since it turns out that Lyman-$\alpha$ scattering lowers the atomic polarization to a level far below the CMB anisotropies. Also, the numerical pre-factors in the transition from 2s$\rightarrow$3p$\rightarrow$1s significantly lower the expected signal. Converting the above phase shift into circular polarization we obtain $V\approx 5.2 \times 10^{-19} \ {\rm K}$. This value is extremely small even in the context of the other signals studied in this paper. Note that in comparison to the signal from spin polarized atoms at low $z$ this signal is a factor of $1000$ smaller.

\section{Photon-photon scattering}
\label{sec:photon}

\begin{figure}[t]
  \centering
  \includegraphics[width=0.5\linewidth]{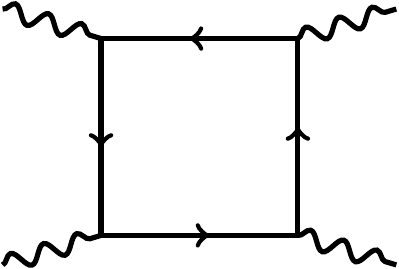}
  \caption{Photon-photon scattering, $\gamma \gamma \rightarrow \gamma \gamma$. The case of interest to us is where the two ingoing photons and two outgoing photons occupy the same pair of wavenumbers, so that the diagrams can interfere with free propagation and result in a phase shift.}
  \label{fig:fig2}
\end{figure}

In this section photon-photon forward scattering is treated as a possible source of cosmic birefringence. This is the non-linear behavior of the vacuum that emerges from the creation and annihilation of virtual electron-positron pairs (see Fig.~\ref{fig:fig2}). We expect this process to be most important shortly after recombination, when the radiation density is high.

\subsection{Order of magnitude}
We start using order of magnitude and dimensional analysis to obtain a rough estimate for the birefringence and subsequent phase shift of the vacuum. The difference in indices of refraction should be proportional to the anisotropic energy density of the CMB ${\cal U}_{\rm ani}$. Also, since the lowest-order photon-photon scattering diagram has 4 vertices and 4 electron propagators, we expect that it is proportional to $\alpha^2/m_e^4$.  The factors of $\hbar$ and $c$ can be inferred from dimensional analysis:
\begin{equation}
\label{eq:nphoton}
n_x - n_y \approx \frac{\alpha^2}{m_e c^2} \left(\frac{\hbar}{m_e c}\right)^3 {\cal U}_{\rm ani}\sim 10^{-35} \textup{,}
\end{equation}
where we have obtained the anisotropic energy density by multiplying the usual blackbody energy density evaluated at $1+z =1000$ by a factor of $10^{-5}$ (${\cal U}_{\rm ani} = 4 \times 10^{-7} \ \textup{J} \ \textup{m}^{-3}$). We have written this equation with the reduced Compton wavelength $\lambdabar_C = \hbar/(m_e c)$ for emphasis.

By means of Eq.~(\ref{eq:phase}), and for a photon of $1 \ \mu \textup{m}$ at $1+z =1000$ with a path length of $100$ kpc physical, we find a phase shift of
\begin{equation}
\label{eq:pphoton}
\Delta \phi \sim 10^{-7} \textup{.}
\end{equation}

\subsection{Detailed calculation}
To obtain a more accurate estimate, we use the Euler-Heisenberg Lagrangian equations of motion to construct an expression for the birefringence via the vacuum current. We are interested in the effect of photon-photon scattering in two beams of radiation with electric fields $\boldsymbol{E}_A$ and $\boldsymbol{E}_B$, and magnetic fields $\boldsymbol{B}_A$ and $\boldsymbol{B}_B$, respectively. The focus of attention is in the effective index of refraction that affects beam $A$, i.e. the birefringence felt by $A$ in the presence of $B$.

The Euler-Heisenberg Lagrangian (see e.g.~Ref.~\cite{Zavattini}) is  
\begin{equation}
\label{eq:pplag}
{\cal L}_{EH} = \frac{A_e}{\mu_0} \left[ \left( \frac{E^2}{c^2} - B^2 \right)^2 + 7 \left( \frac{\boldsymbol{E}\cdot \boldsymbol{B}}{c} \right)^2 \right] \textup{,}
\end{equation}
where $ A_e \equiv 2\alpha^2 \lambdabar^3_C/(45\mu_0 m_e c^2)$. The equations of motion, after including the classical EM Lagrangian, are 
\begin{equation}
\label{eq:vden}
\rho_{\rm vac} = \epsilon_0  \boldsymbol{\nabla} \cdot \boldsymbol{E} =  - \epsilon_0 A_e \Biggl\{ 2 \ \boldsymbol{\nabla} \cdot \left[ \left(\frac{\boldsymbol{E}^2}{c^2} - \boldsymbol{B}^2 \right) \boldsymbol{E} \right] - 7 \left[ \boldsymbol{\nabla} \cdot (\boldsymbol{E} \cdot \boldsymbol{B}) \right] \boldsymbol{B} \Biggr\} 
\end{equation}
and
\begin{eqnarray}
{\boldsymbol J}_{\rm vac} =  \frac1{\mu_0}\boldsymbol \nabla \times\boldsymbol B - \epsilon_0\frac{\partial \boldsymbol E}{\partial t} & = & A_e \Biggl\{ -\frac{2}{\mu_0} \boldsymbol \nabla \times \left[ \left(\frac{{\boldsymbol E}^2}{c^2} - {\boldsymbol B}^2\right) \boldsymbol B \right] + 7\epsilon_0 \boldsymbol \nabla \times \left[ (\boldsymbol E \cdot \boldsymbol B) \boldsymbol E \right] \nonumber \\ 
\label{eq:vcu}
&& -  2\epsilon_0 \frac\partial{\partial t} \left[ \left( \frac{{\boldsymbol E}^2}{c^2} - {\boldsymbol B}^2\right) \boldsymbol E \right] -7\epsilon_0 \frac\partial{\partial t} \left[ (\boldsymbol E \cdot \boldsymbol B) \boldsymbol B \right] \Biggr\} \textup{,} 
\end{eqnarray}
where ``$\rho_{\rm vac}$'' and ``${\boldsymbol J}_{\rm vac}$'' denote the charge and current densities associated with the polarization of the vacuum.

We next treat the vacuum polarization as a perturbation; the {\em unperturbed} electric field is then the incident field of the two interacting photon beams $A$ and $B$:
\begin{equation}
\boldsymbol{E} = \boldsymbol{E}_A e^{i(\boldsymbol{k}_A \cdot \boldsymbol{r}- \omega_A t)} + \boldsymbol{E}_B e^{i(\boldsymbol{k}_B \cdot \boldsymbol{r}- \omega_B t)} + {\rm c.c.,}
\end{equation}
and similarly for the magnetic field.

One can use the current to obtain the birefringence that affects beam $A$. The relevant terms are those with the same spatial and time dependences as beam $A$, i.e.\ $\propto e^{i(\boldsymbol{k}_A \cdot \boldsymbol{r}- \omega_A t)}$. These will also have dependences on the field amplitudes of the form $E_A E_B E^\ast_B$ or $E_A B_B B^\ast_B$. Furthermore, one can split the vacuum current into isotropic and anisotropic contributions:
\begin{equation}
J_i \ni i e^{i(\boldsymbol{k}_A \cdot \boldsymbol{r}- \omega_A t)}\left\{-\epsilon_0\omega_A ( \chi^{\rm iso}_e E^A_i + \chi^{\rm ani,e}_{ij} E^A_j ) + \frac{\chi^{\rm iso}_m \epsilon_{ijk} k^A_j B^A_k + \epsilon_{ijk} k^A_j \chi^{\rm ani,m}_{kl} B^A_l }{\mu_0} \right\}.
\end{equation}
Expanding the electric and magnetic fields in Eq.~(\ref{eq:vcu}), and keeping only the relevant terms for birefringence we obtain the expressions for the anisotropic susceptibilities
\begin{eqnarray}
\chi^{\rm ani,e}_{ij} & \simeq & - \frac{A_e}{c^2} \left[ 4 \left( \bigl< E^B_j E^{B\ast}_i \bigr> + \bigl< E^B_i E^{B\ast}_j \bigr> \right) + 7 \left( \bigl< B^B_j B^{B\ast}_i \bigr>  + \bigl< B^B_i B^{B\ast}_j \bigr> \right) \right] \textup{~and} \nonumber\\
\chi^{\rm ani,m}_{ij} & \simeq & A_e \left[ 4 \left( \bigl< B^B_j B^{B\ast}_i \bigr> + \bigl< B^B_i B^{B\ast}_j \bigr> \right) + 7 \left( \bigl< E^B_j E^{B\ast}_i \bigr> + \bigl< E^B_i E^{B\ast}_j \bigr> \right) \right]\textup{,}
\end{eqnarray}
where $\bigl< ... \bigr>$ indicates an average over the $B$ beam when we replace the monochromatic plane waves with realistic stochastic background radiation.

The anisotropic part of the index of refraction is then given by
\begin{eqnarray}
n_{xx} - n_{yy} & = & \frac{1}{2} \left( \chi^{\rm ani,m}_{xx}-\chi^{\rm ani,m}_{yy} + \chi^{\rm ani,e}_{xx} - \chi^{\rm ani,e}_{yy} \right) \nonumber \\ \label{eq:ggn}
& = & 3 A_e \left[ \frac{1}{c^2} \left( \bigl< E^B_x E^{B\ast}_x \bigr> - \bigl< E^B_y E^{B\ast}_y \bigr> \right) - \left( \bigl< B^B_x B^{B\ast}_x \bigr> - \bigl< B^B_y B^{B\ast}_y \bigr> \right) \right] \textup{.}
\end{eqnarray}

In contrast to the other circular polarization sources, here we will take into consideration the contribution due to both temperature {\em and} polarization, since the contribution of temperature anisotropies vanishes. Hence 
\begin{equation}
\bigl< X^B_i X^{B\ast}_j \bigr> = {\bigl< X^B_i X^{B\ast}_j \bigr>} _{\rm Temp} + {\bigl< X^B_i X^{B\ast}_j \bigr>}_{\rm  Pol} \textup{,}
\end{equation}
where $X \in \{E,B \}$ (no cross terms). Since the temperature contribution is given by 
 \begin{equation}
{\bigl< X^B_i X^{B\ast}_j \bigr>}_{\rm Temp} = C_X \int_{S^2} \frac{d{\cal U}}{d \boldsymbol{\Omega}} \frac{1}{2} \left( \delta_{ij} - \frac{k^B_i k^B_j}{k_B^2} \right) d^2 \boldsymbol{\Omega} \textup{,}
\end{equation}
with $C_E = c^2\mu_0/2$ and $C_B = \mu_0/2$, the contribution due to temperature anisotropies to Eq.~(\ref{eq:ggn}) vanishes because the magnetic and electric terms cancel identically.

However, the polarization contribution does not cancel. In what follows we adopt the polarization vectors used in Ref.~\cite{Venumadhav}:
\begin{equation}
\epsilon_{\pm} = \mp \frac{1}{\sqrt{2}} (\hat{\boldsymbol{\theta}} \pm i \hat{\boldsymbol{\phi}}) \textup{.}
\end{equation}
We expand the electric and magnetic fields using annihilation and creation operators:
\begin{eqnarray}
E_i^B (\boldsymbol x, t) & = & \int \frac{d^3 \boldsymbol k}{(2 \pi)^3} \sum_\alpha \Big( \sqrt{\frac{U_k}{2 \, \epsilon_0}} e^{i (\boldsymbol k \cdot \boldsymbol x - \omega t)} a_\alpha (\boldsymbol k ) \ \epsilon^\alpha_i  +  {\rm h.c.} \Big) \textup{~and}\nonumber \\
B_i^B (\boldsymbol x, t) & = & \int \frac{d^3 \boldsymbol{k}}{(2 \pi)^3} \sum_\alpha \Big( \sqrt{\frac{U_k}{2 \, \epsilon_0 \, c^2}} e^{i (\boldsymbol{k} \cdot \boldsymbol x - \omega t)} a_\alpha (\boldsymbol  k ) \ [\hat{\boldsymbol{k}} \times \epsilon^\alpha]_i  +  {\rm h.c.} \Big) \textup{,}
\end{eqnarray}
where $U_k=\hbar ck$ represents the energy of the photon with wavevector $k$. The annihilation and creation operators obey the standard commutation relations.\footnote{We normalize these so that $[ a_\alpha (\boldsymbol{k}),  a^{\dagger}_\beta (\boldsymbol{k}') ] = (2 \pi)^3 \delta (\boldsymbol{k} - \boldsymbol{k}') \delta_{\alpha \beta} $ and $[ a_\alpha (\boldsymbol{k}), a_\beta (\boldsymbol{k}') ] = [ a^{\dagger}_\alpha (\boldsymbol{k}), a^{\dagger}_\beta (\boldsymbol{k}') ] =0 $.}
The photon occupation, including polarization, is described by a phase-space density matrix $f_{\alpha\beta}$:
\begin{equation}
\langle a^{\dagger}_\alpha (\boldsymbol{k}), a_\beta (\boldsymbol{k}') \rangle =  (2 \pi)^3 \delta (\boldsymbol{k} - \boldsymbol{k}') f_{\beta \alpha} ({\boldsymbol{k}}).
\end{equation}
This can be decomposed in Stokes parameters; again we follow the normalization in Ref.~\cite{Venumadhav}:
\begin{equation}
f_{\alpha \beta} = \left( \begin{array}{cc}
f_I + f_V & -f_Q - i f_U \\
-f_Q + i f_U & f_I - f_V \end{array} \right) \textup{.}
\end{equation}

In order to get the polarization contribution, we need the linear polarization terms $f_Q$ and $f_U$ of beam B (i.e.\ the background CMB). In our case, beam $A$ is propagating on the $z$-axis. The remaining linear polarization contribution to the birefringence  has the form
\begin{eqnarray}
\bigl< E^B_x {E^B_x}^* \bigr> - \bigl<E^B_y {E^B_y}^{*} \bigr>  &\approx  & 2 f_Q  (1 + \cos^2 \theta)\cos 2 \phi - 4 f_U \cos \theta \sin 2 \phi \textup{~and}\nonumber \\
\bigl<B^B_x {B^B_x}^* \bigr> - \bigl< B^B_y {B^B_y}^{*} \bigr> & \approx & -2 f_Q  (1 + \cos^2 \theta)\cos 2 \phi + 4 f_U \cos \theta \sin 2 \phi \textup{.}
\end{eqnarray}
Using the previous relations we can finally construct an expression for the anisotropic index of refraction:
\begin{eqnarray}
n_{xx} - n_{yy} & \simeq & \frac{48 A_e}{\epsilon_0 c^2} \sqrt{\frac{\pi}{5}} \int \sqrt{\frac{U_k}{2}}k^2 dk \int d^2 \hat{k} \left[ f_Q \Re \Big\{{} _2 Y_{22} + ~_2 Y_{2,-2} \Big\} + f_U \Im \Big\{ {}_2 Y_{22} + _2 Y_{2,-2} \Big\} \right] \nonumber \\ \label{eq:ggdeltan}
& \approx & 96 \sqrt{\frac{\pi}{5}}  A_e \mu_0 a_{\rm rad} T^4_{\rm CMB} \Re a^E_{22} \textup{,}
\end{eqnarray}
where $_s Y_{lm}$ are the spin weighted spherical harmonics, $a^E_{2m}$ are the local quadrupole moments ($\ell=2$) of the $E$-mode polarization, and we have used the simplification that $a^{E}_{22} = a^{E \ast}_{2,-2}$.
The rate of change of relative phase per Hubble time due to photon-photon scattering is given by
\begin{eqnarray}
\frac{d\phi}{d \ln a} & = & \frac{\omega}{H} ( n_{xx} - n_{yy}) \nonumber \\
& = & \frac{128 \pi}{15} \sqrt{\frac{\pi}{5}} \frac{\alpha^2 \lambdabar^3_C}{m_e c^2}  \frac{\nu  a_{\rm rad} T^4_{\rm CMB} }{H}  \Re  a^E_{22}  \nonumber \\
\label{eq:ggphi}
& = &  8.7 \times 10^{-8} \left( \frac{\nu_{\rm today}}{100 \ \textup{GHz}}\right) \left(\frac{1+z}{1000} \right)^{7/2} \frac{ \Re a^E_{22} } {10^{-6}} \textup{,}
\end{eqnarray}
where we have estimated the contribution due to the local quadrupole of the polarization  with the expected 
CMB polarization at recombination, i.e of order $10^{-6}$, and substituted in $A_e$.

Note that Eq.~(\ref{eq:ggphi}) is consistent with Eq.~(\ref{eq:pphoton}). However, Eq.~(\ref{eq:ggphi}) is due to the local polarization anisotropies in contrast to the birefringence sourced by temperature anisotropies used in the order of magnitude section. The reason is that the temperature anisotropy contribution to the birefringence vanishes, which could not have been guessed from order-of-magnitude arguments; instead only the local polarization anisotropies generate birefringence. Further, note that the more detailed computation includes the dependence in the specific spin weighted spherical harmonic that causes the birefringence.

The signal from photon-photon scattering is the strongest source of circular polarization in the CMB studied in this work (although it is small), hence there is value in asking what will happen to our signal if we do the full integral over the line of sight and the Fourier modes contributing to both the initial linear polarization and the birefringence, and including geometrical suppression factors. This is particularly important since -- as discussed in \S\ref{ss:prev-pps} -- some previous results have been too optimistic because of order-unity factors.

As mentioned before, this section deviates from the cosmology used in the rest of the paper because it relies on extensive computations performed in an older cosmological model. Here we used the 2013F(CY) cosmology from Ref.~\cite{ade2016planck}, namely $\Omega_{\rm b}h^2 =0.02230$, $\Omega_{\rm cdm} h^2 = 0.1188$, $H_0 = 67.8$, $\Omega_m = 0.308$. The difference in cosmology should have a tiny effect since these parameters differ by $<3\%$ from those used in the rest of this paper.

In order to analyze the evolution of this signal we use our result for the birefringence of photon-photon scattering Eq.~(\ref{eq:ggdeltan}) in Eq.~(\ref{eq:Vr}) as we did in the previous line of sight computation for the birefringence from spin polarized hydrogen atoms from the Cosmic Dawn (\S\S\ref{sec:21z20}). Our objective is the same -- to compute $V_{\rm rms}$ -- but here we focus first on computing the power spectrum of the circular polarization since now both the source of linear polarization and the birefringence are coming from $z\sim 1000$ and hence are strongly correlated (so $V_{\rm rms}$ cannot be obtained by pointwise multiplication of two independent fields). We start by writing the CMB polarization as $P = Q + i \, U$, and by approximating $n_Q \, U - n_U \, Q \simeq \Re a^{\rm E}_{2,-2} \Im P - \Im a^{\rm E}_{2,-2} \Re P$. (We have used rotational symmetry to infer that $n_U$ should be derived from $n_Q$ by replacing $\cos2\phi_B$ with $\sin2\phi_B$, i.e.\ by replacing $\Re a^{\rm E}_{2,-2} $ with $\Im a^{\rm E}_{2,-2} $.) Then we can obtain the new circular polarization of the CMB at a comoving distance $s'$ from the last scattering surface as \begin{equation}
\label{eq:lsvr}
V \simeq A \int_{0}^{s'} ds \, (1 + z )^4 \Im \left\lbrace {a^{\rm E}_{2,-2}}^* P \right\rbrace  ,    
\end{equation}
where $a^{\rm E}_{2m}$ and $P$ are evaluated at position $s$ along the line of sight, and we extract all non-integrating parameters to define
\begin{equation}
A \equiv 96 \sqrt{\frac{\pi}{5}}\, A_e  \mu_0  a_{\rm rad} T^4_0 c^{-1} \omega_0 = 1.11\times 10^{-38} \left( \frac{\nu_0}{100\,\rm GHz} \right)\, {\rm m}^{-1}.
\end{equation}


We write both the polarization and the local quadrupole moments in Fourier space. In what follows we use the distant observer approximation ($kr \ll 1$) and assume that the surface of last scattering is narrow. Under these conditions we can rewrite both the quadrupole moments and the polarization, which are spin-2 quantities, by rotating them with the help of spherical harmonics. Then, for the local quadrupole moment in Eq.~(\ref{eq:lsvr}) we have
\begin{equation}
\label{eq:rotation}
a^{{\rm E}*}_{2,-2} (\boldsymbol{k}_1)
= D^{2*}_{-2,0}(\phi_{\boldsymbol{k}_1}, \theta_{\boldsymbol{k}_1}, 0) \bar a^{{\rm E}*}_{20}  (\boldsymbol{k}_1)
= \sqrt{\frac{4\pi}{5}} Y^*_{22}(\hat{\boldsymbol{k}}_1)
\bar a^{{\rm E}*}_{20} (\boldsymbol{k}_1).
\end{equation}
Here un-barred quantities are represented in the line-of-sight frame (with the $z$-direction toward the observer), and barred quantities are the wave vector frame ($\hat{\boldsymbol{k}}_1$ in the $\bar z$-direction and the direction to the observer in the $\bar x\bar z$-plane, with $\bar x<0$), and $D^{\ell}_{m',m}(\alpha,\beta,\gamma)$ is the usual Wigner rotation matrix \cite{wigner}.\footnote{The Wigner $D$-matrix applies to active rotation by the three Euler angles in the order $\gamma$ around $z$, $\beta$ around $y$, and $\alpha$ around $z$. In this application, we start with a wave vector $\boldsymbol{k}_1$ pointed toward the observer; the wave vector and the associated barred basis vectors are then rotated by $\phi_{\boldsymbol{k}_1}$ and then $\theta_{\boldsymbol{k}_1}$. Due to our choice of the $\bar x$-axis, no further rotation is necessary.} We have used the fact that we have scalar perturbations, so of the $\bar a^{\rm E}_{2m}$, only $m=0$ is non-zero, hence we do not need a summation over $m$ in Eq.~(\ref{eq:rotation}). Note that the $\boldsymbol{k}_1$ mode is responsible for birefringence.

We now need the CMB linear polarization for light propagating in the direction $\hat{\bf e}_z$ (i.e.\ toward the observer) from the last scattering surface. Assuming the last scattering surface was narrow, this will be dominated by the quadrupole ($\ell=2$) just after last scattering. We then write:
\begin{eqnarray}
P(\boldsymbol{k}_2, \hat{\bf e}_z)
& = & 
-\sum_{m=-2}^2 a_{2m}^{\rm E}(\boldsymbol{k}_2)\, {_2}Y_{2m}(\hat{\bf e}_z)
= -\sqrt{\frac{5}{4\pi}} a_{2,-2}^{\rm E}(\boldsymbol{k}_2)
\nonumber \\
& = & 
-\sqrt{\frac{5}{4\pi}} 
\, D^{2}_{-2,0}(\phi_{\boldsymbol{k}_2}, \theta_{\boldsymbol{k}_2}, 0) \bar a^{{\rm E}}_{20}  (\boldsymbol{k}_2)
= -Y_{22}(\hat{\boldsymbol{k}}_2)
\bar a^{{\rm E}}_{20} (\boldsymbol{k}_2) \, ,
\label{eq:rotpop}
\end{eqnarray}
where barred frame is now the wave frame for $\boldsymbol{k}_2$, and we have collapsed the sum using the fact that only the $m=-2$ spin-weighted spherical harmonic is non-zero and it evaluates to $\sqrt{5/4\pi}$. Note that for Eq.~(\ref{eq:rotpop}) to hold the assumption of narrow surface of last scattering is vital. Now it is clear what are the additional difficulties regarding this particular line of sight evolution with respect to the previous line of sight computation from \S\ref{ssec:21z20}. Here we have the CMB quadrupole moments entering from two different avenues: directly from the polarization, and from the birefringence. Furthermore, we switch the local quadrupole moments for appropriate polarization transfer functions\footnote{We follow the convention of CLASS for the transfer functions, e.g. $T^{\rm E}_{\ell} =- \frac{5}{4\sqrt{6}} (G^{(0)} + G^{(2)})$, see Eq.~(B.11) and related expressions in Ref.~\cite{tram2013optimal} or even \cite{ma1995cosmological} for more information.} and primordial curvature perturbations by using $\bar a^{\rm E}_{20} (c \, \eta, k) = T^{\rm E}_{20} (c \, \eta, k) \zeta(k)$. Then, Eq.~(\ref{eq:lsvr}) becomes 
\begin{eqnarray}
\label{eq:ls1}
V(\boldsymbol{r}_{\perp})  =  -\bar{A} \Im  \Big\{ && \int  \frac{d^3 \boldsymbol{k}_1}{(2 \pi)^3} \int \frac{d^3 \boldsymbol{k}_2}{(2 \pi)^3} \int d s \, (1+z)^4 \, T^{\rm E}_{20}(c \, \eta_{\rm LSS} + s, k_1) \, T^{\rm E}_{20}(c \, \eta_{\rm LSS}, k_2)   \nonumber \\
\label{eq:ls1}
&& Y_{22}(\hat{\boldsymbol k}_2) Y^*_{22}(\hat{\boldsymbol k}_1) \zeta^*({\boldsymbol k}_1) \zeta(\boldsymbol{k}_2) \exp{i(\boldsymbol{k}_{2 \perp} - \boldsymbol{k}_{1 \perp})\cdot \boldsymbol{r}_\perp -i k_{1 \parallel} s} \Big\} ,
\end{eqnarray}
where we have defined $ \bar{A} = \sqrt{\frac{4\pi}{5}} A$. Moreover, the complex exponentials come from the Fourier transformations of both the polarization and the quadrupole moment.  Since the $\boldsymbol{k}_2$ is the mode from the polarization, it should be in the $xy$-plane (surface of last scattering) where the line of sight is in the $z$-direction. Consequently, the exponential only has a contribution from $\boldsymbol{k}_{2 \perp} \cdot \boldsymbol{r}_\perp$. On the other hand, the exponential of $\boldsymbol{k}_1$ has a radial term because it cares about the propagation in the medium, i.e. it depends on the distance from the surface of last scattering to the intersection between the wavefront and the line of sight. 

The 2D power spectrum of the circular polarization as seen by a distant observer is
\begin{eqnarray}
(2\pi)^2 \delta (\boldsymbol{k}_\perp - \boldsymbol{k}'_\perp) P_V (k_\perp)
&=& \langle \tilde{V}^*(\boldsymbol{k}_\perp) \tilde{V}(\boldsymbol{k}'_\perp) \rangle
\nonumber \\
& = & \int d^2 \boldsymbol{r}_\perp \int d^2 \boldsymbol{r}'_\perp \langle V^*(\boldsymbol{r}_\perp) V(\boldsymbol{r}'_\perp) \rangle \exp{ i \boldsymbol{k}_\perp \cdot \boldsymbol{r}_\perp - i \boldsymbol{k}'_\perp \cdot \boldsymbol{r}'_\perp} \nonumber \\
\label{eq:lsvv}
&=& I_1 + I_2 + I_3 + I_4,
\end{eqnarray}
where we have expanded the imaginary part as Im$\,X = (X-X^\ast)/(2i)$, and used Eq.~(\ref{eq:ls1}) to write $V(\boldsymbol{r}_\perp)$ in terms of an integral over $\boldsymbol k_1$ and $\boldsymbol k_2$, and $V(\boldsymbol{r}'_\perp)$ in terms of an integral over $\boldsymbol k'_1$ and $\boldsymbol k'_2$.
We then separated the terms in the following way: $I_1$ contains the contribution from $\langle \zeta (\boldsymbol{k}_1) \zeta^* (\boldsymbol{k}_2) \zeta^* (\boldsymbol{k}'_1) \zeta (\boldsymbol{k}'_2) \rangle$, $I_2$ contains $\langle \zeta (\boldsymbol{k}_1) \zeta^* (\boldsymbol{k}_2) \zeta (\boldsymbol{k}'_1) \zeta^* (\boldsymbol{k}'_2) \rangle$, $I_3$ contains $\langle \zeta^* (\boldsymbol{k}_1) \zeta (\boldsymbol{k}_2) \zeta^* (\boldsymbol{k}'_1) \zeta (\boldsymbol{k}'_2) \rangle$ and $I_4$ contains $\langle \zeta^* (\boldsymbol{k}_1) \zeta (\boldsymbol{k}_2) \zeta (\boldsymbol{k}'_1) \zeta^* (\boldsymbol{k}'_2) \rangle$. We explicitly show the form of these integrals in Appendix~\ref{ap:integrals}.  

We use Wick's theorem to rewrite the 4-point functions into the relevant products of two point functions for our line of sight calculation. Generally the 4-point functions simplify to $(4-1)!!=3$ terms, however translation invariance guarantees that one of these will be zero for $\boldsymbol k_\perp\neq 0$. Therefore, every $I$ integral becomes two integrals, which for the sake of clarity we will refer to as $J$ integrals. To illustrate we show steps of the procedure for $I_1$:
\begin{eqnarray}
\label{eq:lswi1}
&&\langle \zeta (\boldsymbol{k}_1) \zeta^* (\boldsymbol{k}_2) \zeta^* (\boldsymbol{k}'_1) \zeta (\boldsymbol{k}'_2) \rangle = \langle \zeta (\boldsymbol{k}_1) \zeta^* (\boldsymbol{k}'_1) \rangle \langle \zeta^*(\boldsymbol{k}_2) \zeta (\boldsymbol{k}'_2) \rangle + \langle \zeta (\boldsymbol{k}_1) \zeta (\boldsymbol{k}'_2) \rangle \langle \zeta^* (\boldsymbol{k}_2) \zeta^*(\boldsymbol{k}'_2) \rangle \nonumber \\
&&= (2\pi)^6 P_\zeta(k_1) P_\zeta(k_2) \left[ \delta^3 (\boldsymbol{k}_1 - \boldsymbol{k}'_1) \delta^3 (\boldsymbol{k}_2 - \boldsymbol{k}'_2) + \delta^3 (\boldsymbol{k}_1 + \boldsymbol{k}'_2) \delta^3 (\boldsymbol{k}_2 + \boldsymbol{k}'_1) \right]. 
\end{eqnarray}
Then the $I_1$ integral becomes
\begin{eqnarray}
 I_1 &=& \frac{\bar{A}^2}{4} \int d^2 \boldsymbol{k}_{1, \perp} \int \frac{d k_{1,\parallel}}{2 \pi} \int \frac{d k_{2,\parallel}}{2\pi} \int ds \int ds' (1+z)^4 (1+z')^4 P_\zeta(k_1) P_\zeta(k_2) \delta^2 (\boldsymbol{k}_\perp -\boldsymbol{k}'_\perp)  \times  \nonumber \\ && \left\{ T^{\rm E}_{20} (c\,\eta_{\rm LSS}, k_1)\, T^{\rm E}_{20} (c \, \eta_{\rm LSS} + s') Y^*_{22}(\hat{\boldsymbol k}_2) Y_{22}(\hat{\boldsymbol k}_1) Y_{22}(\hat{\boldsymbol k}_2) Y^*_{22}(\hat{\boldsymbol k}_1) [T^{\rm E}_{20} (c\,\eta_{\rm LSS}, k_2)]^2 e^{i k_{1,\parallel}(s-s')} \right. \nonumber \\
&& + \, T^{\rm E}_{20} (c\,\eta_{\rm LSS} + s, k_1)\, T^{\rm E}_{20} (c\,\eta_{\rm LSS} + s', k_2) \, T^{\rm E}_{20} (c\,\eta_{\rm LSS},k_1)  T^{\rm E}_{20}(c\,\eta_{\rm LSS}, k_2) e^{ik_{1,\parallel}s} e^{- ik_{2,\parallel}s'} \times \nonumber \\ 
&& \left. [Y^*_{22}(\hat{\boldsymbol k}_2)]^2 [Y_{22}(\hat{\boldsymbol k}_1)]^2 \right\}
\nonumber \\ &=&
J_1 + J_2.
\label{eq:lsjints}
\end{eqnarray}
Note that we have already integrated over the $\boldsymbol{k}_{2,\perp}$ with the 2-D delta functions coming directly from the exponential factors in Eq.~(\ref{eq:lsvv}). For the case of $I_1$ this means that $\boldsymbol{k}_{2,\perp} = \boldsymbol{k}_{1,\perp} + \boldsymbol{k}_{\perp}$. Moreover, we can separate the integrals in Eq.~(\ref{eq:lsjints}) with respect to $s$ and $s'$ since they are decoupled, e.g.\ $J_1$ becomes
\begin{eqnarray}
J_1 &=& (2\pi)^2 \frac{\bar{A}^2}{4} \delta^2 (\boldsymbol{k}_\perp - \boldsymbol{k}'_\perp) \int \frac{d^2 \boldsymbol{k}_{1,\perp}}{(2\pi)^2} \int \frac{d k_{1,\parallel}}{2 \pi} \int \frac{d k_{2,\parallel}}{2\pi} |F(k_1, k_{1,\parallel})|^2 \,[T^{\rm E}_{20}(c\,\eta_{\rm LSS}, k_2)]^2  \nonumber \\ 
\label{eq:lsj1-main} && \times Y^*_{22}(\hat{\boldsymbol k}_2) Y_{22}(\hat{\boldsymbol k}_1) Y_{22}(\hat{\boldsymbol k}_2) Y^*_{22}(\hat{\boldsymbol k}_1) P_\zeta(k_1) P_\zeta(k_2),
\end{eqnarray}
where we define $F(k_{\rm magnitude}, k_{\rm parallel})$ as
\begin{equation}
\label{eq:lsfdef}
F(k_1, k_{1,\parallel}) = \int_{\eta_{\rm LSS}}^{\eta_{z \sim 100}} c \, d\eta \, T^{\rm E}_{20} (\eta,k_1)\, e^{i k_{1,\parallel}c (\eta - \eta_{\rm LSS})} \left( \frac{4 \Omega_{\rm m}}{H_0^2 \Omega_{\rm m}^2 \eta^2 - 4 \Omega_{\rm r}}\right)^4, 
\end{equation}
with $\Omega_{\rm m}$ being the density of matter in the Universe, $\Omega_{\rm r}$ the radiation density, and $H_0$ the Hubble constant. 
The full list of integrals $J_1...J_8$ is provided in Appendix~\ref{ap:integrals}, Eqs.~(\ref{eq:lsj1}--\ref{eq:lsj8}). Note that in the $J_1...J_4$ integrals, $\boldsymbol{k}_{2,\perp} = \boldsymbol{k}_{1,\perp} + \boldsymbol{k}_{\perp}$, but in the $J_5...J_8$ integrals, $\boldsymbol{k}_{2,\perp} = \boldsymbol{k}_{1,\perp} - \boldsymbol{k}_{\perp}$.

As a computational strategy, we focus first in the $F$ functions because we already had separated them from the remaining 4D integral; by pre-computing and tabulating them the 4D integral can be computed much faster. In order to tackle the object in Eq.~(\ref{eq:lsfdef}) one needs to tabulate and interpolate the appropriate polarization transfer function in terms of conformal time and wavenumber. We obtain our table for the interpolation of the transfer function running CLASS \cite{tram2013optimal} for $k$ values ranging from $0.0001  ~ {\rm to} \ 1 \ \textup{Mpc}^{-1}$ with a sampling of $\Delta k = 0.0005 \ \textup{Mpc}^{-1}$. Further, we interpolate with the help of the 2D cubic interpolation function from the GNU Scientific Library \cite{galassi2002gnu} . 

Once the $F$ functions have been obtained one can focus in the remaining 4D integrals. We restrict all the involved wavenumbers\footnote{i.e. $k_{1,\perp}$, $k_{1,\parallel}$, and $k_{2,\parallel}$.} to span from $0.0005 ~ {\rm to} \ 0.3 \ \textup{Mpc}^{-1}$. The choice of the upper limit is justified by CMB observations, in the sense that we want to guarantee that the peak of the CMB linear polarization ($\ell \sim 10^3$) is included. The comoving angular diameter distance at surface of last scattering is $D=1.4\times 10^4 \ \textup{Mpc}$, thus the maximum wavenumber is related to the maximum multipole by $\ell_{\max} = k_{\rm max} D$. Our choice of $k_{\rm max}=0.3$ Mpc$^{-1}$ corresponds to $\ell_{\rm max} = 4200$, well into the CMB damping tail.
We use Cartesian coordinates for simplicity and for the symmetries of the problem to be replicated by the numerical discretization.
We take a sampling of $\Delta k_{1x}=\Delta k_{1y}=0.0005$ Mpc$^{-1}$ with an offset from zero of $0.00025$ Mpc$^{-1}$ in order to avoid sampling the points $\boldsymbol k_{1,\perp}=0$ or $\boldsymbol k_{2,\perp}=0$.\footnote{The integrals are well-behaved as analytic functions at these points, e.g.\ $Y_{22}(\boldsymbol{\hat k}_1)$ goes to zero at $\boldsymbol k_{1,\perp}=0$, however the polar coordinates are ill-defined and the {\tt if} statements needed to handle this case would slow down the computation.}
We only integrate in the upper $(k_{1x},k_{1y})$-plane because most integrands are even in $k_{1y}$, except for the terms that are odd and hence vanish. Of course then one needs to rewrite the spherical harmonics in terms of $k_{1x}$, $k_{1y}$, $k_{1,\parallel}$ and $k_{2,\parallel}$. Moreover, the variance is a function of only the magnitude of $\boldsymbol{k}_\perp$, because the Universe is homogeneous and isotropic. Therefore, we can choose $\boldsymbol{k}_\perp = k_\perp \hat{\boldsymbol x}$ without losing generality, and  $\boldsymbol{k}_{2,\perp} = \boldsymbol{k}_{1,\perp} \pm k_{\perp} \hat{\boldsymbol x}$. 

There is still one significant simplification. By changing variables $\boldsymbol{k}^{\rm new}_1 = \boldsymbol{k}_2$ and $\boldsymbol{k}^{\rm new}_2 = \boldsymbol{k}_1$ for $J_5, J_6, J_7$ and $J_8$, we can analytically cancel the imaginary parts and reduce the number of integrals. Nevertheless, instead of assuming the analytical cancellation we decided to use the imaginary component as a null test, which provides a simple confirmation that our code produces sensible results.

\begin{figure}[t]
  \centering
  \includegraphics[width=0.6\linewidth]{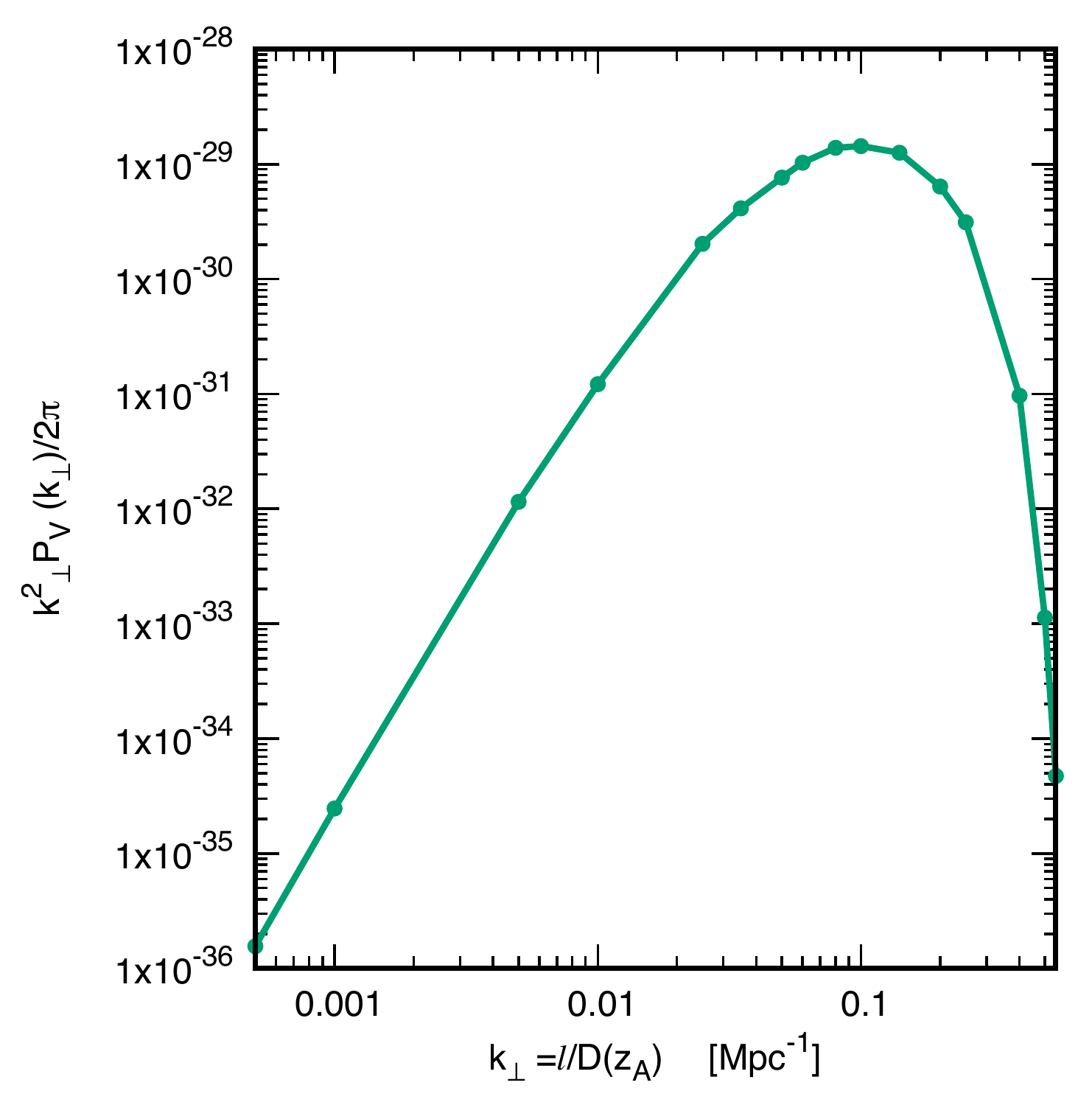}
  \caption{Circular polarization power spectrum from photon-photon scattering.}
  \label{fig:figvplot}
\end{figure}

The power spectrum of the circular polarization is related to the RMS of the circular polarization signal by 
\begin{equation}
V^2_{\rm RMS} = \int \frac{d^2 \boldsymbol{k}_\perp}{(2 \pi)^2} P_V (k_\perp).  
\end{equation}
Nevertheless, this power spectrum has never been studied before, so we decided to sample a few different $k_\perp$ with the objective of learning the behavior of this function both at low and high $k$, see Fig.~\ref{fig:figvplot}. We do not observe acoustic oscillations. However, we do note that the power spectrum has some expected features. At small $k$, we have $P_V(k_\perp)\propto k_\perp^2$. This occurs because we have a convolution integral --which is usually analytic as $k\rightarrow 0$ -- and in this case the circular polarization at $\boldsymbol k_\perp=0$ must be zero since then the observer line of sight, $\boldsymbol k_1$, and $\boldsymbol k_2$ are co-planar, and reflection symmetry prevents circular polarization from being generated. On the other hand, at large $k_\perp$ where we enter the CMB damping tail region, the power spectrum falls off.

Finally, integrating the power spectrum and taking the square root we get 
\comment{From code we get $V_{rms} \simeq 4.64 \times 10^{-15}= \frac{\bar{A}}{4} (\frac{15}{32 \pi}) \frac{2\pi^2}{(2\pi)^2} \frac{1}{2\pi} \int_0^0.55 P_V (k_\perp) k_\perp dk_\perp$, after this you must multiply by $2.73$ K to get units. The term with the $15$ comes from the spherical harmonics, the fraction with pis comes from the $\delta$ power divided by the 2pis in denominators in integrals.} 
\begin{equation}
\label{eq:vrsmphot}
V_{\rm RMS} \simeq 1.3 \times 10^{-14} \ \textup{K}.
\end{equation}
Note that this value is smaller than the estimate from the detailed calculation without the line of sight suppression based on Eq.~(\ref{eq:ggphi}), namely $V_{\rm RMS} \sim 1.4 \times 10^{-13} \ \textup{K}$. The difference of one order of magnitude between both methods is reasonable, especially when taking into account that in Eq.~(\ref{eq:ggphi}) we have ignored the radiation contribution to the Hubble parameter. 

\subsection{Comparison with previous calculations}
\label{ss:prev-pps}

As highlighted before our results are in apparent contradiction with Ref.~\cite{motie2012euler}. Their Eq.~(32) reports an upper limit for the phase shift of $10^{-13}$, six orders of magnitude less than our order of magnitude estimate. However, numerical evaluation of their expression (with factors of $\hbar$ and $c$ inserted) gives
\begin{equation}
\label{eq:oops}
\Delta \phi_{\rm FC} \simeq \frac{\pi^6 \alpha^2}{120 \zeta (3)} \left( \frac{k_{\rm B}T_0}{m_ec^2} \right)^4 \frac{k_{\rm B}T_0}{\hbar} C_Q \int^{1000}_0 \frac{(1+z)^5}{H(z)}dz \approx 1\, C_Q \sim 10^{-6} \textup{,}
\end{equation}
where $C_Q$ is the ratio of linear polarization intensity to the total intensity, and we evaluated the expression with the same parameters as Ref.~\cite{motie2012euler}: $\Omega_m =0.3$, $\Omega_\Lambda = 0.7$ and $H_0 = 75 \ \textup{km}\ \textup{s}^{-1} \ \textup{Mpc}^{-1}$. Equation~(\ref{eq:ggphi}) is consistent with this upper limit.

Our calculation also gives a smaller result than Ref.~\cite{sadegh}. In our notation, their Eq.~(12) for circular polarization generation in beam $A$ by a background beam $B$ reads:
\begin{equation}
\dot f_V({\boldsymbol k}_A) = \frac{8\alpha^2\hbar^4}{45m_e^4c^5\omega_A}f_U({\boldsymbol k}) \int \frac{d^3\boldsymbol k_B}{(2\pi)^3\,2\omega_B} \,(\omega_A\omega_B)^2 f_2(\hat{\boldsymbol k}_B,\hat{\boldsymbol k}_A) f_I({\boldsymbol k}_B),
\label{eq:s12}
\end{equation}
where $f_2$ is a term with trigonometric factors. However, if this equation is to be Lorentz invariant, then so long as $\hat{\boldsymbol k}_A\neq \hat{\boldsymbol k}_B$, one can boost to a frame where $\hat{\boldsymbol k}_A = \hat {\boldsymbol e}_3$ and $\hat{\boldsymbol k}_B=-\hat{\boldsymbol e}_3$. In this frame, we must have $f_2=0$, since the right-hand side is of spin 2 when rotating around the $z$-axis, but the left-hand side is of spin 0. Therefore on the right-hand side one should have the polarization rather than the intensity of the background beam $B$, which is weaker by a factor of $10^{-6}$. Thus the $V\sim 10^{-2}\,\mu$K found in Ref.~\cite{sadegh} should be reduced to $10^{-8}\,\mu$K, in line with the order of magnitude of our estimate.

Our result is also in conflict with the result from Ref.~\cite{sawyer2015photon}. They report a circular polarization of $V \sim 10^{-9} \sim 3\times 10^{-9}\ $K, which is smaller than our $V_{\rm rms} \sim 1.3 \times 10^{-14} \ \textup{K}$, i.e. a difference of $+ \,4.89$ dex. We track down the differences to several different contributing factors. First, we use a reference frequency of $100 \ \textup{GHz}$ while the thermal average frequency for the CMB photons is $\sim 154 \ \textup{GHz}$. This results in a difference of $+ \, 0.19$ dex. We also found some suppression from doing the full line-of-sight integral as opposed to an order of magnitude calculation using the Hubble length; this gives a difference of $+ \, 0.9$ dex because Ref.~\cite{sawyer2015photon} did an order of magnitude calculation. Also, in our order-of-magnitude estimate we use a linear polarization\footnote{The real RMS CMB linear polarization is approximately $ 4.47 \ \mu \textup{K} $.} of $P \sim 10^{-6} \sim 2.73 \ \mu\textup{K}$ while Sawyer used $P > 54.6 \ \mu\textup{K}$, based on the argument that one would select the highest-polarization region; this explains $+ \, 1.3$ dex. In contrast to Sawyer's phase shift of $1.1 \times 10^{-6}$ we found $\Delta \phi \sim 10^{-7}$. Part of this difference comes from their Eq.~(22), which we numerically evaluate as $2.7 \times 10^{-11}$ instead of $4\times 10^{-10}$; this explains a $+\,0.58$ dex difference. (Note that there remains a $+\,0.46$ dex difference in the phase shifts, which are likely due to factors of order unity on both sides.) In addition, there is a factor of $+ \, 1.6$ dex coming from how Ref.~\cite{sawyer2015photon} treated the relation between the standard deviation and the circular polarization, in particular in the jump from their Eq.~(24) to $V\sim 10^{-9}$. In conclusion, we believe the underlying physics in the calculation of Ref.~\cite{sawyer2015photon} is the same as in our calculation, and we can identify the main causes of the difference between both results. The detailed calculation leading to Eq.~(\ref{eq:vrsmphot}) is the most rigorous implementation yet of this physics.

\section{Static non-linear polarizability of hydrogen}
\label{sec:nonlinear}

In the previous section we considered the polarization of the {\em vacuum}. However, the post-recombination Universe was filled with a dilute gas of atoms, which also lead to photon-photon scattering due to their non-linear polarizability. Generally atoms are far more non-linear than the QED vacuum at CMB frequencies, but their filling fraction is tiny. Therefore it is not clear whether their contribution to photon-photon scattering is more or less important.

In this section, we consider the birefringence produced by the static non-linear polarizability of hydrogen given an anisotropic ambient medium. This process is most important shortly after the last scattering epoch, when the CMB has had time to free-stream and hence is anisotropic, but when the radiation density is still high. We will focus on the effect of the second order hyper-polarizability, which is the lowest-order effect that gives a diagram with two photons entering and two photons leaving. The two final-state photons occupy the same quantum states as the initial-state photons. We further focus on hydrogen, which is both more abundant than helium and has lower excitation energies.

\subsection{Order of magnitude}

We start with an order of magnitude estimate of both the birefringence and the phase shift using dimensional analysis. The difference in indices of refraction must be proportional to the number density of hydrogen atoms $n_{\rm H}$, the second order hyper-polarizability\footnote{This is a coefficient in the Taylor expansion of the dipole moment of the hydrogen atom: $\boldsymbol p = \alpha_d \boldsymbol E + \frac16\gamma E^2\boldsymbol E + ...$\;.} $\gamma$ and the anisotropic energy density of the electric field of the CMB, ${\cal U}^E_{\rm ani}$. The electrostatic constant $\epsilon_0$ may also appear. Thus, the order of magnitude of the birefringence is
\begin{equation}
\label{eq:nnonlinear}
n_x - n_y \approx \frac{\gamma n_{\rm H} {\cal U}^E_{\rm ani}}{\epsilon_0^2} \approx 10^{-37} \textup{,}
\end{equation}
where we have substituted at $z=1000$ that $n_{\rm H}\approx 2\times 10^8$ m$^{-3}$, and ${\cal U}^E_{\rm ani} = 4\times 10^{-7}$ J m$^{-3}$ (i.e. the energy density of a blackbody at 2725 K, multiplied by $10^{-5}$ to take the anisotropic part).
The hyper-polarizability $\gamma$ of a hydrogen atom should in principle be of order 1 in Bohr units, or $6.2\times 10^{-65}$ J m$^4$ V$^{-4}$.\footnote{The Bohr unit of electric field is 1 Hartree per elementary charge per Bohr radius, or $E_{\rm at} = 5.1\times 10^{11}$ V m$^{-1}$. The Bohr unit of energy is the Hartree, or $U_{\rm at} = 4.4\times 10^{-18}$ J. Therefore the Bohr unit of second-order hyper-polarizability is $U_{\rm at}/E_{\rm at}^4 = 6.2\times 10^{-65}$ J m$^4$ V$^{-4}$.} In fact, in atomic units, we have $\gamma = 1333.125$ \cite{1949PCPS...45..678S} or $8.3\times 10^{-62}$ J m$^4$ V$^{-4}$.
Using Eq.~(\ref{eq:phase}), and for a photon with wavelength $\sim 1$ mm today (1 $\mu$m at $z=1000$), and a path length of order 100 kpc physical (the horizon size at recombination), we find
\begin{equation}
\label{eq:pnonlinear}
\Delta \phi \approx 10^{-9} \textup{.}
\end{equation}

\subsection{Detailed computation}
\label{ssec:nonlinear}

Next, we do a more accurate order of magnitude estimate of the birefringence produced by the non-linear polarizability of hydrogen. We place a hydrogen atom in two beams of radiation, $A$ and $B$, with electric fields $\boldsymbol E_A$ and $\boldsymbol E_B$ respectively. Our interest is in the effective index of refraction that applies to beam $A$.

One can start with the energy of the hydrogen atom, which can be expanded in even powers of the applied electric field:
\begin{equation}
U = U_0 - \frac{1}{2} \alpha_d E^2 - \frac{1}{24} \gamma E^4 + ...\,;
\end{equation}
then the electric dipole moment is
\begin{equation}
\label{eq:nlpd}
\boldsymbol{p} = -\boldsymbol\nabla_{\boldsymbol E}U = \alpha_d \boldsymbol{E} +\frac{1}{6} \gamma E^2 \boldsymbol{E} +... \textup{,}
\end{equation}
where in this case the applied electric field is
\begin{equation}
\boldsymbol{E} = \boldsymbol{E}_A e^{-i \omega_A t} + \boldsymbol{E}_B e^{-i \omega_B t} + {\rm c.c.}\,.
\label{eq:vecE}
\end{equation}
The first term gives rise to the ordinary linear index of refraction, but no birefringence. For birefringence the important term is the second one in Eq.~(\ref{eq:nlpd}), particularly the terms that have an $e^{-i \omega_A t}$ time dependence but also contain $\boldsymbol E_B$. These terms are
\begin{equation}
p_i \ni
\frac{1}{3} \gamma \Bigl(\bigl< E^B_j E^{B\ast}_j\bigr> E^A_i + \bigl< E^B_j E^{B\ast}_i \bigr> E^A_j + \bigl<E^B_i E^{B\ast}_j\bigr>E^A_j \Bigr)e^{-i \omega_A t} \textup{.}
\end{equation}
Here $\bigl<...\bigr>$ will indicate an average over the beams $B$ when we replace the single monochromatic plane waves $B$ with a stochastic radiation background. This implies an effective nonlinear contribution to the polarizability tensor
\begin{equation}
\alpha^A_{{\rm NL}, ij} = \frac{\partial p_i|_{{\rm NL}, e^{-i\omega_At}}}{\partial E_j^A}
= \frac{1}{3} \gamma \Bigl(\bigl< E^B_k E^{B\ast}_k\bigr> \delta_{ij} + \bigl< E^B_j E^{B\ast}_i\bigr> + \bigl<E^B_i E^{B\ast}_j\bigr> \Bigr).
\end{equation}
The anisotropic part of the index of refraction is
\begin{equation}
n_{xx} - n_{yy} = \frac{n_{\rm HI}}{2\epsilon_0} (\alpha^A_{{\rm NL}, xx} - \alpha^A_{{\rm NL}, yy})
= \frac{\gamma n_{\rm HI}}{3\epsilon_0} \Bigl( \bigl< E^B_x E^{B\ast}_x \bigr> - \bigl<E^B_y E^{B\ast}_y \bigr> \Bigr),
\label{eq:n-aniso}
\end{equation}
where $n_{\rm HI}$ is the density of neutral hydrogen atoms.

In the presence of a background radiation field, we need to compute the expectation values $\bigl< E^B_x E^{B\ast}_x\bigr>$. For any single beam of radiation, the mean square electric field is $\bigl< |\boldsymbol E^B|^2 \bigr> = {\cal U}/(2\epsilon_0)$.\footnote{Since electric field energy density is $\frac12\epsilon_0E^2$, we would at first write down $2{\cal U}/\epsilon_0$. However there are two more factors of two in the denominator: one from the fact that only half of the energy density in an electromagnetic wave is in the electric field, and one from our convention in Eq.~(\ref{eq:vecE}) that ``$\boldsymbol E^B$'' is only the positive-frequency part of the wave.} For unpolarized radiation traveling in direction $\boldsymbol\Omega$, the fraction of the mean square electric field that is in the $x$-component is $\frac12(1-\Omega_x^2)$. Therefore, we have
\begin{equation}
\bigl< E^B_x E^{B\ast}_x\bigr> = \int_{S^2} \frac{1}{2\epsilon_0} \frac{d{\cal U}}{d\boldsymbol\Omega} \frac12(1-\Omega_x^2) \,d^2\boldsymbol\Omega,
\end{equation}
where $d{\cal U}/d\boldsymbol\Omega$ is the energy density in ambient electromagnetic waves per unit solid angle (i.e. in J m$^{-3}$ sr$^{-1}$), and similarly for the $y$-direction. Then Eq.~(\ref{eq:n-aniso}) reduces to
\begin{eqnarray}
n_{xx} - n_{yy} 
&=& -\frac{\gamma n_{\rm HI}}{12\epsilon_0^2} 
\int_{S^2} \frac{d{\cal U}}{d\boldsymbol\Omega} (\Omega_x^2-\Omega_y^2) \,d^2\boldsymbol\Omega
\nonumber \\
&=& -\frac{\gamma n_{\rm HI}}{12\epsilon_0^2} \sqrt{\frac{32\pi}{15}} \Re
\int_{S^2} \frac{d{\cal U}}{d\boldsymbol\Omega} Y_{22}(\boldsymbol\Omega) \,d^2\boldsymbol\Omega
\nonumber \\
&=& -\sqrt{\frac{32\pi}{135}}\,\frac{\gamma n_{\rm HI}}{\epsilon_0^2} a_{\rm rad}T_{\rm CMB}^4 \Re a_{22},
\label{eq:n-aniso2}
\end{eqnarray}
where $a_{\rm rad}$ is the radiation energy density constant, $a_{22}$ is the local quadrupole moment of the CMB, and there is a factor of 4 in the last step coming from converting temperature anisotropies to energy density anisotropies.

The rate of change of relative phase per Hubble time is related to the rate of change per unit comoving distance (Eq.~\ref{eq:phase}):
\begin{eqnarray}
\label{eq:angulo}
\frac{d\phi}{d\ln a}
&=& \frac{c}{aH} \frac{d\phi}{d\chi}
\nonumber \\
&=& \frac{\omega}{H}(n_{xx} - n_{yy})
\nonumber \\
&=& -\sqrt{\frac{32\pi}{135}}\,\frac{(1+z)\omega_{\rm today}}{H}\,\frac{\gamma n_{\rm HI}}{\epsilon_0^2} a_{\rm rad}T_{\rm CMB}^4 \Re a_{22}
\nonumber \\
&=& -1.2\times 10^{-9}\,\left( \frac{\nu_{\rm today}}{100\,\rm GHz} \right) \left( \frac{1+z}{1000} \right)^{13/2} x_{\rm HI} \left( \frac{\Re a_{22}}{10^{-5}} \right),
\end{eqnarray}
where we used the standard baryon abundance and denote by $x_{\rm HI}$ the fraction of the hydrogen that is neutral.

This is in agreement with the order-of-magnitude calculation, but includes the dependence on the specific spherical harmonic components of the radiation that causes the birefringence. The peak of the birefringence effect occurs at recombination, due to the power-law suppression at lower $z$ and the suppression of the neutral fraction and the CMB anisotropy at higher $z$. In any case, the peak effect is only at the $\sim 10^{-9}$ level. In terms of circular polarization Eq.~(\ref{eq:angulo}) converts into $V\approx 1.0 \times 10^{-15} \ \textup{K}$.

\section{Plasma delay: non-linear response of free electrons}
\label{sec:plasma}

Plasma delay is another non-linear polarization process, this time using the free electrons instead of the virtual pairs in the vacuum (\S\ref{sec:photon}) or the hydrogen atoms (\S\ref{sec:nonlinear}). It relies on the fact that recombination is not complete: there are still some free electrons $x_e>0$. These electrons are much less abundant than hydrogen atoms, however their excitation energy is zero so for CMB photon energies much less than the excitation energy of hydrogen we expect that electrons can be much more efficient at producing phase shifts than an equal number of hydrogen atoms.

Plasma delay can be described by two beams of light one of frequency $\omega_A$, and the perturbing beam with frequency $\omega_B$ both incident on a free electron. Plasma delay produces birefringence in the presence of an anisotropic radiation field. This effect should be stronger around recombination since the CMB is already anisotropic, there is still a significant fraction of ionized hydrogen, and the radiation density is still high. Note that this effect also has two incoming photons and two outgoing photons. Again, the interest is in the effective index of refraction that affects beam A.

\subsection{Order of magnitude}

As usual we start by doing the order of magnitude for the birefringence produced by the effect in consideration. In order to compute this, note that by dimensional analysis we expect factors of $e^2/\epsilon_0$, and combinations of factors of $c$, mass of the electron $m_e$ and frequencies of both the perturbing beam $\omega_B$ and the original beam $\omega_A$. Moreover, $n_x - n_y$ must be proportional to the number density of electrons $n_e$ and the anisotropic energy density of the perturbing beam (here the CMB). Birefringence should be generated by the third order susceptibility (in the sense that the responsible term in the current contains three factors of electric field), so we expect a factor of $(e^2/\epsilon_0)^2$ instead of $e^2/\epsilon_0$, and three factors of the mass of the electron. This leads to
\begin{equation}
\label{eq:nplasma}
n_x - n_y \approx \frac{e^4  {\cal U}^E_{\rm ani}  n_e}{\epsilon^2_0 m^3_e \omega^4 c^2} \approx10^{-40} \textup{,}
\end{equation}
where $n_e = x_e n_H= 9.9 \ \textup{cm}^{-3}$ at $1+z =1000$, and we approximate $\omega \sim 10^{15} \ \textup{rad s}^{-1}$, and ${\cal U}^E_{\rm ani} = 4 \times 10^{-7} \ \textup{J} \ \textup{m}^{-3}$.  Note that from dimensional analysis alone one could only say that the denominator must have a factor of frequency to the fourth power; when we derive the equations of motion we will learn what combinations of $\omega_A$ and $\omega_B$ actually appear. We will find that the correct expression contains $(\omega_B-\omega_A)^2$ in the denominator, and deal with the apparent divergence in \S\ref{ssec:plasma}.

Next, we use Eq.~(\ref{eq:phase}) for a photon with $\lambda \sim 1 \ \mu \textup{m}$ at $1 +z = 1000$, and with the usual path length of $100$ kpc (physical) relevant for the recombination epoch, we get
\begin{equation}
\label{eq:pplasma}
\Delta \phi \approx 10^{-11} \textup{.}
\end{equation}

\subsection{Rigorous calculation}
\label{ssec:plasma}

In this subsection we will provide a more rigorous calculation of the CMB circular polarization produced by free electrons. Along the way, we will find and regularize a singularity when $\omega_B=\omega_A$. Again, we are interested in the effective index of refraction felt by $E_A$. 

We start with the equations of motion for the electrons with incoming electric field
\begin{equation}
\boldsymbol E = \boldsymbol{E}^A e^{i( \boldsymbol{k}_A \cdot \boldsymbol{r}-\omega_A t)} + \boldsymbol{E}^B e^{i( \boldsymbol{k}_B \cdot \boldsymbol{r}- \omega_B t)} + {\rm c.c.}
\end{equation}
and magnetic field
\begin{equation}
\boldsymbol B = \boldsymbol{B}^A e^{i( \boldsymbol{k}_A \cdot \boldsymbol{r}-\omega_A t)} + \boldsymbol{B}^B e^{i( \boldsymbol{k}_B \cdot \boldsymbol{r}- \omega_B t)} + {\rm c.c.}\,,
\end{equation}
with $\boldsymbol B^A = \hat{\boldsymbol k}{^A}\times\boldsymbol E^A/c$, etc.
We take $\omega_B = \omega^{\rm R}_B + i \epsilon$ with $\epsilon \rightarrow 0^+$, so that beam $B$ had zero amplitude in the distant past. The displacement of the electrons obeys
\begin{equation}
\label{eq:eom}
\ddot{\xi}_i = -\frac{e}{m_e}\left(E_i+\xi_j E_{i,j} +\frac{1}{2}\xi_j \xi_k E_{i,jk} + \epsilon_{ijk} \dot{\xi}_j B_k + \epsilon_{ijk} \dot{\xi}_j \xi_\ell B_{k,\ell}  \right) \textup{.}
\end{equation}
The solution of this equation of motion can be expanded as
\begin{equation}
\xi_i = \xi^{(1)}_i + \xi^{(2)}_i + \xi^{(3)}_i+...\,,
\end{equation}
where $\xi_i^{(n)}$ denotes terms containing $n$ powers of the electric or magnetic fields.

We can use the solution for the displacement to obtain the current density and hence the susceptibility. The part of the current density we want is that with a $e^{i (\boldsymbol {k}_A \cdot \boldsymbol{r}-\omega_At)}$ dependence, and containing two powers of ${\boldsymbol E}_B$ or ${\boldsymbol B}_B$. This piece can be written in Fourier space as:
\begin{equation}
\label{eq:fucur}
\tilde{J}_i (\boldsymbol {k}_A) = \int d^3 \boldsymbol{r}\, e^{-i \boldsymbol {k}_A \cdot \boldsymbol {r} } J_i(\boldsymbol {r} ) = \int d^3 \boldsymbol {x}\, e^{-i \boldsymbol {k}_A \cdot \boldsymbol{x} }  n_e e \dot{\xi}_i  e^{-i \boldsymbol {k}_A\cdot \boldsymbol{\xi}}\textup{,}
\end{equation}
where we have used $\boldsymbol r$ to denote the Eulerian position of the electrons and $\boldsymbol x$ to denote their Lagrangian position. In the last integral in Eq.~(\ref{eq:fucur}), $n_e$ thus denotes the unperturbed (Lagrangian) electron number density. We define
\begin{equation}
J_i^{\rm L}(\boldsymbol x) = n_e e \dot{\xi}_i  e^{-i \boldsymbol {k}_A\cdot \boldsymbol{\xi}}\textup{,}
\label{eq:JIL}
\end{equation}
so that $\tilde{J}_i (\boldsymbol {k}_A)$ is the Fourier transform of $J_i^{\rm L}$ in Lagrangian space: $\tilde{J}_i (\boldsymbol {k}_A) = \int d^3 \boldsymbol {x}\, e^{-i \boldsymbol {k}_A \cdot \boldsymbol{x} } J_i^{\rm L}(\boldsymbol x)$. Conceptually, we can understand the exponential factor in Eq.~(\ref{eq:JIL}) as being associated with the retarded time for a ``downstream'' observer measuring beam A.

Since the problem is translation-invariant, we may evaluate $J_i^{\rm L}$ at the origin $\boldsymbol x=0$ without any loss of generality. Taking the previous result we can break the current into terms that act as isotropic and anisotropic electric and magnetic susceptibilities; the third order terms (superscript $^{(3)}$) are:
\begin{eqnarray}
\label{eq:pdcu}
 J_i^{\rm L(3)} = e n_e [e^{-i \boldsymbol {k}_A \cdot \boldsymbol{\xi}}\dot{\xi}_i ]^{(3)}
 &=& e n_e \left( \dot{\xi}_i^{(3)} - i k^A_j \xi^{(1)}_j \dot{\xi}_i^{(2)} -\frac{1}{2}k^A_j k^A_k \xi^{(1)}_j \xi^{(1)}_k \dot{\xi}^{(1)}_i - i k^A_j \xi^{(2)}_j \dot{\xi}^{(1)}_i \right) 
\nonumber \\
&=&
\epsilon_0 \chi^{\rm iso}_e \dot{E}_i + \epsilon_0 \chi^{\rm ani,e}_{ij} \dot{E}_j +\frac{1}{\mu_0} \chi^{\rm iso}_{m} \epsilon_{ijk} B^A_{k,j} +\frac{1}{\mu_0}\epsilon_{ijk} \chi^{\rm ani,m}_{k \ell} B^A_{\ell,j} \textup{.}~~~~~~
\end{eqnarray}
The first order solution of Eq.~(\ref{eq:eom}) is simply
\begin{equation}
\xi^{(1)}_i =\frac{e}{m_e} \left[ \frac{E^A_i}{\omega^2_A} e^{i(\boldsymbol{k}_A \cdot \boldsymbol{r} - \omega_A t)} + \frac{E^B_i}{\omega^2_B}e^{i(\boldsymbol{k}_B \cdot \boldsymbol{r} - \omega_B t)} + \frac{{E^A_i}^*}{\omega^2_A} e^{-i(\boldsymbol{k}_A \cdot \boldsymbol{r} - \omega_A t)} + \frac{{E^B_i}^*}{{\omega_B^*}^2}e^{-i(\boldsymbol{k}_B \cdot \boldsymbol{r} - \omega_B^* t)} \right] \textup{,}
\end{equation}
where $k_B \simeq k_B^{\rm R}$. The absence of magnetic fields is due to the electron being static at this order. The terms in the Lagrangian current density, that are relevant for the birefringence studied here, have two electric or magnetic fields from the perturbing beam (no mixed ones) and either a electric or magnetic field from the unperturbed beam. Additionally, these terms must contain $e^{-i \omega_A t }$. Likewise,  the relevant terms for birefringence of the third order term are given by
\begin{eqnarray}
\xi^{(3)}_i & \ni & - \frac{e^3}{m^3_e \omega^2_A} \left[ - \frac{{E^B_i}^* E^A_j E^B_k k^B_j k^A_k}{ \omega_B^2 (\omega_A + \omega_B)^2} + \frac{E^B_i E^A_j {E^B_k}^* k^B_j k^A_k}{ {\omega^*_B}^2 (\omega^*_B -\omega_A)^2} + \frac{E^A_k B^B_k {B^B_i}^*}{\omega_A (\omega_A - \omega^*_B)}  \right. \nonumber \\ 
&+& \left. \frac{E^A_k B^B_i {B^B_k}^*}{\omega_A (\omega_A + \omega_B)} + \epsilon_{ijk} \left( \frac{k^A_\ell E^B_\ell {E^B_j}^* B^A_k}{\omega^2_B \omega^*_B} - \frac{k^A_\ell E^B_j {E^B_\ell}^* B^A_k}{{\omega^*_B}^2 \omega_B} \right) \right. \nonumber \\ &+& \left. \epsilon_{jlk} \left( \frac{E^B_i{E^B_l}^*B^A_k k^B_j}{\omega^*_B (\omega_A -\omega_B)^2} + \frac{{E^B_i}^* E^B_l B^A_k k^B_j}{\omega_B (\omega_A+\omega_B)^2}\right) \right]{e^{i(\boldsymbol{k}_A \cdot \boldsymbol{r} - \omega_A t)}} \textup{.}
\end{eqnarray}
Therefore, the anisotropic electric susceptibility coming from third-order terms, keeping only those with the correct spatial and time dependence to interfere with the original beam $A$, are given by
\begin{eqnarray}
\chi^{\rm ani,e}_{ij}& \simeq & \frac{n_e e^4}{\epsilon_0 m^3_e \omega^2_A } \Biggl\{ k^B_j k^A_k \bigl<E^B_i {E^B_k}^* \bigr>\left[ \frac{1}{\omega_A {\omega_B^*}^2 (\omega_A + \omega_B)}  - \frac{1}{{\omega^*_B}^2 (\omega_B^* - \omega_A)^2}  \right.  \nonumber \\
&&- \frac{1}{\omega_A \omega_B (\omega^*_B -\omega_A)^2}+ \left.\frac{1}{\omega_B^2 (\omega_A + \omega_B)^2} - \frac{1}{\omega_A \omega_B^* (\omega_A + \omega_B)^2} - \frac{1}{\omega_A \omega^2_B (\omega_A - \omega_B^\ast)} \right] \nonumber \\
&& - \frac{\bigl< B^B_i {B^B_j}^* \bigr>}{\omega_A} \left[\frac{1}{\omega_A (\omega_A - \omega^*_B)} + \frac{1}{\omega_A (\omega_A + \omega_B)} \right] \Biggr\} \textup{,}
\end{eqnarray}
where $\bigl< ... \bigr>$ implies an average over the beam $B$, whose amplitude is stochastic since the CMB radiation is thermal. Therefore we need to integrate the susceptibility contributions both over direction of propagation and over frequency. We first perform the angular integration, keeping only the temperature perturbations and neglecting the smaller polarization signal:
\begin{equation}
\bigl<E_i^B E^{B\ast}_k \bigl> = \frac{1}{2 \epsilon_0} \int_{S^2} \frac{d {\cal U}}{d \boldsymbol{\Omega}} \frac{1}{2}\Big( \delta_{ik} - \frac{k^B_i k^B_k}{k^2_B}\Big) d^2 \boldsymbol \Omega \textup{.}
\end{equation}
A similar relation holds for the magnetic fields with the replacement $1/\epsilon_0\rightarrow \mu_0$. Choosing $\boldsymbol{k}_A = (0,0,k_A)$, and still working in the monochromatic case, the anisotropic electric susceptibility is given by
\begin{eqnarray}
\chi^{\rm ani,e}_{xx} - \chi^{\rm ani,e}_{yy} 
&=& \frac{n_e e^4}{\epsilon_0 m_e^3} \Biggl\{ -\frac{k_A k_B}{\epsilon_0 \omega^2_A} \sqrt{\frac{2\pi}{105}} \int_{S^2}  \frac{d {\cal U}}{d \boldsymbol{\Omega}} \Re Y_{32}(\boldsymbol{\Omega})  \left[\frac{1}{\omega_A {\omega^*_B}^2 (\omega_A + \omega_B)}  \right. \nonumber \\
&& \left. - \frac{1}{{\omega^*_B}^2 (\omega_A - \omega^*_B)^2}-\frac{1}{\omega_B \omega_A (\omega_A - \omega^*_B)^2} + \frac{1}{\omega^2_B (\omega_A+\omega_B)^2} 
\right. \nonumber \\
&& \left.
- \frac{1}{\omega_A \omega^*_B (\omega_A + \omega_B)^2} - \frac{1}{\omega^2_B \omega_A (\omega_A -\omega^*_B)}\right] d^2 \boldsymbol{\Omega} \nonumber \\
&&+ \mu_0 \sqrt{\frac{2 \pi}{15}} \int_{S^2}  \frac{d {\cal U}}{d \boldsymbol{\Omega}} \Re Y_{22}(\boldsymbol{\Omega}) \left[\frac{1}{\omega^3 (\omega_A + \omega_B)}+ \frac{1}{\omega_A^3 (\omega_A - \omega^*_B)}\right] d^2 \boldsymbol{\Omega} \Biggr\}\textup{.}~~~~~~~~
\end{eqnarray}
In an analogous way, we obtain the anisotropic magnetic susceptibility
\begin{eqnarray}
\chi^{\rm ani,m}_{xx} - \chi^{\rm ani,m}_{yy} &=& \frac{n_e e^4}{\epsilon_0^2 m_e^3 c^2}\Biggl\{ \sqrt{\frac{2\pi}{15}} \int_{S^2}  \frac{d {\cal U}}{d \boldsymbol{\Omega}} \Re Y_{22}(\boldsymbol{\Omega})  \nonumber \\
&& \times \left[\frac{1}{\omega^*_B \omega_B (\omega_A + \omega_B)^2} + \frac{1}{\omega_B \omega^*_B (\omega_A - \omega^*_B)^2} \right] d^2 \boldsymbol{\Omega} \nonumber \\
&& + \sqrt{\frac{2\pi}{105}} \frac{k_B}{k_A} \int_{S^2}  \frac{d {\cal U}}{d \boldsymbol{\Omega}} \Re Y_{32}(\boldsymbol{\Omega}) \nonumber \\ 
&& \times \left[\frac{2}{\omega_A \omega_B (\omega_A +\omega_B)^2} +\frac{2}{\omega_A \omega_B^\ast (\omega_A -\omega_B^\ast)^2} \right] \Biggr\}  
\end{eqnarray}

Combining both contributions we obtain the anisotropic index of refraction in the monochromatic scenario
\begin{eqnarray}
n_{xx}-n_{yy} &=&  \frac{n_e e^4}{\epsilon_0^2 m_e^3 c^2} \Biggl\{ - \sqrt{\frac{2\pi}{105}} \int_{S^2} \frac{d {\cal U}}{d \boldsymbol \Omega}  \frac{\omega_B}{\omega_A} \Re Y_{32}(\boldsymbol \Omega) W(\omega_A,\omega_B,\omega^\ast_B) \,d^2 \boldsymbol \Omega  \nonumber \\
&&+ \sqrt{\frac{2\pi}{15}} \int_{S^2} \frac{d {\cal U}}{d \boldsymbol \Omega} \Re Y_{22} (\boldsymbol \Omega) W_B(\omega_A,\omega_B,\omega^\ast_B) \,d^2 \boldsymbol \Omega \Biggr\}\textup{,}
\end{eqnarray}
where we define two frequency window functions $W$ and $W_B$:
\begin{eqnarray}
W(\omega_A,\omega_B,\omega^\ast_B) & \equiv &  \frac{1}{\omega_A \omega_B^2 (\omega_A + \omega_B)}   - \frac{1}{{\omega^\ast_B}^2 (\omega_B^\ast - \omega_A)^2} - \frac{1}{\omega_A \omega_B (\omega^\ast_B -\omega_A)^2} \nonumber\\
& & + \frac{1}{\omega_B^2 (\omega_A + \omega_B)^2} - \frac{1}{\omega_A \omega_B^\ast (\omega_A + \omega_B)^2} - \frac{1}{\omega_A \omega^2_B (\omega_A - \omega_B^\ast)} \nonumber\\
& & -\frac{2}{\omega_A \omega_B (\omega_A + \omega_B)^2} - \frac{2}{\omega_A \omega_B^\ast (\omega_A - \omega^\ast_B)^2},
\label{eq:exp-W}
\end{eqnarray}
and
\begin{eqnarray}
W_B (\omega_A,\omega_B,\omega^\ast_B) & \equiv & \frac{1}{\omega^3_A (\omega_A + \omega_B)} +\frac{1}{\omega^3_A (\omega_A - \omega^*_B)} \nonumber \\
&&+ \frac{1}{\omega^*_B \omega_B (\omega_A + \omega_B)^2} + \frac{1}{\omega_B \omega^*_B (\omega_A - \omega^*_B)^2}.
\label{eq:exp-WB}
\end{eqnarray}
Now we consider the full distribution of frequencies instead of treating $B$ as a monochromatic wave; this leads to
\begin{equation}
\label{eq:plasmabi}
n_{xx}-n_{yy} =  \frac{n_{\rm H} x_e e^4}{\epsilon^2_0 m^3_e c^2} \left[-\sqrt{\frac{ 2 \pi}{105}}(\Re a_{32}) \, I_{\rm T}(\omega_A) + \sqrt{\frac{2\pi}{15}}(\Re a_{22}) \, I^B_{\rm T}(\omega_A)\right] \textup{,}
\end{equation}
where $a_{32}$ is the local octopole moment of the CMB, $a_{22}$ is the local quadrupole moment, $x_e$ is the ionization fraction, and we have defined the frequency integrals
\begin{equation}
I_{\rm T}(\omega_A) = \int_0^\infty\frac{\omega_B^{\rm R}}{\omega_A} W(\omega_A,\omega_B,\omega^\ast_B)\,
T\frac{d}{dT} \left[ \frac{\hbar \omega_B^{\rm R\,3}}{4\pi^3 c^3 (e^{\hbar\omega_B^{\rm R}/k_{\rm B}T}-1)}
\right]\,d\omega_B^{\rm R}
\label{eq:ITDef}
\end{equation}
and
\begin{equation}
\label{eq:ITBDef}
I^B_{\rm T} (\omega_A) = \int_0^\infty W_B(\omega_A,\omega_B,\omega^\ast_B)\,  T\frac{d}{dT} \left[ \frac{\hbar \omega_B^{\rm R\,3}}{4\pi^3 c^3 (e^{\hbar\omega_B^{\rm R}/k_{\rm B}T}-1)}
\right]\,d\omega_B^{\rm R}.
\end{equation}
In Eqs.~(\ref{eq:ITDef}, \ref{eq:ITBDef}), the quantity in square brackets is $d{\cal U}/d\boldsymbol\Omega/d\omega_B^{\rm R}$, which depends on the CMB temperature $T$; its derivative $T\,d/dT$ is the conversion factor from CMB anisotropy units ($\Delta T/T$) to energy anisotropy $d{\Delta\cal U}/d\boldsymbol\Omega/d\omega_B^{\rm R}$.

We may furthermore expand $I_{\rm T}$ as $I_{\rm T} = I_1 - I_2 -I_3 +I_4 -I_5 +I_6 -I_7 - I_8$ and $I^B_T = I^B_1 - I^B_2 + I^B_3 + I^B_4$ using the terms in Eqs.~(\ref{eq:exp-W}) and (\ref{eq:exp-WB}). We explicitly show these integrals in Appendix \ref{ap:integrals2}, and show only one of the involved integrals here for pedagogical reasons -- in particular, to illustrate how we handle the $i\epsilon$ terms. For example
\begin{equation}
I_2 =\frac{\hbar^2}{4 \pi^3 c^3 k_{\rm B}T \omega_A} \int_0^\infty \frac{ e^{\hbar\omega^{\rm R}_B/k_{\rm B}T} {\omega^{\rm R}_B}^5 d\omega^{\rm R}_B }{(e^{\hbar\omega^{\rm R}_B/k_{\rm B}T} -1)^2(\omega^{\rm R}_B -i \epsilon)^2 (\omega^{\rm R}_B - \omega_A -i \epsilon)^2 } \textup{,}
\end{equation}
where the exponential factor comes from the black body spectrum. 

In order to simplify the integrals we utilize a combination of integration by parts and principal value, e.g.
\begin{equation}
\label{eq:pi1}
I_2 = - \frac{F(\omega^{\rm R}_B)}{(\omega^{\rm R}_B - \omega_A - i\epsilon)} \Big|^{\infty}_0 +  {\rm PV} \int^\infty_0  \frac{F' (\omega^{\rm R}_B) \ d \omega^{\rm R}_B}{\omega^{\rm R}_B - \omega_A} + i \pi F' (\omega^{\rm R}_B = \omega_A) \textup{,}
\end{equation} 
where the prime indicates differentiation with respect to $\omega^{\rm R}_B$, and 
\begin{equation}
F(\omega^{\rm R}_B) \equiv \frac{\hbar^2 }{4 \pi^3 c^3 k_{\rm B}T \omega_A} \frac{{\omega^{\rm R}_B}^3 e^{\hbar\omega^{\rm R}_B/k_{\rm B}T}}{(e^{\hbar\omega^{\rm R}_B/k_{\rm B}T}-1)^2} \left(1 + \frac{2i \epsilon}{\omega^{\rm R}_B} \right) \textup{.}
\end{equation}

We proceed to numerically integrate the principal value with special care for the redshift dependence in the temperature and frequency of the CMB photons. Furthermore, we include the imaginary terms for completion here, even though these relate to differential absorption on the two axes, which does {\em not} convert linear to circular polarization. We finally take the limit of $\epsilon \rightarrow 0$. We obtain the following values around recombination ($1+z =1000$), for $T=2.73 \ (1+z) \ \rm{K}$ and $\omega^{\rm R}_B = \omega_A = 2 \pi \times 10^{11} (1+z) \ \textup{ rad s}^{-1}$ 
\begin{equation}
I_{\rm T} = (9.23-35.2i)\times 10^{-62}\,{\rm kg}\,{\rm m}^{-1}\,{\rm s}^2
\end{equation}
and
\begin{equation}
I_{\rm T}^B = (-5.78 - 13i)\times 10^{-62}\,{\rm kg}\,{\rm m}^{-1}\,{\rm s}^2.
\end{equation}
We estimate the order of magnitude of the birefringence due to plasma delay by replacing the previous values into Eq.~(\ref{eq:plasmabi}) to get
$\Delta n \approx (-7.31-0.24i)\times 10^{-40}$.
Moreover, the subsequent phase shift is given by 
\begin{eqnarray}
\frac{d \phi} {d \ln a} & = & \frac{\omega}{H} (n_{xx} - n_{yy}) \nonumber \\
& = & \frac{e^4}{ \epsilon^2_0 m^3_e c^2} \frac{n_{\rm H} x_e  \omega }{H} \left(- \sqrt{\frac{2 \pi}{105}} \Re a_{32} \Re I_{\rm T} + \sqrt{\frac{2\pi}{15}} \Re a_{22} \, I^B_{\rm T} \right) \nonumber \\ 
\label{eq:plasmacir}
& = & -2.4 \times 10^{-10} \Big(\frac{1 +z }{1000} \Big)^{5/2} x_e \Big( - \sqrt{\frac{2 \pi}{105}} \frac{\Re a_{32}}{10^{-5}}  + \sqrt{\frac{2\pi}{15}}\frac{ \Re a_{22}}{10^{-5}} \Big) \textup{.}
\end{eqnarray}

Note that this result is consistent with Eq.~(\ref{eq:pplasma}) estimated in the previous section. Nevertheless, Eq.~(\ref{eq:plasmacir}) is more rigorous because it includes the specific spherical harmonics that are involved in this effect. In addition, it addresses the singularity that occurs at second order or higher in the equations of motion when $\omega_A = \omega_B$.

In the end, the circular polarization produced, $V\approx 2.6 \times 10^{-17}\ {\rm K}$, is small even in comparison to the other sources of circular polarization studied in this work.

\begin{table}
\centering
\caption{Summary table for sources of circular polarization at an observed frequency of 100 GHz. The final column shows the level of detail of the calculation (A = detailed numerical evaluation of all line of sight factors and power spectra; B = the RMS was determined with all relevant numerical pre-factors; C = the numerical pre-factors in the birefringence were computed, but the full statistics of the line of sight integral were not computed).}
\label{tab:table2}
\begin{tabular}{l|c|c}
\hline \hline
Birefringence source  & Circular Polarization & Level of
\\ & $V$ $(\mu  \textup{K})$ & detail \rule{0pt}{2.5ex}  \\
\hline 
Photon-Photon scattering & $1 \times 10^{-8}$ & A  \rule{0pt}{2.5ex} \\
Static non-linear polarizability of hydrogen  & $1 \times 10^{-9}$ & C \rule{0pt}{2.5ex} \\
Spin polarized H atoms (low z, aligned by 21 cm)  & $3 \times 10^{-10}$ & B \rule{0pt}{2.5ex}\\
Plasma delay & $3 \times 10^{-11}$ & C \rule{0pt}{2.5ex}\\
Spin polarized H atoms (high z, aligned by H$\alpha$)  & $5 \times 10^{-13}$ & C \rule{0pt}{2.5ex}\\
\hline \hline
\end{tabular}
\end{table}

\section{Conclusion}
\label{sec:conclusion}

We have summarized our results for the circular polarization of the CMB due to the conventional sources of cosmic birefringence considered in this work in Table~\ref{tab:table2}. We see that of the cosmological signals, all are very small compared to the linear polarization. The largest of the cosmological contributions comes from photon-photon scattering during the epoch of recombination ($z\sim 1000$ in our model). The predicted RMS circular polarization for our fiducial model is 13 fK; this may vary somewhat for more accurate computations, but in any case the signal is tiny compared to present observational sensitivities or even compared to dominant foregrounds such as Galactic synchrotron emission, which at $\nu = 10 \ \textup{GHz}$ could potentially reach circular polarization levels of $\sim 10^{-8} \ \textup{K}$ \cite{King}. At higher frequencies, the Galactic foreground situation is improved: the circular polarization from synchrotron is expected to fall as $V\propto\nu^{-3.5}$ but is still large compared to the expected signal, so $\sim 4$ pK at 100 GHz.\footnote{The foreground temperature scaling is in Rayleigh-Jeans units, so requires a factor of 1.29 to convert to blackbody temperature at 100 GHz. The fractional circular polarization $V/I\sim 1/\gamma \propto \nu^{-1/2}$, so circular polarization will have another $-1/2$ power of frequency relative to temperature.} Note that thermal dust emission -- the dominant foreground at $>100$ GHz -- would be expected to have very low circular polarization, even though the grains are aligned, because circularly polarized emission requires a phase delay between the $x$ and $y$ axes. This phase delay and hence a net circular polarization can be imprinted on the thermal radiation if the grain is rotating (see e.g.\ \S4.2 of Ref.~\cite{M72}), but this is very weak (of order $\Omega_{\rm grain}/\omega$) will only lead to circular polarization if the grain angular momenta have net vectorial alignment (e.g.\ spin parallel to the ambient magnetic field more or less likely than anti-parallel), which is not necessarily expected in radiative grain alignment theory \cite{LH07}.\footnote{In the Larmor precession angle-averaged calculation -- \S7 of Ref.~\cite{LH07} -- right-handed grains will prefer to spin one direction and left-handed grains the other direction. A racemic mixture would not show vectorial alignment, but would still have the ``headless vector'' alignment that produces linear polarization.} Thermal magnetic dipole emission from grains \cite{DL99} could yield a stronger circular polarization signal, but again only if vectorial alignment of the grains (this time between the permanent magnetic moment and the ambient magnetic field) is achieved. We leave investigation of all of these possibilities to future work; in any case, it is clear that there is at least one foreground (synchrotron) that is far larger than the conventional cosmological signals.

In conclusion, none of the conventional sources of circular polarization studied in this work are likely to be important in the foreseeable future. However, one could study {\em non-standard} sources of circular polarization as a channel for new physics. For example: Lorentz invariance violations \cite{colladay1998lorentz,Zarei}, primordial magnetic fields \cite{Zarei}, non-commutative gauge theories \cite{aschieri2003noncommutative}, and scattering with the cosmic neutrino background \cite{Mohammadi}; see Table 1 in \cite{King} for a summary of these sources of circular polarization. We view these searches as more promising in light of the tiny contribution of conventional cosmological effects.

\appendix

\section{Source term for polarized atoms from CMB anisotropies}
\label{app:CG}

This appendix considers the source term in Eq.~(\ref{eq:rely}). The main text motivates the ``order of magnitude'' of this term, however a detailed treatment of the transition probabilities to each level in the 2s$\rightarrow$3p$\rightarrow$1s scattering process is needed in order to get the correct numerical pre-factor for this process. Note that the reported pre-factor is $\tfrac1{16\sqrt{10\pi}} \approx 0.011$, so in fact the suppression due to numerical pre-factors is very significant even if the factors are in some sense ``of order unity.'' This is a common phenomenon in polarization problems.

Let us consider a hydrogen atom with the electron in the 2s orbital in a random spin state, located in a CMB background with quadrupole moment $\Theta_{2m}$. For simplicity, we consider here the case where only $\Theta_{20}\neq 0$, so that the radiation field is axisymmetric around the $z$-axis, i.e. $T = T_{\gamma} [ 1 + \Theta_{20} Y_{20}(\theta,\phi) ]$. This axisymmetry also prevents any off-diagonal terms in the density matrix between states of different total magnetic quantum number, which simplifies our arguments. Our objective is to suppose that the atom undergoes 2s$\rightarrow$3p$\rightarrow$1s scattering (by absorbing an H$\alpha$ photon and then emitting a Ly$\beta$ photon), and then compute the density matrix of the final 1s state.

The first step is to recognize that the excitation goes to either the $3p_z$ orbital (orbital quantum number $m_l=0$) or the $3p_x,3p_y$ orbitals ($m_l=\pm 1$), but with a random spin state. The relative probabilities of going to the different orbitals is determined by the phase space densities of photons where the receiving electric dipole antenna is aligned on the different axes (the $z$-axis vs.\ in the $xy$-plane). An electric dipole on the $z$ axis sees a beam-pattern-averaged temperature of
\begin{equation}
T_z = T_\gamma \left[ 1 + \Theta_{20} \int B(\theta,\phi) Y_{20}(\theta,\phi) \,\sin\theta\,d\theta\,d\phi \right] = T_\gamma \left[ 1 - \frac{\Theta_{20}}{\sqrt{20\pi}} \right],
\end{equation}
where $B(\theta,\phi) = \frac{3}{8\pi}\sin^2\theta$ is the beam pattern normalized to integrate to unity. Similarly, electric dipoles on the $T_x$ and $T_y$ axes see a temperature of
\begin{equation}
T_{xy}\equiv T_x=T_y = T_\gamma \left[ 1 + \frac{\Theta_{20}}{\sqrt{80\pi}} \right].
\end{equation}
Now the phase space density for the exciting H$\alpha$ photons for absorption with the electric field on the $z$-axis is
\begin{equation}
f_z = \frac1{e^{h\nu_{{\rm H}\alpha}/k_{\rm B}T_z} - 1}
= \frac1{e^{h\nu_{{\rm H}\alpha}/k_{\rm B}T_\gamma} - 1} \left[ 1 - \frac1{\sqrt{20\pi}}
\frac{1}{1-e^{-h\nu_{{\rm H}\alpha}/k_{\rm B}T_\gamma}}
\frac{h\nu_{{\rm H}\alpha}}{k_{\rm B}T_\gamma} \Theta_{20} \right],
\end{equation}
where we perform a Taylor expansion. The phase space density for excitations with electric fields in the $xy$-plane is similar but with the replacement of the factor $-1/\sqrt{20\pi}$ with $+1/\sqrt{80\pi}$. We thus conclude that the probabilities for excitation to the 3p orbitals are
\begin{equation}
P_{\rightarrow m_l=-1}:P_{\rightarrow m_l=0}:P_{\rightarrow m_l=1}
= \tfrac13-\tfrac12\delta_p:\tfrac13+\delta_p:\tfrac13-\tfrac12\delta_p,
\label{eq:Pratio}
\end{equation}
where
\begin{equation}
\delta_p = - \frac1{3\sqrt{20\pi}}
\frac{1}{1-e^{-h\nu_{{\rm H}\alpha}/k_{\rm B}T_\gamma}}
\frac{h\nu_{{\rm H}\alpha}}{k_{\rm B}T_\gamma} \Theta_{20} .
\end{equation}

Our next step is to note that fine+hyperfine structure splits 3p into 12 quantum states (3 orbital states, 2 electron spin, 2 nuclear spin). The excitation puts the atom into orbital states in proportion to Eq.~(\ref{eq:Pratio}), and random electron and nuclear spin states, but $m_l$ is not a conserved quantum number: instead, we should take our probabilities and project them into the $|j;FM_F\rangle$ basis using the Clebsch-Gordan coefficients. (Here again $j$ is the electron total angular momentum, $F$ is the total angular momentum including nuclear spin, and $M_F$ is its $z$-projection.) The four $m_l=0$ states can be expressed in the $|m_l, m_s, M_I\rangle$ basis as
\begin{eqnarray}
|0,\tfrac12,\tfrac12\rangle &=& \sqrt{\tfrac12} |\tfrac32;21\rangle - \sqrt{\tfrac16} |\tfrac32;11\rangle - \sqrt{\tfrac13}|\tfrac12;11\rangle,
\nonumber \\
|0,\tfrac12, -\tfrac12\rangle &=& \sqrt{\tfrac13}|\tfrac32;20\rangle + \sqrt{\tfrac13}|\tfrac32;10\rangle
  - \sqrt{\tfrac16}|\tfrac12;10\rangle - \sqrt{\tfrac16} |\tfrac12;00\rangle, \nonumber \\
|0, -\tfrac12, \tfrac12\rangle &=& \sqrt{\tfrac13}|\tfrac32;20\rangle - \sqrt{\tfrac13}|\tfrac32;10\rangle
  + \sqrt{\tfrac16}|\tfrac12;10\rangle - \sqrt{\tfrac16} |\tfrac12;00\rangle, ~~~~{\rm and} \nonumber \\
|0,-\tfrac12, -\tfrac12\rangle &=& \sqrt{\tfrac12}|\tfrac32;2-1\rangle + \sqrt{\tfrac16}|\tfrac32;1-1\rangle
  + \sqrt{\tfrac13}|\tfrac12;1-1\rangle .
  \label{eq:CG}
\end{eqnarray}
Because the fine and hyperfine splittings in 3p are larger than the natural state width, the phases of the different $j$ and $F$ states will be randomized before the atom decays. Thus, if the excitation goes to the $m_l=0$ orbital with the 4 spin states chosen at random, then the probabilities for the excited states in the $|j;FM_F\rangle$ basis are:
\begin{list}{$\bullet$}{}
\item Probability $\tfrac16$ for each of $|\tfrac32;20\rangle$ and $|\tfrac32;10\rangle$;
\item Probability $\tfrac18$ for each of $|\tfrac32;21\rangle$ and $|\tfrac32;2-1\rangle$;
\item Probability $\tfrac1{12}$ for each of $|\tfrac12;11\rangle$, $|\tfrac12;10\rangle$, $|\tfrac12;00\rangle$, and $|\tfrac12;1-1\rangle$;
\item Probability $\tfrac1{24}$ for each of $|\tfrac32;11\rangle$ and $|\tfrac32;1-1\rangle$;
\item Probability 0 for each of $|\tfrac32;22\rangle$ and $|\tfrac32;2-2\rangle$.
\end{list}

We now consider the Lyman-$\beta$ decay to the 1s level.
The Clebsch-Gordan coefficents and the Wigner-Eckart theorem can then be used to determine the amplitudes for decay of each of these states to the four $|1s: FM_F\rangle$ states. The decay rate for $|(3p)j;FM_F\rangle \rightarrow |(1s)\tfrac12;F'M'_F\rangle$ is proportional to (see Eq.~B2 of Ref.~\cite{Hirata2005} and use the Wigner-Eckart theorem):
\begin{eqnarray}
{\rm Prob}[|(3p)j;FM_F\rangle \rightarrow |(1s)\tfrac12;F'M'_F\rangle]
&\propto & (2j+1)(2F'+1)
\left\{ \begin{array}{ccc} 1 & j & \tfrac12 \\ \tfrac12 & 0 & 1 \end{array} \right\}^2
\nonumber \\
&& \times
\left\{ \begin{array}{ccc} j & F & \tfrac12 \\ F' & \tfrac12 & 1 \end{array} \right\}^2
 \sum_{q=-1}^1 \left| \langle F'M'_F,1q|FM_F\rangle \right|^2,
\end{eqnarray}
where the last object is a Clebsch-Gordan coefficient. A lengthy but straightforward calculation gives the final probabilities for the 1s states starting from excitation to 3p $m_l=0$:
\begin{eqnarray}
P(F'=0,M'_F=0) &=& \tfrac14, 
\nonumber \\
P(F'=1,M'_F=-1) &=& P(F'=1,M'_F=1) = \tfrac{5}{24}, ~~~ {\rm and} 
\nonumber \\
P(F'=1,M'_F=0) &=& \tfrac13. 
\end{eqnarray}
The polarization of the final atomic state in this case is given by Eq.~(13) of Ref.~\cite{Venumadhav}
\begin{equation}
{\mathcal P}_{20}^{\rm final} = \frac1{\sqrt2} [ P(F'=1,M'_F=1) - 2P(F'=1,M'_F=0) + P(F'=1,M'_F=-1)]
  = -\frac1{12\sqrt2}.
\end{equation}
Note that this is the polarization for pure excitation to 3p $m_l=0$, i.e.\ if $\delta_p=\tfrac23$; since the final polarization must be proportional to $\delta_p$ given Eq.~(\ref{eq:Pratio}), we must have
\begin{equation}
{\mathcal P}_{20}^{\rm final} = -\frac1{8\sqrt2} \delta_p
= \frac1{24\sqrt{40\pi}}
\frac{1}{1-e^{-h\nu_{{\rm H}\alpha}/k_{\rm B}T_\gamma}}
\frac{h\nu_{{\rm H}\alpha}}{k_{\rm B}T_\gamma} \Theta_{20}.
\label{eq:x1Pa}
\end{equation}
Finally, the rate of sourcing polarized atoms through 2s$\rightarrow$3p$\rightarrow$1s scattering is ${\mathcal P}_{20}^{\rm final}$ times the rate at which atoms scatter through this channel (in units of atoms per available H nucleus per unit time) is:
\begin{equation}
[{\rm Rate\,of\,}2s\rightarrow3p\rightarrow1s] = x_{2s} 3A_{3p\rightarrow 2s} \frac{1}{e^{h\nu_{{\rm H}\alpha}/k_{\rm B}T_\gamma}-1}
\frac{A_{3p\rightarrow 1s}}{A_{3p\rightarrow 1s} + A_{3p\rightarrow 2s}/(1-e^{-h\nu_{{\rm H}\alpha}/k_{\rm B}T_\gamma})},
\label{eq:x1Pb}
\end{equation}
where the factor of 3 comes from the ratio of statistical weights of 3p:2s; the first exponential factor is the blackbody phase space density of exciting photons; and the last factor is the branching fraction for 3p$\rightarrow$1s. Finally, multiplying Eq.~(\ref{eq:x1Pa}) by Eq.~(\ref{eq:x1Pb}), and taking the Wien limit $e^{-h\nu_{{\rm H}\alpha}/k_{\rm B}T_\gamma}\ll 1$ (relevant for H$\alpha$ at the surface of last scattering), we get the source term in Eq.~(\ref{eq:rely}):
\begin{equation}
S_{20} = \frac1{16\sqrt{10\pi}} x_{2s} 
\frac{ A_{3p\rightarrow 2s} A_{3p\rightarrow 1s}}{A_{3p\rightarrow 1s} + A_{3p\rightarrow 2s}}
\frac{h\nu_{{\rm H}\alpha}}{k_{\rm B}T_\gamma} e^{-h\nu_{{\rm H}\alpha}/k_{\rm B}T_\gamma} \Theta_{20}.
\end{equation}
(Note that since the choice of $z$-axis did not matter, this equation must hold for all $m=-2,-1,0,1,2$.)

\section{Supplemental information for line of sight of photon-photon scattering}
\label{ap:integrals}

Here we show the exact form of the integrals $I_1$, $I_2$, $I_3$ and $I_4$ used for the line of sight suppression computation of the circular polarization from photon-photon scattering in \S\ref{sec:photon}. Also, we show the expansion of the 4-point functions into 2-point functions used for obtaining the $J$ integrals.   

We write the 2-point function of the CMB circular polarization $\tilde V (\boldsymbol{k}_\perp)$, Eq.~(\ref{eq:lsvv}), as four contributions from different combinations of the 4-point functions of the curvature perturbations involved. Also, we make use of Eq.~(\ref{eq:lsvr}) to get
{\allowdisplaybreaks
\begin{eqnarray}
I_1  &=& \frac{\bar{A}^2}{4} \int \frac{d^3 \boldsymbol{k}_1}{2\pi} \int \frac{d^3 \boldsymbol{k}_2}{2\pi} \int ds \int \frac{d^3 \boldsymbol{k}'_1}{2\pi} \int \frac{d^3 \boldsymbol{k}'_2}{2\pi} \int ds' \, (1+z)^4 \, (1+z)'^4 \, \exp{-i k'_{1,\parallel}s' + i k_{1,\parallel} s}  \nonumber \\* && \times T^{\rm E}_{20}(c\,\eta_{\rm LSS} + s, k_1)\, T^{\rm E}_{20}(c\,\eta_{\rm LSS}, k_2) \, T^{\rm E}_{20}(c\,\eta_{\rm LSS} + s', k_1')\, T^{\rm E}_{20}(c\,\eta_{\rm LSS}, k_2') \,Y^*_{22} (\hat{\boldsymbol k}_2) Y_{22} (\hat{\boldsymbol k}_1) Y_{22} (\hat{\boldsymbol k}'_2) \nonumber \\* \label{eq:ai1}
&& \times Y^*_{22} (\hat{\boldsymbol k}'_1) \langle \zeta (\boldsymbol{k}_1) \zeta^* (\boldsymbol{k}_2) \zeta^* (\boldsymbol{k}'_1) \zeta (\boldsymbol{k}'_2) \rangle  \delta^2 (\boldsymbol{k}_\perp + \boldsymbol{k}_{1,\perp} - \boldsymbol{k}_{2,\perp}) \delta^2 (\boldsymbol{k}'_{2,\perp} - \boldsymbol{k}'_{1,\perp} - \boldsymbol{k}'_\perp),   \\ \label{eq:ai2}
I_2  &=&- \frac{\bar{A}^2}{4} \int \frac{d^3 \boldsymbol{k}_1}{2\pi} \int \frac{d^3 \boldsymbol{k}_2}{2\pi} \int ds \int \frac{d^3 \boldsymbol{k}'_1}{2\pi} \int \frac{d^3 \boldsymbol{k}'_2}{2\pi} \int ds' \, (1+z)^4 \, (1+z)'^4 \, \exp{i k'_{1,\parallel}s' + i k_{1,\parallel} s}  \nonumber \\* && \times T^{\rm E}_{20}(c\,\eta_{\rm LSS} + s, k_1)\, T^{\rm E}_{20}(c\,\eta_{\rm LSS}, k_2) \, T^{\rm E}_{20}(c\,\eta_{\rm LSS} + s', k_1')\, T^{\rm E}_{20}(c\,\eta_{\rm LSS}, k_2') \,Y^*_{22} (\hat{\boldsymbol k}_2) Y_{22} (\hat{\boldsymbol k}_1) Y_{22}^* (\hat{\boldsymbol k}'_2) \nonumber \\*
&& \times Y_{22} (\hat{\boldsymbol k}'_1) \langle \zeta (\boldsymbol{k}_1) \zeta^* (\boldsymbol{k}_2) \zeta (\boldsymbol{k}'_1) \zeta^* (\boldsymbol{k}'_2) \rangle  \delta^2 (\boldsymbol{k}_\perp + \boldsymbol{k}_{1,\perp} - \boldsymbol{k}_{2,\perp}) \delta^2 (\boldsymbol{k}'_{1,\perp} - \boldsymbol{k}'_{2,\perp} - \boldsymbol{k}'_\perp),  \\ \label{eq:ai3}
I_3  &=&- \frac{\bar{A}^2}{4} \int \frac{d^3 \boldsymbol{k}_1}{2\pi} \int \frac{d^3 \boldsymbol{k}_2}{2\pi} \int ds \int \frac{d^3 \boldsymbol{k}'_1}{2\pi} \int \frac{d^3 \boldsymbol{k}'_2}{2\pi} \int ds' \, (1+z)^4 \, (1+z)'^4 \, \exp{-i k'_{1,\parallel}s' - i k_{1,\parallel} s}  \nonumber \\* &&  \times T^{\rm E}_{20}(c\,\eta_{\rm LSS} + s, k_1)\, T^{\rm E}_{20}(c\,\eta_{\rm LSS}, k_2) \, T^{\rm E}_{20}(c\,\eta_{\rm LSS} + s', k_1')\, T^{\rm E}_{20}(c\,\eta_{\rm LSS}, k_2') \,Y_{22} (\hat{\boldsymbol k}_2) Y^*_{22} (\hat{\boldsymbol k}_1) Y_{22} (\hat{\boldsymbol k}'_2) \nonumber \\*
&& \times Y^*_{22} (\hat{\boldsymbol k}'_1) \langle \zeta^* (\boldsymbol{k}_1) \zeta (\boldsymbol{k}_2) \zeta^* (\boldsymbol{k}'_1) \zeta (\boldsymbol{k}'_2) \rangle  \delta^2 (\boldsymbol{k}_\perp + \boldsymbol{k}_{2,\perp} - \boldsymbol{k}_{1,\perp}) \delta^2 (\boldsymbol{k}'_{2,\perp} - \boldsymbol{k}'_{1,\perp} - \boldsymbol{k}'_\perp), ~{\rm and} \\ \label{eq:ai4}
I_4  &=& \frac{\bar{A}^2}{4} \int \frac{d^3 \boldsymbol{k}_1}{2\pi} \int \frac{d^3 \boldsymbol{k}_2}{2\pi} \int ds \int \frac{d^3 \boldsymbol{k}'_1}{2\pi} \int \frac{d^3 \boldsymbol{k}'_2}{2\pi} \int ds' \, (1+z)^4 \, (1+z)'^4 \, \exp{i k'_{1,\parallel}s' - i k_{1,\parallel} s}  \nonumber \\* && \times T^{\rm E}_{20}(c\,\eta_{\rm LSS} + s, k_1)\, T^{\rm E}_{20}(c\,\eta_{\rm LSS}, k_2) \, T^{\rm E}_{20}(c\,\eta_{\rm LSS} + s', k_1')\, T^{\rm E}_{20}(c\,\eta_{\rm LSS}, k_2') \,Y_{22} (\hat{\boldsymbol k}_2) Y^*_{22} (\hat{\boldsymbol k}_1) Y_{22}^* (\hat{\boldsymbol k}'_2) \nonumber \\*
&& \times Y_{22} (\hat{\boldsymbol k}'_1) \langle \zeta^* (\boldsymbol{k}_1) \zeta (\boldsymbol{k}_2) \zeta (\boldsymbol{k}'_1) \zeta^* (\boldsymbol{k}'_2) \rangle  \delta^2 (\boldsymbol{k}_\perp + \boldsymbol{k}_{2,\perp} - \boldsymbol{k}_{1,\perp}) \delta^2 (\boldsymbol{k}'_{1,\perp} - \boldsymbol{k}'_{2,\perp} - \boldsymbol{k}'_\perp). 
\end{eqnarray}}

Furthermore, using Wick's theorem the 4-point functions can be expanded in to the terms that connect the two triangles $\boldsymbol{k}_\perp = \boldsymbol{k}_{1} - \boldsymbol{k}_{2}$ and $\boldsymbol{k}'_\perp = \boldsymbol{k}'_1 - \boldsymbol{k}'_2$, which are 
\begin{eqnarray}
\langle \zeta (\boldsymbol{k}_1) \zeta^* (\boldsymbol{k}_2) \zeta^* (\boldsymbol{k}'_1) \zeta (\boldsymbol{k}'_2) \rangle &=&\langle \zeta^* (\boldsymbol{k}_1) \zeta (\boldsymbol{k}_2) \zeta (\boldsymbol{k}'_1) \zeta^* (\boldsymbol{k}'_2) \rangle =
(2 \pi)^6 P_\zeta(k_1)P_\zeta(k_2)  \nonumber \\
&& \label{eq:a4i1-4} \times \left[ \delta^3 (\boldsymbol{k}_1 - \boldsymbol{k}'_1) \delta^3 (\boldsymbol{k}_2 - \boldsymbol{k}'_2) +  \delta^3 (\boldsymbol{k}_1 + \boldsymbol{k}'_2) \delta^3 (\boldsymbol{k}_2 + \boldsymbol{k}'_1)\right]
\end{eqnarray}
and
\begin{eqnarray}
\langle \zeta (\boldsymbol{k}_1) \zeta^* (\boldsymbol{k}_2) \zeta (\boldsymbol{k}'_1) \zeta^* (\boldsymbol{k}'_2) \rangle &=&\langle \zeta^* (\boldsymbol{k}_1) \zeta (\boldsymbol{k}_2) \zeta^* (\boldsymbol{k}'_1) \zeta (\boldsymbol{k}'_2) \rangle =
(2 \pi)^6 P_\zeta(k_1)P_\zeta(k_2) \nonumber \\
&& \label{eq:a4i2-3} \times \left[ \delta^3 (\boldsymbol{k}_1 + \boldsymbol{k}'_1) \delta^3 (\boldsymbol{k}_2 + \boldsymbol{k}'_2) +  \delta^3 (\boldsymbol{k}_1 - \boldsymbol{k}'_2) \delta^3 (\boldsymbol{k}_2 - \boldsymbol{k}'_1)\right]. 
\end{eqnarray}
Thus we obtained the $I$'s integrals and expanded the 4-point functions in terms of the primordial curvature power spectrum. Now, we can finally integrate out the perpendicular component of $k_2$ with the help of the 2D $\delta$-functions to obtain the eight $J$ integrals. The first four of these are
{\allowdisplaybreaks
\begin{eqnarray}
J_1 &=& (2\pi)^2 \frac{\bar{A}^2}{4} \delta^2 (\boldsymbol{k}_\perp - \boldsymbol{k}'_\perp) \int \frac{d^2 \boldsymbol{k}_{1,\perp}}{(2\pi)^2} \int \frac{d k_{1,\parallel}}{2 \pi} \int \frac{d k_{2,\parallel}}{2\pi} |F(k_1, k_{1,\parallel})|^2 \,[T^{\rm E}_{20}(c\,\eta_{\rm LSS}, k_2)]^2  \nonumber \\* \label{eq:lsj1}
&& \times Y^*_{22}(\hat{\boldsymbol k}_2) Y_{22}(\hat{\boldsymbol k}_1) Y_{22}(\hat{\boldsymbol k}_2) Y^*_{22}(\hat{\boldsymbol k}_1) P_\zeta(k_1) P_\zeta(k_2),\\
J_2 & = &  (2\pi)^2 \frac{\bar{A}^2}{4} \delta^2(\boldsymbol{k}_\perp - \boldsymbol{k}'_\perp) \int \frac{d^2 \boldsymbol{k}_{1,\perp}}{(2\pi)^2} \int \frac{d k_{1,\parallel}}{2 \pi} \int \frac{d k_{2,\parallel}}{2\pi} F(k_1,k_{1,\parallel}) F(k_2, k_{2,\parallel}) \nonumber \\* \label{eq:lsj2}
&& \times T^{\rm E}_{20}(\eta_{\rm LSS}, k_2)\, T^{\rm E}_{20}(\eta_{\rm LSS}, k_1) Y^*_{22}(\hat{\boldsymbol k}_2) Y_{22}(\hat{\boldsymbol k}_1) Y^*_{22}(\hat{\boldsymbol k}_2) Y_{22}(\hat{\boldsymbol k}_1) P_\zeta(k_1)P_\zeta(k_2), \\
J_3 & = & - (2\pi)^2 \frac{\bar{A}^2}{4} \delta^2(\boldsymbol{k}_\perp - \boldsymbol{k}'_\perp) \int \frac{d^2 \boldsymbol{k}_{1,\perp}}{(2\pi)^2} \int \frac{d k_{1,\parallel}}{2 \pi} \int \frac{d k_{2,\parallel}}{2\pi} |F(k_1,k_{1,\parallel})|^2 [T^{\rm E}_{20}(\eta_{\rm LSS}, k_2)]^2 \nonumber \\* \label{eq:lsj3}
&& \times Y^*_{22}(\hat{\boldsymbol k}_2) Y_{22}(\hat{\boldsymbol k}_1) Y^*_{22}(\hat{\boldsymbol k}_2) Y_{22}(\hat{\boldsymbol k}_1) P_\zeta(k_1)P_\zeta(k_2), ~{\rm and}\\
J_4 & = & - (2\pi)^2 \frac{\bar{A}^2}{4} \delta^2(\boldsymbol{k}_\perp - \boldsymbol{k}'_\perp) \int \frac{d^2 \boldsymbol{k}_{1,\perp}}{(2\pi)^2} \int \frac{d k_{1,\parallel}}{2 \pi} \int \frac{d k_{2,\parallel}}{2\pi} F(k_1,k_{1,\parallel}) F(k_2, k_{2,\parallel}) \nonumber \\*  \label{eq:lsj4}
&& \times T^{\rm E}_{20}(\eta_{\rm LSS},k_2) \, T^{\rm E}_{20} (\eta_{\rm LSS},k_1)\, Y^*_{22}(\hat{\boldsymbol k}_2) Y_{22}(\hat{\boldsymbol k}_1) Y^*_{22}(\hat{\boldsymbol k}_1) Y_{22}(\hat{\boldsymbol k}_2) P_\zeta(k_1)P_\zeta(k_2),
\end{eqnarray}}
which have $\boldsymbol{k}_{2,\perp} = \boldsymbol{k}_{1,\perp} + \boldsymbol{k}_\perp$. The remaining integrals are
{\allowdisplaybreaks
\begin{eqnarray}
J_5 & = & - (2\pi)^2 \frac{\bar{A}^2}{4} \delta^2(\boldsymbol{k}_\perp - \boldsymbol{k}'_\perp) \int \frac{d^2 \boldsymbol{k}_{1,\perp}}{(2\pi)^2} \int \frac{d k_{1,\parallel}}{2 \pi} \int \frac{d k_{2,\parallel}}{2\pi} |F(k_1,k_{1,\parallel}) |^2 [T^{\rm E}_{20}(\eta_{\rm LSS},k_2)]^2 \nonumber \\*  \label{eq:lsj5}
&& \times Y_{22}(\hat{\boldsymbol k}_2) Y_{22}^*(\hat{\boldsymbol k}_1) Y^*_{22}(\hat{\boldsymbol k}_1) Y_{22}(\hat{\boldsymbol k}_2) P_\zeta(k_1)P_\zeta(k_2), \\
J_6 & = & - (2\pi)^2 \frac{\bar{A}^2}{4} \delta^2(\boldsymbol{k}_\perp - \boldsymbol{k}'_\perp) \int \frac{d^2 \boldsymbol{k}_{1,\perp}}{(2\pi)^2} \int \frac{d k_{1,\parallel}}{2 \pi} \int \frac{d k_{2,\parallel}}{2\pi} F^*(k_1,k_{1,\parallel}) F^*(k_2,k_{2,\parallel}) \nonumber \\* 
&& \times T^{\rm E}_{20}(\eta_{\rm LSS},k_2) T^{\rm E}_{20}(\eta_{\rm LSS},k_1) \label{eq:lsj6} Y_{22}(\hat{\boldsymbol k}_2) Y_{22}^*(\hat{\boldsymbol k}_1) Y_{22}(\hat{\boldsymbol k}_1) Y^*_{22}(\hat{\boldsymbol k}_2) P_\zeta(k_1)P_\zeta(k_2), \\
J_7 & = &  (2\pi)^2 \frac{\bar{A}^2}{4} \delta^2(\boldsymbol{k}_\perp - \boldsymbol{k}'_\perp) \int \frac{d^2 \boldsymbol{k}_{1,\perp}}{(2\pi)^2} \int \frac{d k_{1,\parallel}}{2 \pi} \int \frac{d k_{2,\parallel}}{2\pi} |F(k_1,k_{1,\parallel}) |^2 [T^{\rm E}_{20}(\eta_{\rm LSS},k_2)]^2 \nonumber \\*  \label{eq:lsj7}
&& \times Y_{22}(\hat{\boldsymbol k}_2) Y_{22}^*(\hat{\boldsymbol k}_1) Y_{22}(\hat{\boldsymbol k}_1) Y^*_{22}(\hat{\boldsymbol k}_2) P_\zeta(k_1)P_\zeta(k_2), {\rm ~and} \\
J_8 & = &  (2\pi)^2 \frac{\bar{A}^2}{4} \delta^2(\boldsymbol{k}_\perp - \boldsymbol{k}'_\perp) \int \frac{d^2 \boldsymbol{k}_{1,\perp}}{(2\pi)^2} \int \frac{d k_{1,\parallel}}{2 \pi} \int \frac{d k_{2,\parallel}}{2\pi} F^*(k_1,k_{1,\parallel}) F^*(k_2,k_{2,\parallel}) \nonumber \\* 
&& \times T^{\rm E}_{20}(\eta_{\rm LSS},k_2) T^{\rm E}_{20}(\eta_{\rm LSS},k_1) \label{eq:lsj8} Y_{22}(\hat{\boldsymbol k}_2) Y_{22}^*(\hat{\boldsymbol k}_1) Y^*_{22}(\hat{\boldsymbol k}_1) Y_{22}(\hat{\boldsymbol k}_2) P_\zeta(k_1)P_\zeta(k_2),
\end{eqnarray}}
where $\boldsymbol{k}_{2,\perp} = \boldsymbol{k}_{1,\perp} - \boldsymbol{k}_\perp$.

\comment{
Finally, for the purpose of illustration, we include here the results for the $J$ integrals of a run of our code with $k_\perp = 0.1 \ \textup{Mpc}^{-1}$:
\begin{eqnarray}
J_1 & = & J_7 = (2\pi)^2 \delta(\boldsymbol{k}_\perp - \boldsymbol{k}'_\perp) \frac{A^2 (2\pi^2)}{4 (2 \pi)^4} \left(\frac{15}{32 \pi}\right)^2  1.131644 \times 10^7  ,\\
J_2 & = & (2\pi)^2 \delta(\boldsymbol{k}_\perp - \boldsymbol{k}'_\perp) \frac{A^2 (2\pi^2)}{4 (2 \pi)^4} \left(\frac{15}{32 \pi}\right)^2  (-6.021963 + i\ 3.770973) \times 10^4 ,\\
J_3 & = & J_5 = (2\pi)^2 \delta(\boldsymbol{k}_\perp - \boldsymbol{k}'_\perp) \frac{A^2 (2\pi^2)}{4 (2 \pi)^4} \left(\frac{15}{32 \pi}\right)^2 2.265793 \times 10^5,\\
J_4 & = & (2\pi)^2 \delta(\boldsymbol{k}_\perp - \boldsymbol{k}'_\perp) \frac{A^2 (2\pi^2)}{4 (2 \pi)^4} \left(\frac{15}{32 \pi}\right)^2  (1.092155 - i \ 1.103256) \times 10^{5},\\
J_6 & = & (2\pi)^2 \delta(\boldsymbol{k}_\perp - \boldsymbol{k}'_\perp) \frac{A^2 (2\pi^2)}{4 (2 \pi)^4} \left(\frac{15}{32 \pi}\right)^2  (1.092155 + i \ 1.103256) \times 10^5,\\
J_8 & = & (2\pi)^2 \delta(\boldsymbol{k}_\perp - \boldsymbol{k}'_\perp) \frac{A^2 (2\pi^2)}{4 (2 \pi)^4} \left(\frac{15}{32 \pi}\right)^2 (-6.021963 - i \ 3.770973) \times 10^4,
\end{eqnarray}
which translates into a power spectrum   $P_{V} = \frac{A^2 (2\pi^2)}{4 \, (2 \pi)^4} \left(\frac{15}{32 \pi}\right)^2 (J_1 + J_2 - J_3 - J_4 - J_5 - J_6 + J_7 + J_8 \sim  3.58 \times 10^{-27} \  \textup{Mpc}^2$.
}

\section{Supplemental information for plasma delay}
\label{ap:integrals2}
In this section we present the integrals involved in the detailed calculation from section \S\S\ref{ssec:plasma}. Concretely, the integrals  stem from Eqs.~(\ref{eq:ITDef},\ref{eq:ITBDef}). 

We start with $I_T$, which we expanded as $I_T = I_1 - I_2 - I_3 + I_4 - I_5 + I_6 -I_7 -I_8$. The integrals are
{\allowdisplaybreaks
\begin{eqnarray}
I_1 &=& \frac{\hbar^2}{4 \pi^3 c^3 k_{\rm B} T \omega_A^2} \int_0^\infty \frac{ e^{\hbar\omega^{\rm R}_B/k_{\rm B}T} {\omega^{\rm R}_B}^5  d\omega^{\rm R}_B}{(e^{\hbar\omega^{\rm R}_B/k_{\rm B}T} -1)^2 ( \omega^{\rm R}_B + i \epsilon)^2 (\omega^{\rm R}_B +\omega_A +i\epsilon)} \textup{,}\\
I_2 &=&\frac{\hbar^2}{4 \pi^3 c^3 k_{\rm B}T \omega_A} \int_0^\infty \frac{ e^{\hbar\omega^{\rm R}_B/k_{\rm B}T} {\omega^{\rm R}_B}^5 d\omega^{\rm R}_B }{(e^{\hbar\omega^{\rm R}_B/k_{\rm B}T} -1)^2(\omega^{\rm R}_B -i \epsilon)^2 (\omega^{\rm R}_B - \omega_A -i \epsilon)^2 } \textup{,}\\
I_3 &=& \frac{\hbar^2}{4 \pi^3 c^3 k_{\rm B}T \omega_A^2} \int_0^\infty \frac{ e^{\hbar\omega^{\rm R}_B/k_{\rm B}T}{\omega^{\rm R}_B}^5 d\omega^{\rm R}_B }{(e^{\hbar\omega^{\rm R}_B/k_{\rm B}T} -1)^2(\omega^{\rm R}_B + i \epsilon) (\omega^{\rm R}_B - \omega_A -i \epsilon)^2 } \textup{,}\\ 
I_4 &=& \frac{\hbar^2}{4 \pi^3 c^3 k_{\rm B}T \omega_A} \int_0^\infty \frac{ e^{\hbar\omega^{\rm R}_B/k_{\rm B}T}{\omega^{\rm R}_B}^5 d\omega^{\rm R}_B }{(e^{\hbar\omega^{\rm R}_B/k_{\rm B}T} -1)^2(\omega^{\rm R}_B + i \epsilon)^2 (\omega^{\rm R}_B + \omega_A +i \epsilon)^2 } \textup{,}\\
I_5 &=& \frac{\hbar^2}{4 \pi^3 c^3 k_{\rm B} T \omega^2_A} \int_0^\infty \frac{e^{\hbar\omega^{\rm R}_B/k_{\rm B}T} {\omega^{\rm R}_B}^5 d\omega^{\rm R}_B }{(e^{\hbar\omega^{\rm R}_B/k_{\rm B}T} -1)^2(\omega^{\rm R}_B - i \epsilon) (\omega^{\rm R}_B + \omega_A +i \epsilon)^2 } \textup{,}\\
I_6 &=& \frac{\hbar^2}{4 \pi^3 c^3 k_{\rm B} T \omega^2_A} \int_0^\infty \frac{ e^{\hbar\omega^{\rm R}_B/k_{\rm B}T}{\omega^{\rm R}_B}^5 d\omega^{\rm R}_B }{(e^{\hbar\omega^{\rm R}_B/k_{\rm B}T} -1)^2(\omega^{\rm R}_B + i \epsilon)^2 (\omega^{\rm R}_B - \omega_A -i \epsilon)} \textup{,} \\
I_7 &=& \frac{ \hbar^2}{2 \pi^3 c^3 k_{\rm B} T \omega^2_A} \int_0^\infty \frac{ e^{\hbar\omega^{\rm R}_B/k_{\rm B}T}{\omega^{\rm R}_B}^5 d\omega^{\rm R}_B }{(e^{\hbar\omega^{\rm R}_B/k_{\rm B}T} -1)^2(\omega^{\rm R}_B + i \epsilon)^ (\omega^{\rm R}_B + \omega_A +i \epsilon)^2} \textup{,~and} \\
I_8 &=& \frac{\hbar^2}{2 \pi^3 c^3 k_{\rm B} T \omega^2_A} \int_0^\infty \frac{ e^{\hbar\omega^{\rm R}_B/k_{\rm B}T}{\omega^{\rm R}_B}^5 d\omega^{\rm R}_B }{(e^{\hbar\omega^{\rm R}_B/k_{\rm B}T} -1)^2(\omega^{\rm R}_B - i \epsilon) (\omega^{\rm R}_B - \omega_A -i \epsilon)^2} \textup{,}
\end{eqnarray}}
where the exponential factor comes from the black body spectrum.

Similarly, for the case of $I^B_T = I^B_1 - I^B_2 + I^B_3 + I^B_4$ we have
{\allowdisplaybreaks
\begin{eqnarray}
I^B_1 &=& \frac{\hbar^2}{4 \pi^3 c^3 k_{\rm B} T \omega^3_A} \int_0^\infty \frac{ e^{\hbar\omega^{\rm R}_B/k_{\rm B}T} {\omega^{\rm R}_B}^4  d\omega^{\rm R}_B}{(e^{\hbar\omega^{\rm R}_B/k_{\rm B}T} -1)^2 (\omega^{\rm R}_B +\omega_A +i\epsilon)} \textup{,}\\
I^B_2 &=& \frac{\hbar^2}{4 \pi^3 c^3 k_{\rm B} T \omega^3_A} \int_0^\infty \frac{ e^{\hbar\omega^{\rm R}_B/k_{\rm B}T}{\omega^{\rm R}_B}^4  d\omega^{\rm R}_B}{(e^{\hbar\omega^{\rm R}_B/k_{\rm B}T} -1)^2 (\omega^{\rm R}_B -\omega_A -i\epsilon)} \textup{,}\\
I^B_3 &=& \frac{\hbar^2}{4 \pi^3 c^3 k_{\rm B} T} \int_0^\infty \frac{ e^{\hbar\omega^{\rm R}_B/k_{\rm B}T} {\omega^{\rm R}_B}^4  d\omega^{\rm R}_B}{(e^{\hbar\omega^{\rm R}_B/k_{\rm B}T}-1)^2 (\omega_B^{\rm R} +i \epsilon)(\omega_B^{\rm R} -i \epsilon)(\omega^{\rm R}_B +\omega_A +i\epsilon)^2} \textup{,~and}\\
I^B_4 &=& \frac{\hbar^2}{4 \pi^3 c^3 k_{\rm B} T} \int_0^\infty \frac{ e^{\hbar\omega^{\rm R}_B/k_{\rm B}T} {\omega^{\rm R}_B}^4  d\omega^{\rm R}_B}{(e^{\hbar\omega^{\rm R}_B/k_{\rm B}T} -1)^2 (\omega^{\rm R}_B -i\epsilon)(\omega^{\rm R}_B +i \epsilon) (\omega^{\rm R}_B -\omega_A -i\epsilon)^2} \textup{.}
\end{eqnarray}}
In general the strategy for handling such integrals involved a combination of integration by parts and principal value while being careful about the redshift dependence of both temperature and frequency of CMB photons. Finally, one should take the limit as $\epsilon \rightarrow 0 $.

\section{Anisotropic 21 cm radiation in expanding media}
\label{ap:vgrad}

Peculiar velocities and the expansion of the universe modify the emission and absorption process for photons in a narrow spectral line such as 21 cm. In what follows we want to illustrate the way these particular pieces are present in the calculation done in \S \ref{sec:21z20}. The purpose here is to provide a simplified calculation with only the essential physical ingredients, and with the minimal amount of formalism. For the full detailed calculation an interested reader should consult Ref.~\cite{Venumadhav}.

The key physics to comprehend here are what happen to a spectral line in a moving medium, and how velocity gradients imply anisotropic 21 cm radiation. For simplicity, we will compute the alignment of the hydrogen spins for the special case of $\hat{\boldsymbol k}$ in the $\hat{\boldsymbol z}$ direction (other choices are related by symmetry).
Spectral lines are usually represented by $\delta$-functions; nevertheless, there are broadening effects, and these must be taken into account when doing line radiative transfer to obtain a finite result. For 21 cm radiation, the most important local broadening source is Doppler broadening due to the thermal motion of the hydrogen atoms, and the line center is shifted in accordance with the (position-dependent) bulk velocity of the gas. We work in the optically thin limit $\tau\ll 1$, where we can approximate the emission of 21 cm radiation as isotropic, and then obtain the direction-dependent absorption using line radiative transfer. (It is true that the resulting polarized atoms will produce anisotropic 21 cm emission, which in turn imprints an additional anisotropy in the absorption, but this is at the next higher order in $\tau$.)

Without loss of generality, let us consider the absorption of 21 cm radiation by gas at the origin, and work in the reference frame where the bulk velocity of the gas at the origin is zero. We need to know the cross section for absorption of radiation that was emitted by gas at a position ${\boldsymbol r}$. The emitting gas has a bulk velocity ${\boldsymbol v}$ given by $v_i = (H \delta_{ij} + \partial_j v_{{\rm p},i}) r_j$ and random velocities of $\sigma = \sqrt{k_{\rm B}T_{\rm kin}/m_{\rm H}}$ (root-mean-square per axis, for a hydrogen atom, at kinetic temperature $T_{\rm kin}$). This cross section (units: m$^2$) is given by
\begin{equation}
\label{eq:apcrossabs}
\sigma_{\rm cs} = N \frac{c}{\nu_{21}} \frac{1}{\sqrt{2\pi (2\sigma^2)}} \exp\left\{{-\frac{l^2}{2(2\sigma^2)}[(H \delta_{ij} + \partial_j v_{{\rm p},i})\hat{n}_i \hat{n}_j ]^2}\right\} \, \textup{,}
\end{equation}
where $N = 3c^2A_{\rm hf}/(8 \pi \nu_{21}^2) = \int \sigma_{\rm abs}(\nu)\,d\nu$ is the normalization (units: m$^2$\,s$^{-1}$), $A_{\rm hf}$ is the hyperfine transition Einstein coefficient, and we have broken the position into a path length ($l$) and a direction $\hat{\boldsymbol n}$: ${\boldsymbol r} = l\hat{\boldsymbol n}$. The extra factor of two in the variance comes from taking into account that there is thermal broadening in both the emission and absorption processes, i.e., $ \sigma_{\rm em}^2 + \sigma_{\rm abs}^2 = 2 \sigma^2$. The factor of $c/\nu_{21}$ is the conversion factor from frequency to velocity (since the Gaussian is normalized when integrated over velocity).

On the other hand, the rate of absorption per unit volume per unit solid angle $d\Gamma_{\rm abs}/d\Omega$ (units: cm$^{-3}$ s$^{-1}$) for an absorption cross section $\sigma_{\rm cs}$ is
\begin{equation}
\label{eq:apgrate}
\frac{d\Gamma_{\rm abs}}{d \Omega}  \approx  \frac{1}{4\pi}\int_0^\infty dl \left(n_{\rm HI} x_{0} -\frac{n_{\rm HI} x_{1}}{3}\right) \Gamma_{\rm em}  \sigma_{cs} \, \textup{,}
\end{equation}
where $x_i$ is the fraction of hydrogen atoms in the $i$ state (either ground state $F=0$ or excited state $F=1$). The second term takes into account stimulated emission (it is treated as negative absorption). To lowest order in $T_\star$, we have $x_0 = \frac14 + \frac{3}{16}T_\star/T_s$ and $x_1 = \frac34 - \frac3{16}T_\star/T_s$.\footnote{These are the solutions to the total abundance constraint $x_0+x_1=1$ and the Boltzmann condition $x_1/x_0 = 3e^{-T_\star/T_s}$.} The net emission rate (units: cm$^{-3}$ s$^{-1}$) is approximately given by
\begin{equation}
\label{eq:apgem}
\Gamma_{\rm em} 
= n_{\rm HI} A_{\rm hf} \left[ x_1 \left( 1 + \frac{1}{e^{T_\star/T_{\rm CMB}}-1} \right) - 3x_0 \frac{1}{e^{T_\star/T_{\rm CMB}}-1} \right]
\approx \frac{3}{4}n_{\rm HI}A_{\rm hf}\left(1 - \frac{T_{\rm CMB}}{T_s}\right) \, \textup{,}
\end{equation}
where we included spontaneous and stimulated (by the CMB) emission minus absorption, and carried out the usual algebraic simplifications.

We now make the assumption that a particular hydrogen atom only sees a small part of the total perturbations, i.e., we consider only modes of wavelengths much larger than the Jeans length, which in this case is of order $\sigma/H$. (This will be true for the modes where the gas is actually clustering with the dark matter.) We further recall that for linear growth, the velocity gradient is related to $\delta$ via $\partial_j v_{{\rm p},i} = - H \hat k_i \hat k_j \,\delta$. Finally, we recall that the Sobolev optical depth is
\begin{equation}
\tau = \frac{3c^3A_{\rm hf}}{8\pi \nu_{21}^3 H} \left(n_{\rm HI} x_{0} -\frac{n_{\rm HI} x_{1}}{3}\right) = \frac NH \frac{c}{\nu_{21}} \left(n_{\rm HI} x_{0} -\frac{n_{\rm HI} x_{1}}{3}\right).
\end{equation}
With all of these replacements, Eq.~(\ref{eq:apgrate}) becomes
\begin{eqnarray}
\frac{d\Gamma_{\rm abs}}{d\Omega} &=&
\frac{1}{4\pi}\int_0^\infty dl \left(n_{\rm HI} x_{0} -\frac{n_{\rm HI} x_{1}}{3}\right) \frac{3}{4}n_{\rm HI}A_{\rm hf}\left(1 - \frac{T_{\rm CMB}}{T_s}\right) N \frac{c}{\nu_{21}} \frac{1}{\sqrt{2\pi (2\sigma^2)}}
\nonumber \\ && \times \exp\left\{{-\frac{l^2}{2(2\sigma^2)}[(H \delta_{ij} + \partial_j v_{{\rm p},i})\hat{n}_i \hat{n}_j ]^2}\right\}
\nonumber \\
&=&
\frac{1}{8\pi} \left(n_{\rm HI} x_{0} -\frac{n_{\rm HI} x_{1}}{3}\right) \frac{3}{4}n_{\rm HI}A_{\rm hf}\left(1 - \frac{T_{\rm CMB}}{T_s}\right) N \frac{c}{\nu_{21}} 
\frac1{(H \delta_{ij} + \partial_j v_{{\rm p},i})\hat{n}_i \hat{n}_j }
\nonumber \\
&=&
\frac{1}{8\pi} \tau \frac{3}{4}n_{\rm HI}A_{\rm hf}\left(1 - \frac{T_{\rm CMB}}{T_s}\right)
\frac1{( \delta_{ij} + \partial_j v_{{\rm p},i}/H)\hat{n}_i \hat{n}_j }
\nonumber \\
& \approx & \frac{3}{4\pi} \frac{\tau n_{\rm HI} A_{\rm hf}}{8} \left(1 - \frac{T_{\rm CMB}}{T_s}\right)\Big(1 + \delta (\hat{\boldsymbol k}\cdot \hat{\boldsymbol n})^2\Big) \, \textup{.}
\end{eqnarray}
(The last line uses the first-order expansion of the reciprocal.)
With the help of Eq.~(73) from \cite{Venumadhav} and with $\hat{\boldsymbol n}$ in the $z$ axis we expand the dot product and obtain the perturbation to the absorption rate
\begin{equation}
\label{eq:apaniabs}
\frac{d \Gamma_{\rm abs}}{d \Omega_n} \Big|_{\rm pert} = \frac{3\tau n_{\rm HI} A_{\rm hf}}{8} \left(1 - \frac{T_{\rm CMB}}{T_s}\right)\cos^2 \theta\, \delta  \sqrt{\frac{4\pi}{5}} Y_{20}(\hat{\boldsymbol k}) \, \textup{.}
\end{equation}

We next need the excitation rate to each of the three excited states, $m=-1$, $m=0$, and $m=+1$. To obtain this, we integrate the anisotropic absorption rate with the appropriate electric dipole antenna patterns. In the case of $m=0$, the electric dipole is aligned with the $z$ axis and we have
\begin{equation}
\label{eq:apbeamm0}
\Gamma_{\rm abs}(m=0) = \frac13 \int \frac{d \Gamma_{\rm abs}}{d\Omega}\,\frac{3}{8\pi}\sin^2 \theta\, d\Omega = \frac{n_{\rm HI} \tau A_{\rm hf}}{8} \left(1 - \frac{T_{\rm CMB}}{T_s}\right) \delta  \sqrt{\frac{4\pi}{5}} Y_{20}(\hat{\boldsymbol k}) \frac{1}{5} \, \textup{.}
\end{equation}
Here $\frac3{8\pi}\sin^2\theta$ is the beam pattern (normalized to integrate to 1), and the factor of $\frac13$ is introduced because for an isotropic distribution $\frac13$ of the excitations go to $m=0$. Similarly for $m=\pm 1$ the angular integrals give a factor of $\frac25$.

Finally, we need to take the excitation rate to each state, and determine the resulting spin alignment. From Eq.~(13) in \cite{Venumadhav}, the ``20'' polarization moment is ${\mathcal P}_{20} = \frac{1}{\sqrt{2}}(\rho_{11} - 2\rho_{00} + \rho_{-1-1})$. There is a conversion from excitation rate (cm$^{-3}$ s$^{-1}$) to dimensionless probability of $t_{\rm life}/n_{\rm HI}$, where $t_{\rm life}$ is the lifetime of the atom in the excited state before it is disturbed. Thus:
\begin{equation}
\label{eq:apalip}
{\mathcal P}_{20} \sim \frac{1}{\sqrt{2}} \frac{t_{\rm life}}{n_{\rm HI}} \Big[\Gamma_{\rm abs}(m=+1)  -2\Gamma_{\rm abs}(m=0) + \Gamma_{\rm abs}(m=-1)\Big] \, \textup{,}
\end{equation}
where the lifetime is given by the stimulated emission adjusting for small corrections from both collisions and Lyman-$\alpha$ pumping, e.g. $t_{\rm life}^{-1} = A_{\rm hf} \frac{T_{\rm CMB}}{T_*} (1 + x_\alpha + x_c)$.\footnote{The factors $x_{\alpha,(2)}$ and $x_{c,(2)}$ differ from $x_\alpha$ and $x_c$ because they take into account the fact that Lyman-$\alpha$ scattering or collisions can change the polarization state of an excited hydrogen atom, while leaving it in the excited hyperfine state. This is treated in detail in Ref.\cite{Venumadhav} and will not be repeated here.}

Hence combining the absorption rates into Eq.~(\ref{eq:apalip}) we get
\begin{equation}
\label{eq:app20}
{\mathcal P}_{20}(\boldsymbol{k}) =  \frac{1}{20\sqrt{2}} \frac{\tau }{1 + x_\alpha + x_c} \frac{T_*}{T_\gamma} \left(1 - \frac{T_\gamma}{T_s}\right) \delta(\boldsymbol{k})  \sqrt{\frac{4\pi}{5}} Y_{20}(\hat{\boldsymbol k}) \, \textup{.}
\end{equation}
As expected, the velocity gradient enters Eq.~(\ref{eq:21zali}), but does so through the factor of density perturbations since we used the linear relation between velocity and density modes.

\acknowledgments
We thank Xiao Fang, Benjamin Buckman, Daniel Martens, and Todd Thompson for useful discussions and comments. PMC is grateful to Thomas Tram for useful suggestions and guidance with CLASS. 
PMC is supported by the Simons Foundation.
CMH is supported by the Simons Foundation, the US Department of Energy, the Packard Foundation, and NASA.

\bibliographystyle{JHEP.bst}
\bibliography{birefringence.bib}

\end{document}